\documentclass[prd,altaffilletter,twocolumn,superscriptaddress,floatfix,nofootinbib]{revtex4-2} 
\usepackage{amsfonts,amsmath,graphicx,bm}

\usepackage{tikz-feynman}
\usepackage{ifpdf}
\usepackage{multirow}
\usepackage[colorlinks=true, linkcolor=red, urlcolor=red, citecolor=red]{hyperref}
\usepackage{url}
\usepackage{todonotes}
\usepackage{float}
\usepackage{graphicx}
\usepackage{mathtools}
\usepackage[yyyymmdd,hhmmss]{datetime}
\usepackage{color}
\usepackage{bbold}

\newcommand{\VEC}[1]{{\bf \bm{#1}}} 

\def\today{\number\day\space\ifcase\month\or
January\or February\or March\or April\or May\or June\or
July\or August\or September\or October\or November\or December\fi
\space\number\year}
\newcount\mins \newcount\hours
\def\now{\hours=\time \mins=\time
	\divide\hours by60 \multiply\hours by60 \advance\mins by-\hours
	\divide\hours by60 
	\number\hours:\ifnum\mins<10 0\fi\number\mins }

\usepackage{footmisc}

\usepackage{braket}
\usepackage{slashed}

\newcommand{\glasgow}{SUPA, School of Physics and Astronomy, University of Glasgow, Glasgow, G12 8QQ, UK}
\newcommand{\cornell}{Laboratory of Elementary Particle Physics, Cornell University, Ithaca, New York 14853, USA}

\begin{document}
\title{Precise determination of decay rates for $\eta_c \to \gamma \gamma$, $J/\psi \to \gamma  \eta_c$ and $J/\psi \to \eta_c e^+e^-$ from lattice QCD}

\author{Brian~\surname{Colquhoun}} 
\email[]{Brian.Colquhoun@glasgow.ac.uk}

\author{Laurence~J.~\surname{Cooper}} 

\author{Christine~T.~H.~\surname{Davies}} 
\email[]{Christine.Davies@glasgow.ac.uk}
\affiliation{\glasgow}%

\author{G.~Peter~\surname{Lepage}} 
\affiliation{\cornell}

\collaboration{HPQCD Collaboration}
\email[URL: ]{http://www.physics.gla.ac.uk/HPQCD}


\newcommand{\fitF}{F (0,0) = 0.08793(29)\, \text{GeV}^{-1}}
\newcommand{\finalF}{F (0,0) = 0.08793(29)_{\text{fit}}(26)_{\text{syst}} \, \text{GeV}^{-1}}
\newcommand{\finalError}{0.4}
\newcommand{\finalWidthEtacGammaGamma}{\Gamma (\eta_c \to \gamma \gamma) = 6.788(45)_{\text{fit}}(41)_{\text{syst}} \: \mathrm{keV}}
\newcommand{\finalBrEtacGammaGamma}{\mathcal{B} (\eta_c \to \gamma \gamma) = 2.121(14)_{\text{fit}}(13)_{\text{syst}}(46)_{\text{expt}} \times 10^{-4}}
\newcommand{\finalWidthEtacGammaGammaUncertainty}{0.9\%}
\newcommand{\fitrat}{0.4786(57)_{\text{fit}}(14)_{\text{syst}}\, }
\newcommand{\diffpdgsigma}{4.6}


\newcommand{\fitVhat}{\hat{V} (0) = 1.8649(73)}
\newcommand{\fitVhatUncertainty}{0.4\%}
\newcommand{\finalVhat}{\hat{V} (0) = 1.8649(73)_{\text{fit}}(75)_{\text{syst}}}
\newcommand{\finalVhatUncertainty}{0.56\%}

\newcommand{\finalWidthJpsiToGammaEtac}{\Gamma (J/\psi \to \gamma \eta_c) = 2.219(17)_{\text{fit}}(18)_{\text{syst}}(24)_{\text{expt}}(4)_{\text{QED}} \; \mathrm{keV}}
\newcommand{\finalGammaUncertainty}{1.6\%}
\newcommand{\finalBrJpsiToGammaEtac}{\text{Br} (J/\psi \to \gamma \eta_c) = 2.40(3)_{\text{latt}}(5)_{\text{expt}}\% .}

\newcommand{\finalRee}{R_{ee\gamma} = 0.00607958(34)}
\newcommand{\ReenoVhat}{0.006078421(43) }

\newcommand{\finalDiffReeFromVhatUnity}{$0.02\%$}

\newcommand{\finalDecayWidthJpsiEtacee}{\Gamma (J/\psi \to \eta_c e^+ e^-) = 0.01349(15)_{\text{latt}}(15)_{\text{expt}}(13)_{\text{QED}} \; \mathrm{keV}}
\newcommand{\finalDecayBrJpsiEtacee}{\text{Br} (J/\psi \to \eta_c e^+ e^-) = 1.457(16)_{\text{latt}}(15)_{\text{QED}}(31)_{\text{expt}} \times 10^{-4} }
\newcommand{\finalshif}{0.893(5)_{\hat{V}}(11)_{R_{fF}}}


\begin{abstract}
\noindent We calculate the decay rates for $\eta_c \to \gamma \gamma$, $J/\psi \to \gamma \eta_c$ and $J/\psi \to \eta_c e^+e^-$ in lattice QCD with $u$, $d$, $s$ and $c$ quarks in the sea for the first 
time. We improve significantly on previous theory calculations to achieve accuracies of 1--2\%, giving lattice QCD results that are now more accurate than the experimental values.
In particular our results transform the theoretical picture for $\eta_c\to\gamma\gamma$ decays. 
We use gluon field configurations generated by the MILC collaboration that include $n_f=2+1+1$ flavours of Highly Improved Staggered (HISQ) sea quarks at four lattice spacing values from 0.15 fm to 0.06 fm and with sea u/d masses down to their physical value. We also implement the valence $c$ quarks using the HISQ action. 
We find $\finalWidthEtacGammaGamma$, in good agreement with experimental results using $\gamma\gamma \to \eta_c \to K\overline{K}\pi$ but in 4$\sigma$ tension with the Particle Data Group global fit result~\cite{Workman:2022ynf}; we suggest this fit is revisited. We also calculate $\finalWidthJpsiToGammaEtac$, in good agreement with results from CLEO, and predict the Dalitz decay rate $\finalDecayWidthJpsiEtacee$. We use our results to calibrate other theoretical approaches and to test simple relationships between the form factors and $J/\psi$ decay constant expected in the nonrelativistic limit.

\end{abstract}

\maketitle


\section{Introduction}
\label{sec:intro}

Decay rates of mesons via annihilation to photons, or radiative transitions with emission of a photon, can in principle provide stringent tests of our understanding of the internal structure of these mesons from strong interaction physics. The strong interaction effects are parameterised by a decay constant or a form factor and these can be calculated from first-principles using the techniques of lattice QCD. The decay rates are free from the uncertainties that arise in weak decay processes from sometimes poorly-known Cabibbo-Kobayashi-Maskawa matrix elements. This means that the combination of accurate lattice QCD and experimental results can directly test both QCD and the Standard Model. An example is that of the leptonic decay rate of the $J/\psi$ meson via a photon. The decay constant of the $J/\psi$ was recently calculated with an uncertainty of 0.4\% (including effects from the electric charges of the valence $c$ quarks) giving a value for $\Gamma(J/\psi \to e^+e^-)$ accurate to 0.9\%~\cite{Hatton:2020qhk}. The lattice QCD result agrees well with the experimental average which has an uncertainty of 1.8\%. 
Here we study two further processes of this kind for ground-state charmonium mesons, $\eta_c \to \gamma \gamma$ and $J/\psi \to \gamma \eta_c$. 

There are a few experimental results for the decay width for $J/\psi \to \gamma \eta_c$~\cite{Gaiser:1985ix,CLEO:2008pln,Anashin:2014wva,Haddadi:2017ezv}; the Particle Data Group (PDG)~\cite{Workman:2022ynf} gives a branching fraction of 1.7(4)\% as an average of results from the Crystal Ball~\cite{Gaiser:1985ix} and CLEO~\cite{CLEO:2008pln}, with the uncertainty increased by a factor of 1.5 to allow for the tension between them. The average corresponds to a partial decay width, $\Gamma(J/\psi \to \gamma \eta_c)$, of 1.57(37) keV. 

Following early work in lattice QCD in the quenched approximation~\cite{Dudek:2006ej} and including $u/d$ quarks in the sea~\cite{Chen:2011kpa,Becirevic:2012dc}, a result with a more realistic $n_f=2+1$ quark sea was obtained~\cite{Donald:2012ga}. Despite the varying numbers of sea quarks the lattice QCD calculations consistently give values for the partial decay width that are higher than the PDG average~\cite{Workman:2022ynf} of the experimental results.
Here we aim to shed further light on this issue by improving the accuracy from lattice QCD and including now $u/d$, $s$ and $c$ quarks in the sea. 

For $\eta_c\rightarrow \gamma\gamma$ the experimental and theoretical picture is less clear. The PDG~\cite{Workman:2022ynf} combines multiple sets of products of branching fractions involving $\eta_c\to \gamma\gamma$ to obtain a fit value for that branching fraction with a 7\% uncertainty ($1.61(12) \times 10^{-4}$). Individual experimental results are typically much less accurate than this, however. The fitted branching fraction corresponds to a partial decay width of 5.15(35) keV.

Lattice QCD calculations for $\eta_c \to \gamma\gamma$ also show an uncertain picture on the theoretical side. Early results in the quenched approximation~\cite{Dudek:2006ut} and subsequent results including $u/d$ quarks in the sea~\cite{Chen:2016yau,Chen:2020nar} gave results for the decay rate with a central value much less than the PDG fit value above. Further recent results from lattice QCD including $u/d$ sea quarks~\cite{Liu:2020qfz,Meng:2021ecs} give larger values for the decay rate, in agreement with the PDG fit value~\cite{Liu:2020qfz} or exceeding it~\cite{Meng:2021ecs}. Here we significantly improve the theoretical understanding of this decay rate by performing the first 1\%-accurate lattice QCD calculation of it, and we include a realistic sea quark content ($u/d$, $s$ and $c$). 

The accurate determination of $\Gamma(\eta_c \rightarrow \gamma\gamma)$ here, along with HPQCD's previous accurate determination of $\Gamma(J/\psi \to e^+e^-)$~\cite{Hatton:2020qhk}, allows us to test the relationship between these two quantities expected at leading order (LO) in nonrelativistic QCD (NRQCD). The ratio of rates in NRQCD is~\cite{Czarnecki:2001zc}
\begin{equation}
\label{eq:LONRQCD}
\frac{\Gamma (J/\psi \rightarrow e^+e^-)}{\Gamma (\eta_c \rightarrow \gamma\gamma )} = \frac{1}{3Q_c^2}(1+\mathcal{O}(\alpha_s) +\mathcal{O}(v^2/c^2)) \approx \frac{3}{4} .
\end{equation}
Here $Q_c$ is the electric charge of the $c$ quark in units of $e$. Such a simple formula is possible because the hadronic parameters, here the `wavefunction at the origin', cancel out at leading order. Sizeable radiative and relativistic corrections could be expected to this ratio but there is evidence in~\cite{Czarnecki:2001zc}, calculating through $\mathcal{O}(\alpha_s^2)$, that there is some cancellation between these corrections.  Here we can determine the ratio of these two decay rates accurately and fully nonperturbatively, including the complete relativistic dynamics of the $c$ quarks inside the mesons, using lattice QCD. This allows us to assess how closely the relationship of Eq.~\eqref{eq:LONRQCD} is followed in full QCD. 
 
For $J/\psi \rightarrow \eta_c$ decay our lattice QCD calculation involves calculating a form factor as a function of the 4-momentum transfer, $q$, between parent and daughter meson. The value of the form factor at $q^2=0$ is the appropriate one for the radiative decay of a $J/\psi$ to $\eta_c$ accompanied by a real photon. Here we calculate the form factor across the full $q^2$ range and so can also provide predictions for the case with an off-shell photon, i.e. $J/\psi \rightarrow \eta_c e^+e^-$. The rate for the equivalent process for the $D_s$ meson, $D_s^*\rightarrow D_s e^+e^-$ has been measured experimentally~\cite{CLEO:2011mla}; these decays provide an additional test of our understanding of meson structure in QCD. 

A further simple leading-order relationship between $\Gamma(J/\psi \to \gamma \eta_c)$, $\Gamma(J/\psi \to e^+e^-)$ and $\Gamma(\eta_c \to \gamma \gamma)$ was suggested many years ago by Shifman in~\cite{Shifman:1979nx}, based on the approximation of $J/\psi$ dominance of the vector $c\overline{c}$ current.  He gives
\begin{eqnarray}
\label{eq:shifman}
\Gamma(J/\psi \to \gamma \eta_c) &=& \frac{\Gamma(\eta_c \to \gamma\gamma)}{\Gamma(J/\psi \to e^+e^-)}\frac{2\alpha M^4_{J/\psi}}{9M^3_{\eta_c}} \times  \\
&&\hspace{3.0em}(1-\frac{M^2_{\eta_c}}{M^2_{J/\psi}})^3 (1 + \mathcal{O}(\alpha_s))\, . \nonumber
\end{eqnarray}
By combining the $J/\psi \rightarrow \eta_c \gamma$ and $\eta_c \rightarrow \gamma \gamma$ results we obtain here with HPQCD's earlier values for $\Gamma(J/\psi \to e^+e^-)$~\cite{Hatton:2020qhk}, we can also test how well this works in full QCD. 

We are able to perform these calculations accurately in lattice QCD because we use the Highly Improved Staggered Quark (HISQ) discretisation of the Dirac equation~\cite{Follana:2006rc}. The HISQ action has particularly small discretisation effects and this means that $c$ quark physics can be handled accurately in lattice QCD on lattices with moderate values of the lattice spacing~\cite{Donald:2012ga, Hatton:2020qhk}. This in turn means that a wide range of lattice spacings can be covered for accurate extrapolation to the continuum $a\rightarrow 0$ limit. We use gluon field configurations generated by the MILC collaboration that include $u/d$, $s$ and $c$ quarks in the sea with lattice spacing values ranging from 0.15 fm to 0.06 fm. 

The paper is laid out as follows. In Section~\ref{sec:etac2gamma} we discuss the calculation of the rate for $\eta_c \rightarrow \gamma \gamma$ using our lattice QCD determination of the amplitude. 
This includes first a discussion of the method, followed by a description of the lattice calculation with HISQ quarks and then a discussion of our results, including comparison to earlier lattice calculations and to experiment along with tests of expectations in the nonrelativistic limit. 
In Section~\ref{sec:JpsiGammaEtac} we follow the same path through the calculation of the rate for $J/\psi \rightarrow \gamma \eta_c$ and $J/\psi\rightarrow \eta_c e^+e^-$. 
Section~\ref{sec:conclusions} summarises our results and gives our conclusions. 

\vspace{2mm}
\section{Calculating $\Gamma (\eta_c \to \gamma \gamma)$} \label{sec:etac2gamma}

\subsection{Method} \label{sec:method}

To determine the decay rate of the $\eta_c$ to two photons, we need to calculate the matrix element between the $\eta_c$ and two on-shell photons (each with squared 4-momentum, $q^2=0$). The procedure for the calculation is similar to that for a meson-to-meson transition form factor except that we must take a weighted integral over the time insertion point for one of the vector currents to fix the energy of the final state to that of the required photon. 

\begin{figure}
		\includegraphics[width=0.45\textwidth]{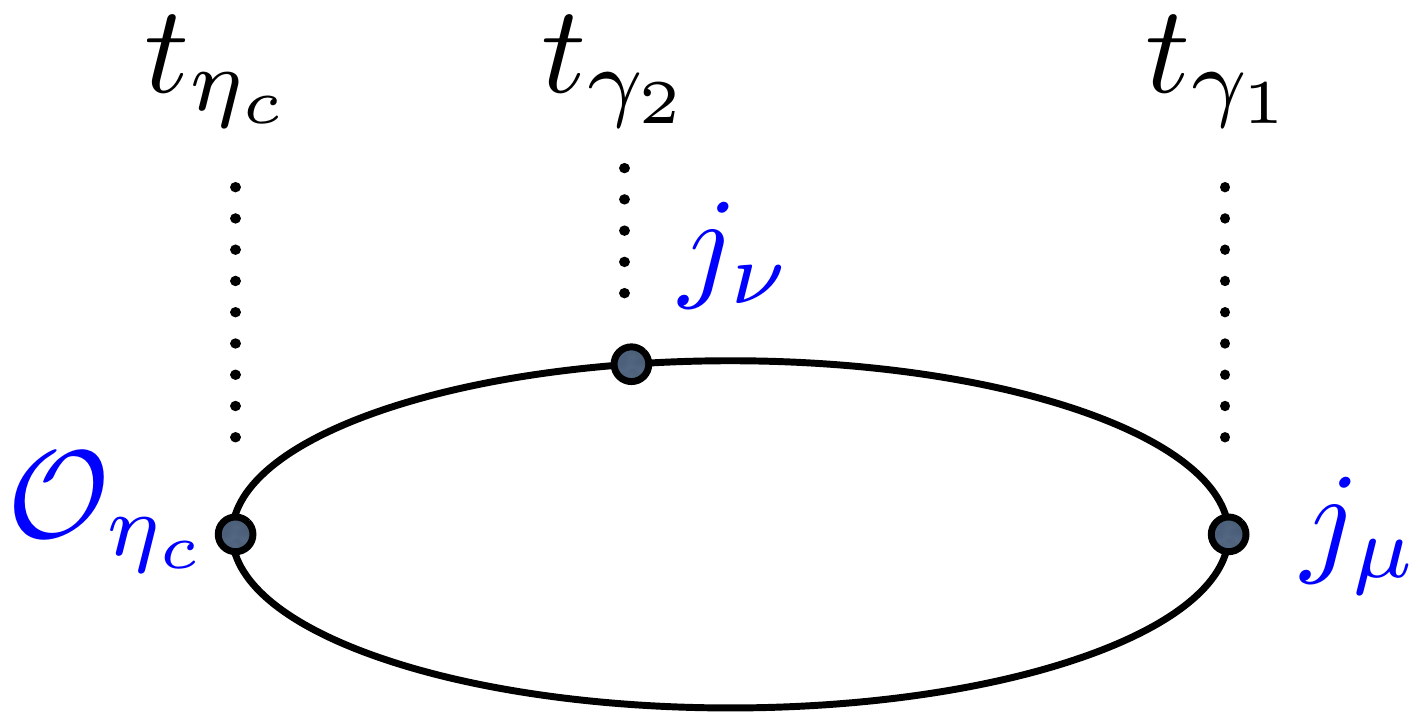}
		\caption{Schematic diagram of the connected 3-point correlation function between $O_{\eta_c}$ and two vector currents, see Eq.~\eqref{eq:etac-3pt}. The lines between the operators represent $c$ quark propagators. We do not include any quark-line disconnected correlation functions in our calculation. }
		\label{fig:3pt_etacgg}
\end{figure}

We start from a standard lattice QCD 3-point function constructed from $c$ quark propagators on each gluon field configuration and averaged over the ensemble, see Fig.~\ref{fig:3pt_etacgg}, 
\begin{eqnarray}
\label{eq:etac-3pt}
C_{\mu\nu}(t_{\gamma_1},t_{\gamma_2},t_{\eta_c}) &=& \\
&&\hspace{-3.0em}\langle 0 | \overline{j}_{\mu}(\VEC{q}_1,t_{\gamma_1})   j_{\nu}(0,t_{\gamma_2})  \overline{O}_{\eta_c}(t_{\eta_c} )  | 0 \rangle \, . \nonumber
\end{eqnarray}
Here 
\begin{eqnarray}
\label{eq:currdefs}
 \overline{j}_{\mu}(\VEC{q}_1,t_{\gamma_1}) &\equiv& a^3\sum_{\VEC{x}}e^{i\VEC{q}_1\cdot\VEC{x}}j_{\mu}(\VEC{x},t_{\gamma_1}) \nonumber \\
 \overline{O}_{\eta_c}(t_{\eta_c} ) &\equiv& a^3\sum_{\VEC{x}}O_{\eta_c}(\VEC{x},t_{\eta_c}) 
\end{eqnarray}
project out 3-momenta $\VEC{q}_1$ and 0, respectively, and $a$ is the lattice spacing. $j_{\mu}$ and $j_{\nu}$ are lattice vector currents, $\overline{c}\gamma_{\mu}c$ and $\overline{c}\gamma_{\nu}c$, and $O_{\eta_c}$ is a pseudoscalar or temporal axial current operator that couples to pseudoscalar charmonium states. 

We can couple an external photon field to $\overline{j}_{\mu}$ by multiplying by a photon propagator~\cite{Ji:2001wha, Ji:2001nf} so that 
\begin{eqnarray}
\label{eq:add-photon}
\langle 0 | A_{\mu}(\VEC{q}_1,t_0)j_{\nu}(t_{\gamma_2})\overline{O}_{\eta_c}(t_{\eta_c})|0\rangle &=& \\
&&\hspace{-5.0em} a\sum_{t_{\gamma_1}}D(t_{\gamma_1}-t_0) C_{\mu\nu}(t_{\gamma_1},t_{\gamma_2},t_{\eta_c}) \, . \nonumber 
\end{eqnarray}
\begin{equation}
\label{eq:Ddef}
D(t_{\gamma_1}-t_0) = \frac{e^{-\omega_1 (t_{\gamma_1}-t_0)}}{2\omega_1}
\end{equation}
is the photon propagator in Euclidean time for a photon with 3-momentum $\VEC{q}_1$, where $\omega_1 = |\VEC{q}_1|$. The sum in Eq.~\eqref{eq:add-photon} is over all $t_{\gamma_1}$ on the lattice. 
The construction of Eq.~\eqref{eq:add-photon} yields a 3-point function between a tower of $\eta_c$ states (at rest) created by $O_{\eta_c}$ at $t_{\eta_c}$ and a photon at $t_0$ induced by a vector current at $t_{\gamma_1}$. The 4-momentum transferred by the current is constrained by energy-momentum conservation. If we choose 
\begin{equation}
\label{eq:on-shell}
|\VEC{q}_1|=\omega_1=\frac{M_{\eta_c}}{2} \, ,
\end{equation}
where $M_{\eta_c}$ is the $\eta_c$ mass, then for the ground-state $\eta_c$ this 3-point function encapsulates an $\eta_c \to \gamma \gamma$ transition with two real photons in the final-state. The ground-state contribution to the 3-point function is 
\begin{equation}
\label{eq:3ptamp}
\frac{e^{-\omega_1 (t_{\gamma_2}-t_0)}}{2\omega_1}\mathcal{F}_{\mu\nu}(\eta_c \to \gamma\gamma)\langle \eta_c|O_{\eta_c}| 0\rangle \frac{e^{-M_{\eta_c}(t_{\eta_c}-t_{\gamma_2})}}{2M_{\eta_c}} \, . 
\end{equation}
$\mathcal{F}_{\mu\nu}$ is the matrix element between $\eta_c$ and $\gamma\gamma$ states that will allow us to determine the decay rate. 

To obtain $\mathcal{F}_{\mu\nu}$ from the 3-point function it is convenient to first peel off the final-state photon by dividing by the left-most factor in Eq.~\eqref{eq:3ptamp}. 
Instead of Eq.~\eqref{eq:add-photon} we construct in practice 
\begin{eqnarray}
\label{eq:Ctilde}
\tilde{C}_{\mu\nu} &=& \frac{ a\sum_{t_{\gamma_1}}e^{-\omega_1(t_{\gamma_1}-t_0)}/(2\omega_1) C_{\mu\nu}(t_{\gamma_1},t_{\gamma_2},t_{\eta_c}) }{e^{-\omega_1(t_{\gamma_2}-t_0)}/(2\omega_1)}  \nonumber \\
&=&a\sum_{t_{\gamma_1}}e^{-\omega_1(t_{\gamma_1}-t_{\gamma_2})} C_{\mu\nu}(t_{\gamma_1},t_{\gamma_2},t_{\eta_c}) \, .
 \end{eqnarray}
 $\tilde{C}_{\mu\nu}(t_{\gamma_2}, t_{\eta_c})$ is now a 2-point function. Note that $t_{\gamma_1}-t_{\gamma_2}$ varies from $-N_t/2$ to +$N_t/2$ as $t_{\gamma_1}$ is varied. At the same time we construct the standard 2-point function 
 \begin{equation}
\label{eq:etac-2pt}
C_{\eta_c}(t, t_{\eta_c}) =
\langle 0 | \overline{O}_{\eta_c}(t)  \overline{O}_{\eta_c}(t_{\eta_c} )  | 0 \rangle \, . 
\end{equation}
 By fitting $\tilde{C}_{\mu\nu}$ and $C_{\eta_c}$ simultaneously we can determine the contribution of the ground-state $\eta_c$ to both 2-point functions and use this to obtain $\mathcal{F}_{\mu\nu}(\eta_c \to \gamma \gamma)$. 
 
 The fit form for $C_{\eta_c}$ can be written as 
 \begin{equation}
 \label{eq:Ceta-fit}
 C_{\eta_c}(t,t_{\eta_c}) = \sum_n a_n^2 f(E_n,t-t_{\eta_c}) 
 \end{equation}
 where 
 \begin{equation}
 \label{eq:fdef}
 f(E,t) = e^{-Et}+e^{-E(N_t-t)}
 \end{equation}
 and the ground-state energy, corresponding to $n=0$, is $E_0=M_{\eta_c}$. The ground-state amplitude, 
 \begin{equation}
 \label{eq:a0def}
 a_0= \frac{\langle \eta_c|O_{\eta_c}| 0\rangle}{\sqrt{2M_{\eta_c}}} \, .
 \end{equation}
 Likewise 
  \begin{equation}
 \label{eq:Cmunu-fit}
 \tilde{C}_{\mu\nu}(t_{\gamma_2},t_{\eta_c}) = \sum_n a_n b_n f(E_n,t_{\gamma_2}-t_{\eta_c}) \, 
 \end{equation}
with (see Eq.~\eqref{eq:3ptamp})
 \begin{equation}
 \label{eq:b0def}
 b_0= \frac{\mathcal{F}_{\mu\nu}(\eta_c\to\gamma\gamma)}{\sqrt{2M_{\eta_c}}} \, .
 \end{equation}
For $n > 0$, the $b_n$ correspond to matrix elements between excited $\eta_c$ states and one on-shell and one off-shell photon. 

Using parity and Lorentz invariance we can define a transition form factor $F(q_1^2,q_2^2)$ by\footnote{Note that~\cite{Dudek:2006ej} uses a normalisation for $F$ that differs by a factor of $M_{\eta_c}$.}
\begin{equation}
\label{eq:Fdef}
\mathcal{F}_{\mu\nu} \equiv \epsilon_{\mu\nu\rho\sigma}p_{\eta_c}^{\rho}q_1^{\sigma}F\, .
\end{equation}
This equation makes clear that a non-zero result will only be obtained (for an $\eta_c$ at rest) if there is a component of the (equal-and-opposite) spatial momentum of the two photons that is orthogonal to the polarisation vectors of both photons (which must themselves be orthogonal). The specific configurations of momenta and polarisations that we use will be described below. Here we are interested in determining $F(0,0)$ (i.e. with two on-shell photons) and so the kinematic factors in Eq.~\eqref{eq:Fdef} mean that $F$ is obtained from $b_0$ via
\begin{equation}
 \label{eq:Fb0def}
 F(0,0)= b_0 \frac{2\sqrt{2M_{\eta_c}}}{M_{\eta_c}^2} \, .
 \end{equation}

The decay amplitude is given by 
\begin{eqnarray}
\label{eq:decayamp}
\mathcal{M}(\eta_c \to \gamma\gamma)&=& 2e^2 Q_c^2 {{\varepsilon}}_1^* \cdot \mathcal{F} \cdot {{\varepsilon}}_2^* \nonumber \\ &=&e^2Q_c^2M_{\eta_c}^2 |\VEC{\varepsilon}_1\times \VEC{\varepsilon}_2| F(0,0)\, ,
\end{eqnarray}
allowing for interchange of the two photons and inserting electric charge factors ($Q_c=2/3$ for the $c$ quark) and polarisation vectors. 
The decay rate is then 
\begin{eqnarray}
\label{eq:decayrate}
\Gamma(\eta_c \to \gamma \gamma) &=& \frac{1}{2}\sum_{\text{spins}}\int d\Omega \frac{|\mathcal{M}|^2}{64\pi^2M_{\eta_c}} \nonumber \\
&=& \pi \alpha^2 Q_c^4 M_{\eta_c}^3 F^2 \, . 
\end{eqnarray}
The factor of 1/2 above avoids double-counting the identical photons and there are two spin combinations for the final-state, both with $|\VEC{\varepsilon}_1\times\VEC{\varepsilon}_2|=1$. 

In the next section we give more details of how we set up our lattice calculation to determine $F(0,0)$. 

\subsection{Lattice Calculation}\label{sec:lattice}

\begin{table*}
	\centering
	\caption{Parameters for the MILC ensembles of gluon field configurations.
		The sets are numbered but also given a `handle' to distinguish them in the second column. The lattice spacing is determined from the Wilson flow parameter, $w_0$~\cite{Borsanyi:2012zs}, and 
		values of $w_0/a$ are given in column 4, following the gauge coupling, $\beta$, in column 3. 
		The physical value $w_0 = 0.1715(9) \: \mathrm{fm}$ was fixed from $f_{\pi}$ in \cite{Dowdall:2013rya}. 
		Sets $1$ and $2$, $3$ and $4$, $5$, and $6$ have $a \approx 0.15,0.12,0.09, 0.06 \: \mathrm{fm}$ respectively. The number of lattice points in space, $N_x$, and time, $N_t$, are given in column 5.
		Sea quark masses in lattice units are given in columns 6, 7 and 8. All the configuration sets have equal-mass $u$ and $d$ quarks with $m_u=m_d=m_l$.
		Sets $1$, $3$, $5$, and $6$ have heavier-than-physical mass $m_l$ such that $m_s/m_l = 5$ and sets 2 and 4 have $m_l$ close to the physical average of $u$ and $d$ quark masses. 
		As described in the text, we use valence $c$ quark masses, given in column 9, that differ from the sea $c$ quark masses, being more closely tuned to the physical value. 
		The $\epsilon_{\mathrm{Naik}}$ parameter that accompanies $am_c^{\text{val}}$ in the HISQ action~\cite{Follana:2006rc,Monahan:2012dq} are given in the next column. 
		On set 3 we also include results from a deliberately mis-tuned valence $c$ quark mass of 0.654, and denote this calculation as `3A'. 
		We use 1000 gluon field configurations from each set, with 2 time-sources on each configuration to increase statistics, except for set 6 where we include only one time-source per configuration. 
	}
	\begin{tabular}{c c c c c c c c | c c } 
		\hline\hline
		set & handle & $\beta$ & $w_0/a$ & $N_x^3 \times N_t$ & $am_l^{\text{sea}}$ & $am_s^{\text{sea}}$ & $am_c^{\text{sea}}$ & $am_c^{\text{val}}$ & $\epsilon_{\mathrm{Naik}}$  \\ [0.1ex]  
		\hline
		1 & very-coarse & $5.80$ & $1.1119(10)$ & $16^3 \times 48$ & $0.013$ & $0.065$ & $0.838$ & $0.888$ & $-0.3820$  \\ 
		2 & very-coarse-physical & $5.80$ & $1.1367(5)$ & $32^3 \times 48$ & $0.00235$ & $0.0647$ & $0.831$ & $0.863$ & $-0.3670$  \\ 
		3 & coarse & $6.00$ & $1.3826(11)$ & $24^3 \times 64$ & $0.0102$ & $0.0509$ & $0.635$ & $0.664$ & $-0.2460$  \\ 
		3A & coarse & $6.00$ & $1.3826(11)$ & $24^3 \times 64$ & $0.0102$ & $0.0509$ & $0.635$ & $0.654$ & $-0.2402$  \\ 
		4 & coarse-physical & $6.00$ & $1.4149(6)$ & $48^3 \times 64$ & $0.00184$ & $0.0507$ & $0.628$ & $0.643$ & $-0.2336$  \\
		5 & fine & $6.30$ & $1.9006(20)$ & $32^3 \times 96$ & $0.0074$ & $0.037$ & $0.440$ & $0.450$ & $-0.1250$  \\
		6 & superfine & $6.72$ & $2.8941(48)$ & $48^3 \times 144$ & $0.0048$ & $0.0240$ & $0.286$ & $0.274$ & $-0.0491$  \\
		\hline\hline
	\end{tabular}
	\label{tab:params}
\end{table*}

\subsubsection{Ensembles and parameters}

We use ensembles with $N_f = 2 + 1 + 1$ flavours of dynamical HISQ sea quarks from the MILC collaboration~\cite{Bazavov:2010ru, Bazavov:2012xda}.
Details of the ensembles are tabulated in Table~\ref{tab:params}. The six ensembles that we use give us four different lattices spacings, $a \approx 0.06, 0.09, 0.12$ and $0.15 \: \mathrm{fm}$. The sea $u$ and $d$ quark masses are taken to be the same and denoted $m_l$ ($l$ for light). 
 Two ensembles, sets $2$ and $4$, have $l$ quarks in the sea with the physical mass, $m_l=(m_u+m_d)/2$, and the other four have heavier-than-physical light quarks. This allows us to test the dependence of our results on the mass of the sea light quarks. 
 
 On these ensembles we calculate correlation functions constructed from valence $c$ quark propagators, also using the HISQ formalism. 
 The valence $c$ quark mass values that we use here are the same as those given in Table I of~\cite{Hatton:2020qhk}. These masses are more accurately tuned than those of the $c$ quark in the sea. The tuning is done by comparing the result for the mass of the $J/\psi$ meson obtained on each ensemble to its physical mass from experiment, as discussed in~\cite{Hatton:2020qhk}.

\subsubsection{Correlation functions $C_{\mu\nu}$ and $C_{\eta_c}$} \label{sec:etac2gamma_corrfs}

On the ensembles of Table~\ref{tab:params} we calculate 2-point and 3-point correlation functions, as described in Section~\ref{sec:method}. 
The 3-point correlation function is that between a source current that couples to the $\eta_c$ and its excitations and two vector currents, depicted in Fig.~\ref{fig:3pt_etacgg}. Note that we construct only one combination of quark propagators on the lattice and determine $F$ from that; the factor of 2 for interchanging the photons is allowed for in Eq.~\eqref{eq:decayamp}. We include only the quark-line connected diagram of Fig.~\ref{fig:3pt_etacgg}. There are quark-line disconnected diagrams (quark loops connected only by gluons) that contribute but we expect their contribution to be small because the $c$ quark mass is large. We will estimate the systematic uncertainty from neglecting quark-line disconnected diagrams in Section~\ref{sec:etac2gamma_uncertainties}.  

The correlation function $C_{\mu\nu}(t_{\gamma_1},t_{\gamma_2},t_{\eta_c})$ (Eq.~\eqref{eq:etac-3pt}) is calculated using the standard sequential source technique. 
On timeslice $t_{\eta_c}$, we set up a random wall source~\cite{Aubin:2004fs} from which two $c$ quark propagators are calculated. The first propagator, evaluated at timeslice $t_{\gamma_1}$ and multiplied by $\gamma_{\mu}$, is used as the source vector for another $c$ quark propagator calculation. 
The result of this calculation, the extended/sequential propagator, is contracted with the second $c$ quark propagator from $t_{\eta_c}$, inserting a $\gamma_{\nu}$ before summing over indices, to obtain the 3-point function. The second $c$ propagator from $t_{\eta_c}$ is calculated with an additional $\gamma_t\gamma_5$ multiplying the random wall source to achieve the quantum numbers of the $\eta_c$ when contracted. The 3-point correlation functions obtained are averaged over all gluon fields in the ensemble. We use 2 time-sources for $t_{\eta_c}$ on each configuration (except for set 6) to improve statistical accuracy. We take $\mu,\nu=z,x$ and take the photon momentum (discussed below) to be in the $y$-direction to satisfy the requirements for a non-zero result from Eq.~\eqref{eq:Fdef}. 

Because we are using HISQ valence $c$ quarks, the $\gamma$ matrices in the paragraph above become spatial-position-dependent phases with which the sources and sinks are patterned to achieve the required spin. We also have to consider operator `taste'~\cite{Follana:2006rc}, also represented by $\gamma$ matrices, and we must choose tastes for the different operators such that the product of the taste $\gamma$ matrices gives 1, for a non-zero correlation function. We want the spin-taste representation of the two vector currents to be the same (except for $z \leftrightarrow x$) for symmetry, and we want to avoid point-splitting of the vector currents along the $y$-direction in which the spatial momentum flows. Our preferred set-up uses a temporal axial current for $O_{\eta_c}$ with spin-taste in the standard notation (see, for example,~\cite{Follana:2006rc}) $\gamma_5 \gamma_t \otimes \gamma_x \gamma_z$ and vector currents with spin-taste $\gamma_x  \otimes \gamma_x$ and $\gamma_z \otimes \gamma_z$. In this case the two vector currents are local and this has the advantage that they have no tree-level discretisation errors. The temporal axial current is one-link point-split in the $y$ direction. Use of the temporal axial current (rather than the pseudoscalar current) avoids any temporal point-splitting. We will call this the `LOCAL' set-up, because of the nature of the vector currents used. An alternative, that we will use as a test on a subset of ensembles (sets 1, 3 and 5) is to take all of the operators to be `tasteless' i.e. with spin-tastes $\gamma_5\gamma_t \otimes 1$, $\gamma_x \otimes 1$ and $\gamma_z \otimes 1$. Now the vector currents have a one-link point-splitting along the $x$ and $z$ directions respectively and the temporal axial current has a 3-link point-splitting (from one corner to the opposite of a cube). This is the `ONE-LINK' set-up. 

Calculating HISQ $c$ quark propagators is numerically relatively inexpensive and so we cover the full range of values of $t_{\gamma_1}$ by calculating $C_{\mu\nu}$ (Eq.~\eqref{eq:etac-3pt}) for $t_{\gamma_1}$ from $t_{\eta_c}$ to $t_{\eta_c}-N_t/2$, obtaining the other values by periodicity. This allows us to test the behaviour of the integral/sum over $t_{\gamma_1}$ in Eq.~\eqref{eq:Ctilde}. Results at all values of $t_{\gamma_2}$ are obtained in the final contraction of the sequential propagator with that from $t_{\eta_c}$. This amount of calculation is not in fact necessary, as we discuss below.

\begin{table}
	\centering
	\caption{Twist $\theta$ used for the sequential propagator in $C_{\mu\nu}$ (Eq.~\eqref{eq:etac-3pt} and Fig.~\ref{fig:3pt_etacgg}) from which $\eta_c \to \gamma\gamma$ form factor is extracted. The corresponding 3-momentum component is given by $a q_1^y = \theta \pi / N_x$ and given in column 4. For the LOCAL set-up (local vector currents, top table) the twist is chosen to achieve $q_1^y = M^{\text{phys,latt}}_{\eta_c}/2$ with $M^{\text{phys,latt}}_{\eta_c}$ given in Eq.~(\ref{eqn:etac_mass}). For the ONE-LINK set-up (one-link vector currents, lower table) the twist was chosen so that $aq_1^y=aM^{\text{latt}}_{\eta_c}/2$, i.e. half the $\eta_c$ meson mass on that ensemble.  }
	\begin{tabular}{l c c c} 
		\hline\hline
		LOCAL & set & $\theta$  & $aq_1^y$ \\ [0.1ex] 
		\hline
		& 1 & $5.928$ &   1.1640    \\
		& 2 & $11.598$ &  1.1386     \\
		& 3 & $7.151$ &  0.9361     \\
		& 4 & $13.976$ &  0.9147    \\
		& 5 & $6.936$ &  0.6809     \\
		& 6 & $6.833$ & 0.4472      \\
		\hline\hline
		ONE-LINK & Set & $\theta$ & $aq^y_1$\\
    \hline
    & 1 & $6.0759$ & $1.1930$ \\
    & 3 & $7.2399$ & $0.9477$ \\
    & 5 & $6.9866$ & $0.6859$ \\
    \hline
    \hline

	\end{tabular}
	\label{tab:twists_etac}
\end{table}

In $\eta_c \to \gamma \gamma$ decay each photon carries away spatial momentum with magnitude $M_{\eta_c}/2$ in the $\eta_c$ rest frame. In our calculation spatial momentum is inserted into the $c$ quark propagator that connects the operators at timeslices $t_{\gamma_1}$ and $t_{\gamma_2}$ (the sequential propagator, see Fig.~\ref{fig:3pt_etacgg}) using twisted boundary conditions~\cite{Sachrajda:2004mi, Guadagnoli:2005be}.
This enables us to choose any value of the spatial momentum, $\VEC{q}_1$, using a twist angle $\theta$.  One photon will have momentum $\VEC{q}_1$ and the other $-\VEC{q}_1$. Given our vector current polarisations, $\VEC{q}_1$ must have a $y$-component and we simply take $\VEC{q}_1$ to be in the $y$-direction. Its value is then related to $\theta$ by 
\begin{align}
\label{eq:q1theta}
a q_1^y = \frac{\theta \pi}{N_x} 
\end{align}
where $N_x$ is the number of lattice points in a spatial direction. The kinematics that are specified in Eq.~\eqref{eq:on-shell} then require a twist angle of 
\begin{equation}
\label{eq:theta-val}
\theta = \frac{aN_x M_{\eta_c} }{ 2 \pi} \, . 
\end{equation}

Since $\omega_1=|\VEC{q}_1|$ and both of these quantities are set in lattice units, photon 1 will be exactly on-shell. It is harder to arrange for photon 2 to be exactly on-shell and there will inevitably be a slight mis-tuning of the on-shell condition for this photon. Its spatial momentum has magnitude $a|\VEC{q}_1|$ and its energy is $aM^{\text{latt}}_{\eta_c}-a\omega_1$ in lattice units, where $M^{\text{latt}}_{\eta_c}$ is the mass of the $\eta_c$ obtained on that ensemble from the lattice calculation. Photon 2 will only be on-shell if $a\omega_1$ is exactly $aM^{\text{latt}}_{\eta_c}/2$.

The mis-tuning depends on what value of $M_{\eta_c}$ is used in determining the twist angle for $\VEC{q}_1$. Various approaches that are equivalent in the continuum limit are possible. Here, for the LOCAL set-up,  we choose to fix the $M_{\eta_c}$ value used in $\VEC{q}_1$ (Eq.~\eqref{eq:theta-val}) to the value we obtain from connected correlation functions in lattice QCD in the physical continuum limit, since this corresponds to the value to which our $\eta_c$ masses will converge in that limit. This value was obtained by calculating the charmonium hyperfine splitting in~\cite{Hatton:2020qhk} for both pure QCD and QCD plus quenched QED. Here we use the pure QCD result, 
\begin{align}
	M^{\mathrm{phys,latt}}_{\eta_c} = 2.9783(11) \: \mathrm{GeV} . \label{eqn:etac_mass}
\end{align}
This mass differs slightly (by 5.6 MeV or 0.2\% of the mass) from the experimental value~\cite{Workman:2022ynf}. The most likely explanation for this is that it represents the impact of quark-line disconnected correlation functions (not included in the lattice calculation) that allow the $\eta_c$ to mix with lighter flavour-singlet mesons such as the $\eta$ and $\eta^{\prime}$. 
The twists from Eqs.~(\ref{eq:q1theta}) and~(\ref{eqn:etac_mass}) used on each set are given in Table~\ref{tab:twists_etac} (top section). For the ONE-LINK set-up we chose to test a different tuning, equivalent in the continuum limit. In that case we determined $aM^{\text{latt}}_{\eta_c}$ for that taste of $\eta_c$ on that ensemble in a separate calculation and then used that value in our choice for $\theta$ (Eq.~\eqref{eq:theta-val}), so that both photons 1 and 2 should be exactly on-shell for each ensemble. The values for the twists used in that case are given in the lower section of Table~\ref{tab:twists_etac}. 

The accuracy with which photon 2 is tuned to the on-shell point will be discussed further in Section~\ref{sec:etacggfits}. The mis-tuning is small in both cases here (very small for the ONE-LINK case) but we will take account of it in our continuum/chiral extrapolation for $F(0,0)$. We will also estimate and include an error associated with the fact that $M^{\text{phys,latt}}_{\eta_c}$ is not equal to the experimental value.

As well as $C_{\mu\nu}$ (Eq.~\eqref{eq:etac-3pt}) we also calculate the 2-point correlation function $C_{\eta_c}$ (Eq.~\eqref{eq:etac-2pt}) constructed from the same $O_{\eta_c}$ operator used in $C_{\mu\nu}$. Fitting $C_{\eta_c}$ and $\tilde{C}_{\mu\nu}$ (derived from $C_{\mu\nu}$, see Eq.~\eqref{eq:Ctilde}) simultaneously allows us to determine the form factor for $\eta_c \to \gamma\gamma$. 

\subsubsection{Vector current renormalisation}\label{sec:ZV}

We use a local vector current, $\gamma_i\otimes\gamma_i$ in our LOCAL set-up and a 1-link $\gamma_i\otimes 1$ current in our ONE-LINK set-up. Neither of these currents is conserved, so we must multiply both by their multiplicative renormalisation factors to match them to their continuum counterparts. Since the vector current appears twice, this means multiplying the raw lattice data for $C_{\mu\nu}$ by $Z_V^2$ before determining $\tilde{C}_{\mu\nu}$.  

We use $Z_V$ values calculated in the symmetric MOM scheme (RI-SMOM) in~\cite{Hatton:2019gha}. These are calculated for groups of ensembles with the same value of $\beta$ (rather than individually for each set) and are reproduced in Table~\ref{tab:ZV}. The values are very close to 1 for the HISQ action and are obtained with an uncertainty of less than 0.4\%.

\begin{table}
	\centering
	\caption[kgjh]{Vector current renormalisation constants, $Z_{V}(\mu)$, using the RI-SMOM scheme.  Values are taken from Tables III and VI in~\cite{Hatton:2019gha} where they were calculated in pure QCD for each $\beta$ value corresponding to a group of ensembles in Table~\ref{tab:params}. 
We use the values given at $\mu=2 \; \mathrm{GeV}$ but note that the small $\mu$-dependence in this quantity is purely a lattice artefact, vanishing in the continuum limit. Here we give values for each set (listed in column 1), but those with the same $\beta$ value are the same. $Z_V^{\gamma_i\otimes \gamma_i}$ (column 3) is the renormalisation constant for the local current used in the LOCAL set-up and $Z_V^{\gamma_i\otimes 1}$ (column 4) is the 1-link current renormalisation needed for the ONE-LINK set-up. Note, in the latter case, that this is for the 1-link tadpole-improved current constructed with a `thin-link' included. The tadpole-improvement factor $u_0$, given by the mean value of the gluon field $U_{\mu}$ in Landau gauge, is listed in column 5. Since we use a 1-link vector current that is not tadpole-improved, the renormalisation factor for the ONE-LINK case is $Z_V^{\gamma_i\otimes 1}/u_0$.
	}
\begin{tabular}{c c c c c}
	\hline \hline set & $\beta$ & $Z^{\gamma_i\otimes\gamma_i}_{V}$  & $Z_V^{\gamma_i\otimes 1}$ & $u_0$ \\
	\hline 
	1 & $5.80$ & $0.95932(18)$ & $0.93516(16)$  & 0.81960 \\
	2 & $5.80$ & $0.95932(18)$ & $0.93516(16)$ &  0.82042 \\
	3 & $6.00$ & $0.97255(22)$ & $0.94966(20)$ & 0.83461 \\
	4 & $6.00$ & $0.97255(22)$ & $0.94966(20)$ & 0.83505 \\
	5 & $6.30$ & $0.98445(11)$ & $0.96695(11)$ & 0.85248 \\
	6 & $6.72$ & $0.99090(36)$ & $0.97996(34)$ & 0.87094 \\
	\hline \hline
\end{tabular} \label{tab:ZV}
\end{table}

\subsubsection{Determining $\tilde{C}_{\mu\nu}$}
\label{sec:Ctilde}

\begin{figure}
	\begin{center}
		\includegraphics[width=0.45\textwidth]{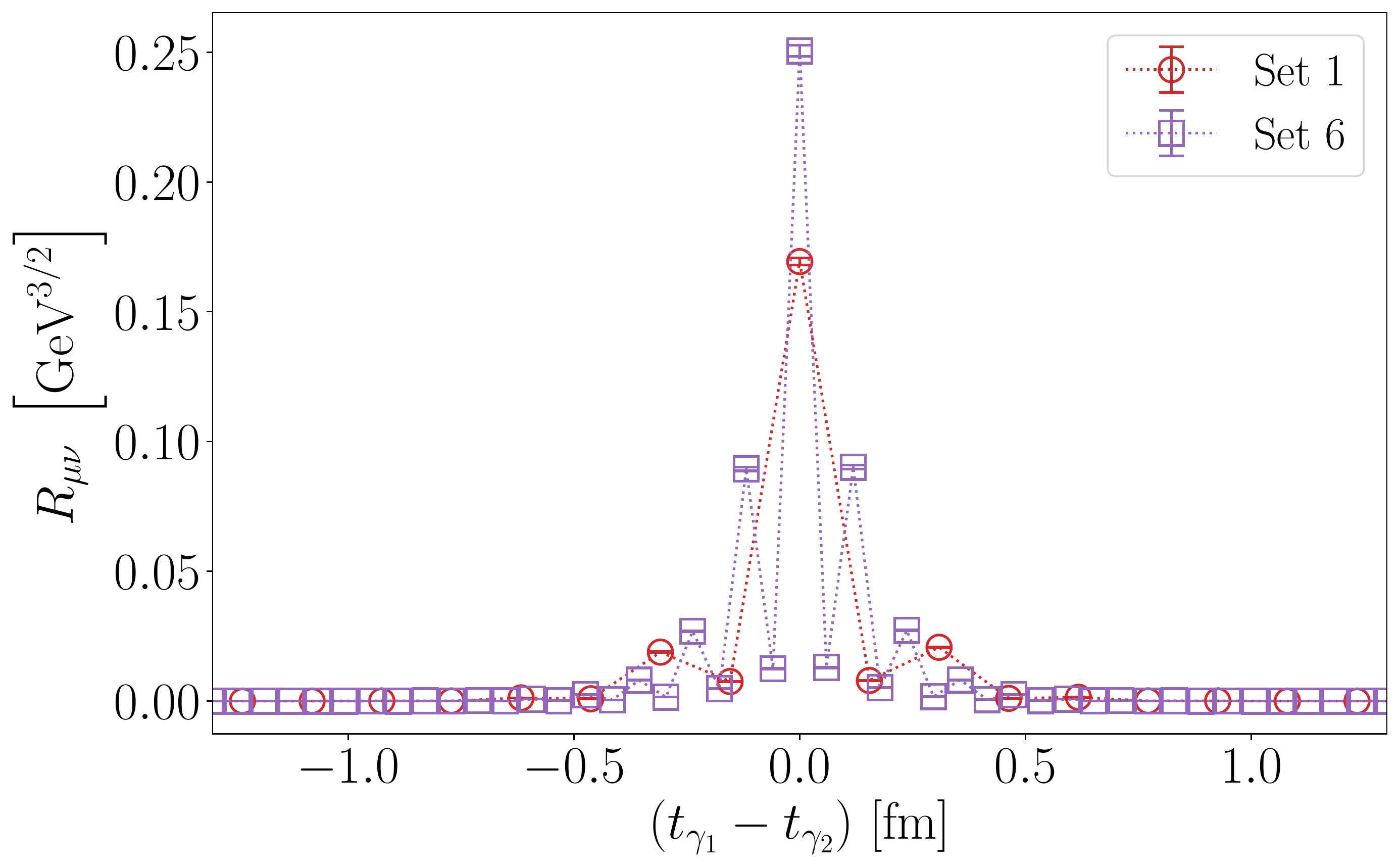}
		\caption{$R_{\mu\nu}$, defined in Eq.~\eqref{eq:intratio}, as a function of $t_{\gamma_1}-t_{\gamma_2}$. We use a fixed value for the separation between $t_{\eta_c}$ and $t_{\gamma_2}$ of $N_t/4$. Results are given for the LOCAL set-up and plotted in physical units for two values of the lattice spacing, corresponding to very-coarse (set 1: $a\approx $ 0.15 fm) and superfine (set 6: $a \approx$ 0.06 fm). Integrating $R_{\mu\nu}$ gives $b_0$ and hence $F$ (see Eq.~\eqref{eq:Fb0def});  note the narrow width of the region of support for the integral. Oscillations are a result of using staggered quarks. Their impact on the integral is a discretisation effect - see Appendix~\ref{sec:osc-sum}. } 
		\label{fig:integrand}
	\end{center}
\end{figure}
\begin{figure}
	\begin{center}
		\includegraphics[width=0.45\textwidth]{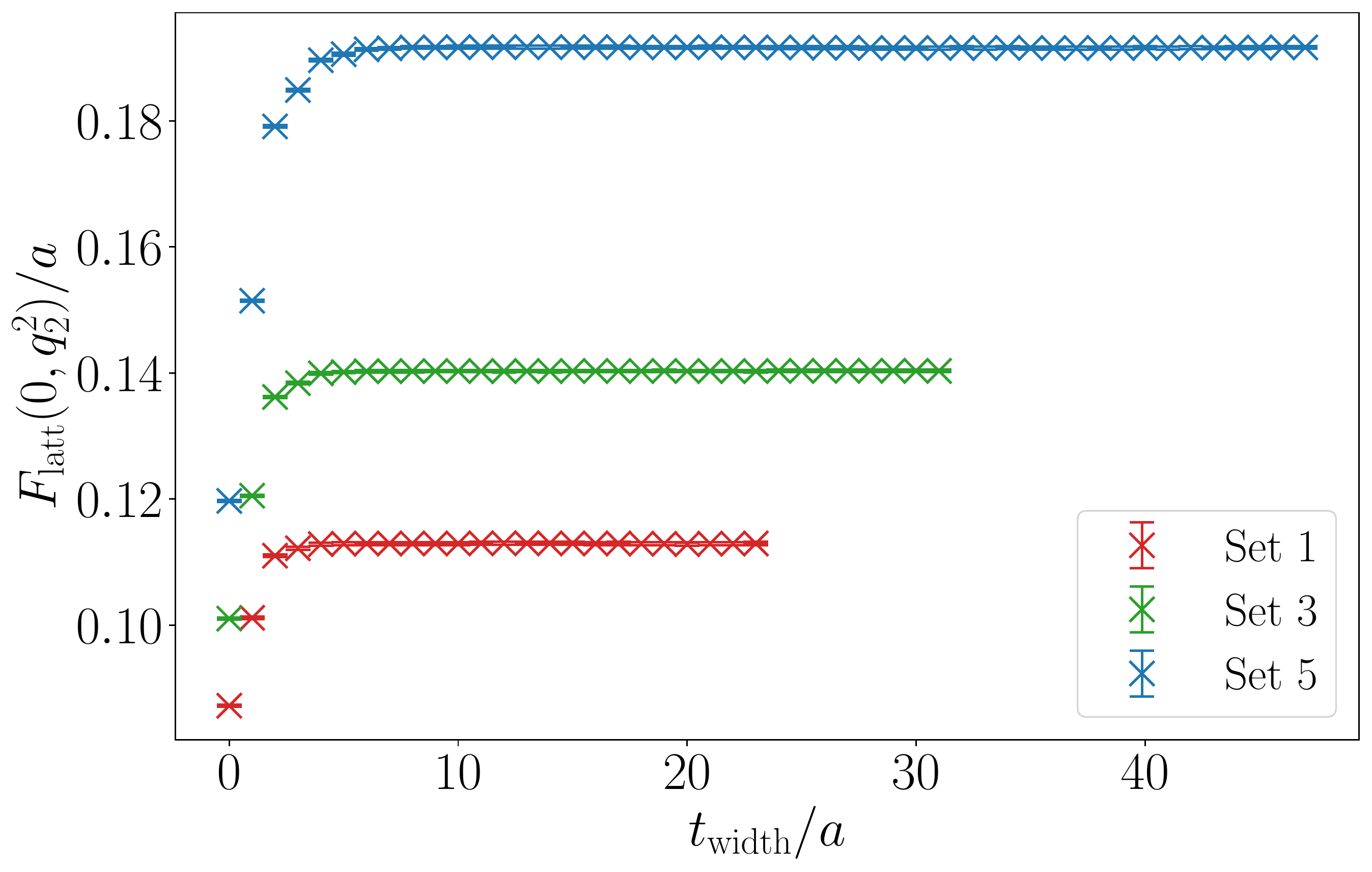}
		\caption{Fitted results for $F_{\text{latt}}(0,q_2^2)/a$ obtained from $\tilde{C}(t\equiv t_{\eta_c}-t_{\gamma_2})$ as a function of $t_{\mathrm{width}}$, the half-width of the region of time integration over $t_{\gamma_1}$ either side of $t_{\gamma_2}$. Results are given for the LOCAL set-up on set 1 (very-coarse, $a \approx$ 0.15 fm, fitting time range $t=6\rightarrow 20$), set 3 (coarse, $a \approx$ 0.12 fm, $t=7\rightarrow 27$) and 5 (fine, $a \approx$ 0.09 fm, $t=9\rightarrow 40$). Note the steep rise of the results to a plateau. }
		\label{fig:vary-width}
	\end{center}
\end{figure}

The 2-point function $\tilde{C}_{\mu\nu}(t_{\gamma_2},t_{\eta_c})$ is constructed from a weighted sum over all lattice time slices of $C_{\mu\nu}$ as given in Eq.~\eqref{eq:Ctilde}, using $a\omega_1 = aq_1^y$ as discussed in Section~\ref{sec:etac2gamma_corrfs}. It is normalised so that the lattice vector currents match those in the continuum as discussed in Section~\ref{sec:ZV}. 
In Fig.~\ref{fig:integrand} we plot a quantity proportional to the summand of Eq.~\eqref{eq:Ctilde} to illustrate how the sum works. The quantity plotted is 
\begin{equation}
\label{eq:intratio}
R_{\mu\nu}(t_{\gamma_1},t_{\gamma_2}) = \frac{e^{-\omega_1(t_{\gamma_1}-t_{\gamma_2})} C_{\mu\nu}(t_{\gamma_1},t_{\gamma_2},t_{\eta_c})}{a_0e^{-M_{\eta_c}(t_{\eta_c}-t_{\gamma_2})}} \, .
\end{equation} 
This is plotted as a function of the time separation between vector current operators for a fixed value of $t_{\eta_c}-t_{\gamma_2}$ of $N_t/4$. We divide by $\exp(-M_{\eta_c}(t_{\eta_c}-t_{\gamma_2}))$ to remove the time-dependence related to the $\eta_c$ mass expected from Eq.~\eqref{eq:Cmunu-fit}. The $\eta_c$ mass used here is the one from the fit to  $C_{\eta_c}$ (Eq.~\eqref{eq:Ceta-fit}) (i.e. $M_{\eta_c}^{\text{latt}}$) . We also divided by $a_0$, which is the ground-state amplitude from this fit. This means that the integral of $R_{\mu\nu}$ is equal to $b_0$ (Eq.~\eqref{eq:Cmunu-fit}) up to excited-state contamination, which is very small at this large value for $t_{\eta_c}-t_{\gamma_2}$.  Indeed the figure is unchanged over a wide range of $t_{\eta_c}-t_{\gamma_2}$ values since excited state contamination falls off on distance scales of $\mathcal{O}(0.5\,\text{fm})$ and $N_t/4 \approx 2\,\text{fm}$.

We see from Fig.~\ref{fig:integrand} that $R_{\mu\nu}$ is strongly dominated by the region of $t_{\gamma_1}$ very close to $t_{\gamma_2}$. This is because, once a photon is emitted from the $\eta_c$, the resulting system is far off-shell and decays exponentially fast in $t$. The region in $|t_{\gamma_1}-t_{\gamma_2}|$ for which the integrand is non-zero is less than about 0.5 fm. 

Because we use staggered quarks the integrand has a component that oscillates in time. On summing/integrating, however, this component reduces to an $a^2$ discretisation effect, because the oscillations get closer together on finer lattices, as we show in Appendix~\ref{sec:osc-sum}. Discretisation effects of this kind are allowed for in our extrapolation of our results for $F$ to the $a\to 0$ continuum limit. 

 We construct $\tilde{C}_{\mu\nu}(t_{\gamma_2},t_{\eta_c})$ for all values of $t_{\eta_c}-t_{\gamma_2}$ and then fit it as described in the next section, in conjunction with $C_{\eta_c}$, to determine $b_0$. Our fits take full account of excited-state contamination.   

\subsubsection{Fitting the correlators $\tilde{C}_{\mu\nu}$ and $C_{\eta_c}$}\label{sec:etacggfits}

We fit $\tilde{C}_{\mu\nu}$ and $C_{\eta_c}$ simultaneously to standard staggered-quark fit forms for two-point correlation functions. These are the same as those given in Eqs.~\eqref{eq:Cmunu-fit} and~\eqref{eq:Ceta-fit} except that there are additional terms that oscillate in time. The fit form for $C_{\eta_c}$ becomes 
\begin{equation} \label{corrfitform_2pt}
	C_{\eta_c} (t) = \sum_n^{N_{\mathrm{n}}} a_{\mathrm{n}}^2 f (E_{\mathrm{n}},t) - (-1)^t\sum_o^{N_{\mathrm{o}}}  a_{\mathrm{o}}^2  f (E_{\mathrm{o}},t) 
\end{equation}
with $f(E,t)$ given in Eq.~\eqref{eq:fdef}. The oscillating terms arise from opposite parity states, all of which are heavier than the ground-state $\eta_c$ and decay faster in the large-time limit. The fit-form for $\tilde{C}_{\mu\nu}$ given in Eq.~\eqref{eq:Cmunu-fit} is likewise extended to include oscillating terms from the same opposite parity states. 

We use the corrfitter package~\cite{corrfitter} to fit our two-point correlation functions, taking as parameters the logarithms of the energy differences (to keep the energy levels ordered) and the logarithms of the amplitudes. We take the prior on the energy differences to be 0.5(2) GeV and the prior width on the amplitudes $a_n$ to be 20\% and $b_n$ (see Eq.~\eqref{eq:Cmunu-fit}) 50\%. All prior widths are orders of magnitude larger than the uncertainties returned by the fit for ground-state values. We find $\chi^2/\mathrm{dof} \lesssim 1$ for all fits.\footnote{We apply a standard procedure to avoid underestimating the low eigenvalues of the correlation matrix and hence the uncertainty. This is described in Appendix D of~\cite{Dowdall:2019bea} which also discusses how to determine $\chi^2$ reliably in that case by including additional noise; we apply that method for $\chi^2$ here.} Our final fits use 4 exponentials, i.e. $N_n=N_o=4$ in Eq.~\eqref{corrfitform_2pt}. We do not include results for values of $t$ below $t_{\text{min}}$ in our fits where $t_{\text{min}}/a$ varies between 6 and 9 depending on lattice spacing for $\tilde{C}_{\mu\nu}$ and is $N_t/8$ for $C_{\eta_c}$. 

The aim of the fits is to determine the ground-state parameters corresponding to $n=0$: $E_0$, $a_0$ and $b_0$. From $b_0$ we can determine the form-factor for $\eta_c \to \gamma\gamma$ using Eq.~\eqref{eq:Fb0def}.  This equation is derived from Eqs.~\eqref{eq:b0def} and~\eqref{eq:Fdef} and assumes that both photons are exactly on-shell. As discussed in Section~\ref{sec:etac2gamma_corrfs}, photon 1 is exactly on-shell but photon 2 is not. We must therefore modify our determination of $F$ from $b_0$ to take account of this, following Eq.~\eqref{eq:Fdef}. Instead of Eq.~\eqref{eq:Fb0def} we must use 
\begin{equation}
\label{eq:Fb0true}
\frac{F_{\text{latt}}(0,q_2^2)}{a} = b_0 \frac{\sqrt{2aM^{\text{latt}}_{\eta_c}}}{aM^{\text{latt}}_{\eta_c}aq_1^y} \, ,
\end{equation}
where $q_2^2$ is the $q^2$ value for photon 2 and $b_0$ is in lattice units. $aM^{\text{latt}}_{\eta_c}$ is the value of the $\eta_c$ mass in lattice units given by $E_0$. 

\begin{table}
  \centering
   \caption{Results for $F_{\text{latt}}(0,q_2^2)$ and $M_{\eta_c}$ in lattice units obtained from our correlator fits (Eqs.~\eqref{corrfitform_2pt} and~\eqref{eq:Fb0true}). The top table shows results from our LOCAL set-up (with local vector current), the lower table those from the ONE-LINK set-up (with 1-link vector current). Column 5 gives the value for $q_2^2$ ($q^2$ for photon 2) in lattice units from Eq.~\eqref{eq:q2val} (these values are close to zero for the ONE-LINK set-up from the way that the momentum twists were chosen). Set 3A for the LOCAL case corresponds to a deliberately mistuned valence $c$ quark mass (see Table~\ref{tab:params}). We fit the correlators for sets 3 and 3A simultaneously so that correlations between them are fully taken into account. }
  \label{tab:f00_metac}
 \begin{tabular}{lcccc}
  \hline
  \hline
 LOCAL &Set & $F_{\mathrm{latt}}(0,q^2_2)/a$ & $aM^{\text{latt}}_{\eta_c}$ & $a^2q^2_2$\\
  \hline
  & 1  & $0.11294(24)$  & $2.36833(41)$    & $0.0964$ \\
  & 2  & $0.115433(94)$ & $2.321816(62)$   & $0.1036$ \\
  & 3  & $0.14031(19)$  & $1.888089(94)$   & $0.0300$ \\
  & 3A & $0.14096(20)$  & $1.868562(95)$   & $-0.0067$ \\
  & 4  & $0.143704(93)$ & $1.845042(31)$   & $0.0288$\\
  & 5  & $0.19163(22)$  & $1.369825(55)$   & $0.0110$ \\
  & 6  & $0.29303(54)$  & $0.897086(57)$   & $0.0024$ \\
  \hline
  \hline
  ONE-LINK &Set & $F_{\mathrm{latt}}(0,q^2_2)/a$ & $aM^{\text{latt}}_{\eta_c}$ & $a^2q^2_2$\\
  \hline
  & 1  & $0.08416(21)$ & $2.38605(24)$    & $0.0002$ \\
  & 3  & $0.11649(22)$ & $1.89533(13)$    & $-0.0001$ \\
  & 5  & $0.17396(36)$ & $1.371840(93)$   & $0.0000$ \\
   \hline
    \hline
\end{tabular}
 \end{table}

Table~\ref{tab:f00_metac} gives our results for $aM^{\text{latt}}_{\eta_c}$ and $F_{\text{latt}}(0, q_2^2)/a$ on each set of gluon field configurations and for both the LOCAL and ONE-LINK set-ups. The results also include those for a mistuned valence $c$ mass on set 3, denoted set 3A. Notice the small statistical uncertainties, well below 1\% in $F/a$ and $aM_{\eta_c}$, typical of lattice QCD calculations with heavy quarks. We include values for the off-shellness of photon 2, $q_2^2$, given by
\begin{equation}
\label{eq:q2val}
q_2^2 = ((M^{\text{latt}}_{\eta_c})^2-q_1^y)^2 - (q_1^y)^2 = M^{\text{latt}}_{\eta_c}(M^{\text{latt}}_{\eta_c}-2q_1^y) \, .
\end{equation}
$q_2^2$ is very close to zero in all cases, being at most 2\% of $M_{\eta_c}^2$ on the coarsest lattices for the LOCAL set-up. We will discuss how we fit the results for $F$ (allowing for the off-shellness $q_2^2$) to obtain a value for $F(0,0)$ in the physical continuum limit in Section~\ref{sec:fit_form_F}. 

First we demonstrate that results can be obtained with a sub-set of the correlation functions that we have calculated here. Figure~\ref{fig:vary-width} shows the results for $F$ if we restrict the range of integration over $t_{\gamma_1}$ to a time distance of $t_{\text{width}}$ either side of $t_{\gamma_2}$. In keeping with Fig.~\ref{fig:integrand}, we see that the result for $F$ reaches its final value very quickly as a function of $t_{\text{width}}$ (in less than 1 fm). The sum over $t_{\gamma_1}$ could be truncated in this case with no loss of accuracy. We perform the full sum here, however. 

In Appendix~\ref{sec:T-test} we further show that we can restrict the fit of the two-point function $\tilde{C}_{\mu\nu}(t)$ to a set of specific $t\equiv t_{\gamma_2}-t_{\eta_c}$ values rather than fitting the full $t$ range.  Here, however, we use the results from our full fit. 

\begin{figure}
	\begin{center}
		\includegraphics[width=0.45\textwidth]{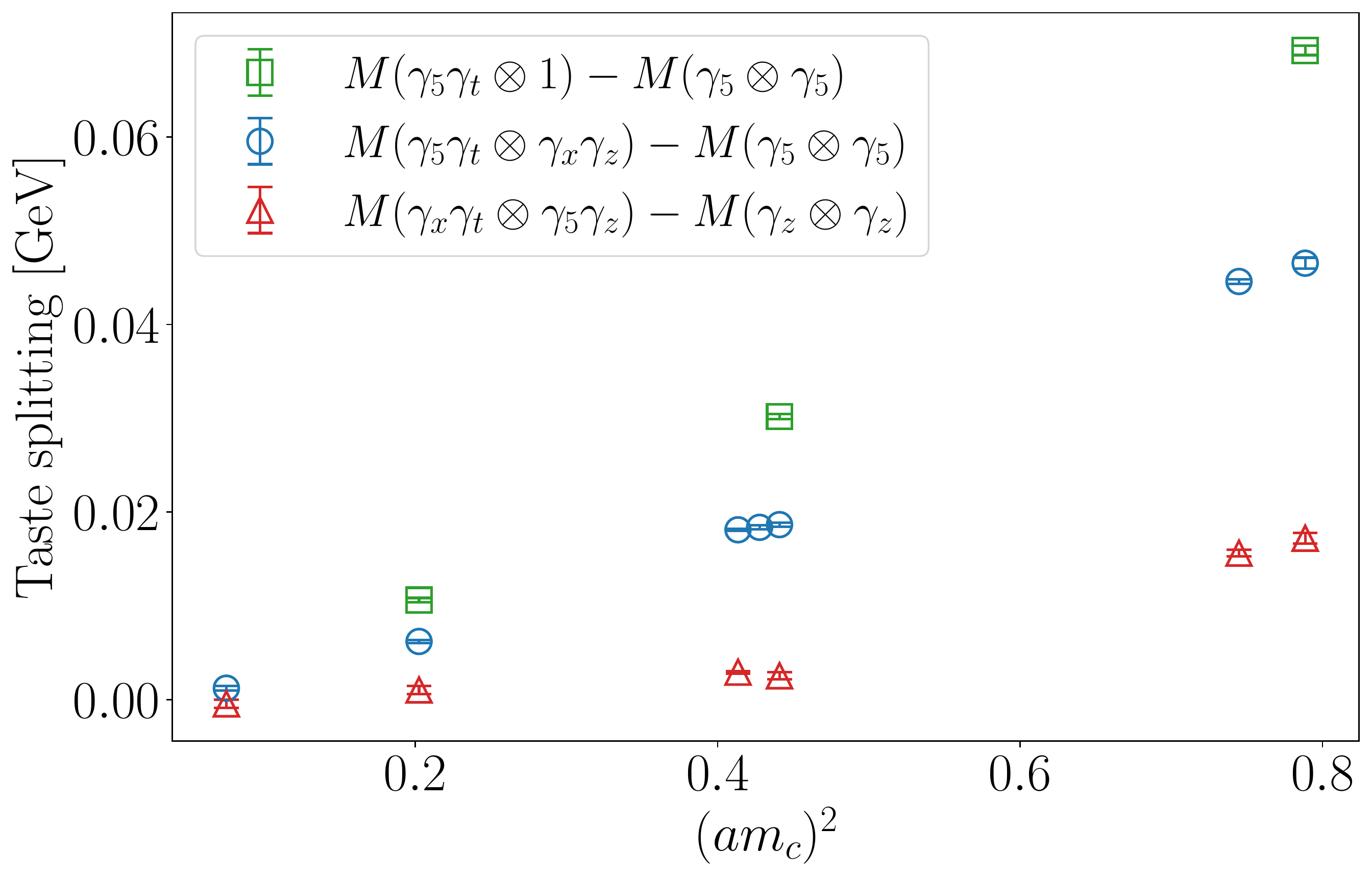}
		\caption{Mass differences between different tastes of charmonium mesons. The green squares show the difference between the $\eta_c$ meson used in our ONE-LINK set-up for $\eta_c \to \gamma\gamma$ and the `Goldstone' (spin-taste $\gamma_5 \otimes \gamma_5$) meson used in our $J/\psi \to \gamma \eta_c$ calculation. The blue circles show the same difference for the $\eta_c$ used in our LOCAL set-up. We also show in red triangles the mass difference between the $J/\psi$ interpolated by $\gamma_x \gamma_t \otimes \gamma_5\gamma_z$ (used for $J/\psi \to  \gamma\eta_c$, see Section~\ref{sec:JpsiGammaEtac}) and the $J/\psi$ interpolated by $\gamma_z \otimes \gamma_z$ whose mass in lattice units we obtain from~\cite{Hatton:2020qhk}. In all cases the mass differences are small (less than 2\% of the meson mass) and fall to zero as $a\to 0$ as a discretisation effect. }
		\label{fig:etac_taste_split}
	\end{center}
\end{figure}

As discussed in Sec.~\ref{sec:etac2gamma_corrfs}, we use different staggered spin-taste representations for the mesons in different parts of our calculations. Here we test that the different representations agree in mass in the continuum limit. We use an interpolating operator with spin-taste $\gamma_5 \gamma_t \otimes \gamma_x \gamma_z$ for the pseudoscalar $\eta_c$ meson in our LOCAL set-up and $\gamma_5 \gamma_t \otimes 1$ for our ONE-LINK set-up.
In the study of $J/\psi  \to \gamma \eta_c$, we instead take a $\gamma_5 \otimes \gamma_5$ (Goldstone) interpolator for the $\eta_c$ meson.
This latter spin-taste corresponds to the lightest $\eta_c$ in the taste-multiplet and the one used, for example, in~\cite{Hatton:2020qhk}. 
In Fig.~\ref{fig:etac_taste_split}, we compare the masses of the different $\eta_c$ tastes determined from our fits to the suite of correlation functions that we have.
The plot shows that the mass differences between the different tastes are very small, at most a few tens of MeV for a meson with mass of 3 GeV. The mass differences vanish in the continuum limit as expected for a discretisation effect. In the next Section we show the impact that the different spin-taste set-ups have on the determination of $F(0,0)$ as a function of lattice spacing.

\subsubsection{Taking the physical-continuum limit} \label{sec:fit_form_F}

\begin{figure}
	\begin{center}
		\includegraphics[width=0.45\textwidth]{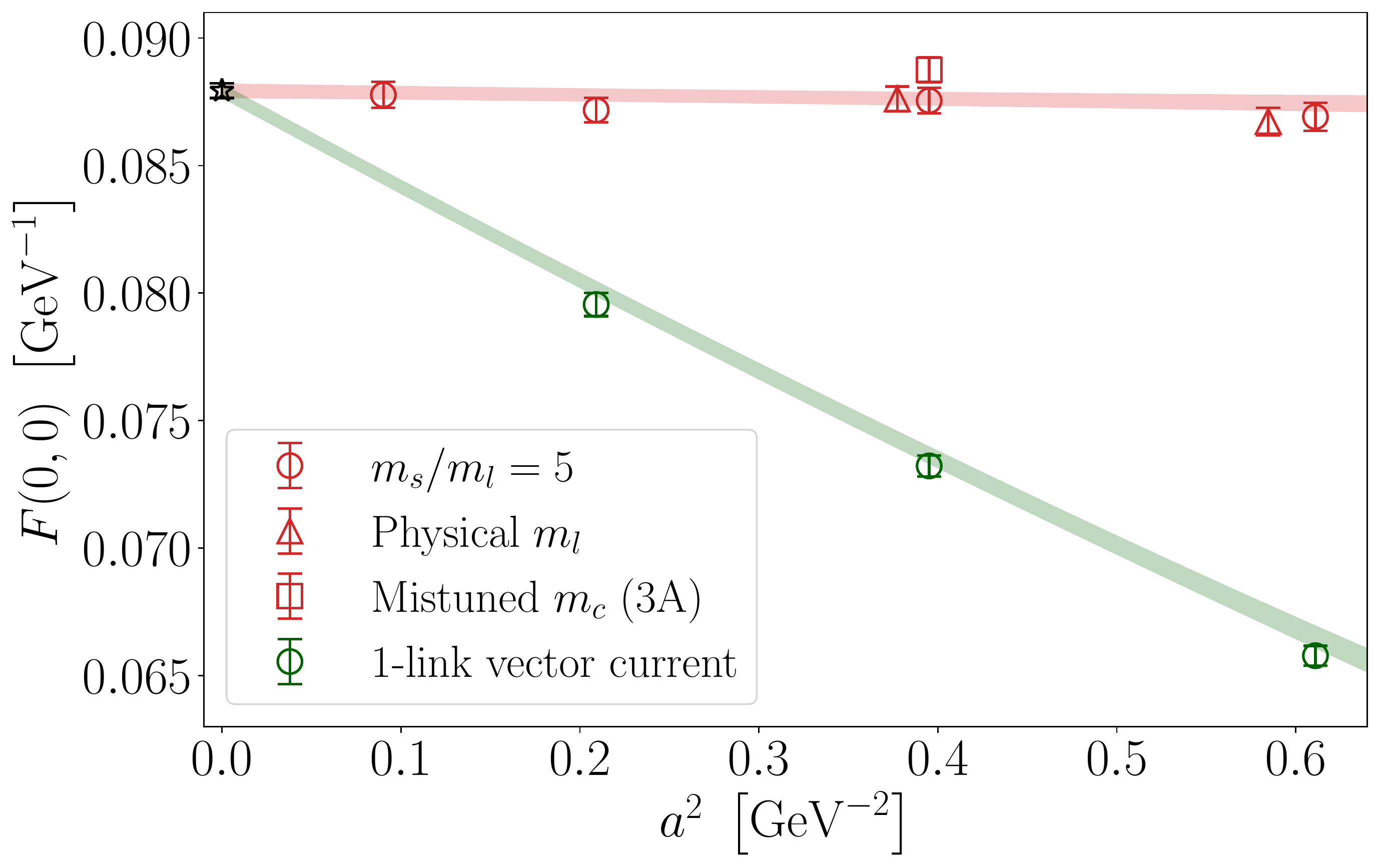}
		\caption{Data points show our lattice QCD results for the $\eta_c \to \gamma\gamma$ form factor, in red for the LOCAL set-up (with local vector current) and green for the ONE-LINK set-up (with 1-link vector current). The LOCAL points have been corrected for the slight off-shellness of one photon (see text). The LOCAL points include one at a deliberately mistuned $c$ quark mass (open red square). The blue and pink bands show our chiral/continuum fit to the points, applying Eq.~\eqref{eqn:F_fit_form} simultaneously for the LOCAL and ONE-LINK points. The fit bands are plotted as a function of $a$ at the physical quark mass point. The physical result in the continuum limit is then shown by the black star.}
		\label{fig:continuum_extrap}
	\end{center}
\end{figure}

To obtain a physical result for the form factor $F(0,0)$, we fit our lattice data for $F(0,q_2^2)$ from Eq.~(\ref{eq:Fb0true}) (given in Table~\ref{tab:f00_metac} and plotted in Fig.~\ref{fig:continuum_extrap}) to a function that accounts for discretisation effects and mistunings of the sea and valence quark masses, as well as allowing for the small amount by which photon 2 is off-shell.
The data is fit using the lsqfit package~\cite{lsqfit} and our fit form is 
\begin{align}
	&\frac{F_{\mathrm{latt}}^{(t)}(0,q_2^2)}{a} = \frac{F(0,0)}{(1-\frac{q_2^2}{M_{\text{pole}}^2})} \times \nonumber \\
	& \hspace{0mm} \Bigg[1+\sum_{i=1}^{i_{\text{max}}} \kappa_{a\Lambda}^{(i,t)}\left(a \Lambda^{(t)}\right)^{2 i} + \kappa_{\mathrm{val}, c} \delta^{\mathrm{val}, c} + \kappa_{\mathrm{sea}, c} \delta^{\mathrm{sea}, c} \nonumber \\
	& \hspace{3mm} + \kappa_{\mathrm{sea}, u d s}^{(0)} \delta^{\mathrm{sea}, u d s}\left\{1+\kappa_{\mathrm{sea}, u d s}^{(1,t)}(a\tilde{\Lambda} )^{2}+\kappa_{\mathrm{sea}, u d s}^{(2,t)}(a\tilde{\Lambda} )^{4}\right\}\Bigg] \label{eqn:F_fit_form}
\end{align}
where $F_{\mathrm{latt}}^{(t)}(0,q_2^2)$ is the lattice data with superscript $t$ denoting the taste, i.e. either the LOCAL (local vector) or ONE-LINK (1-link vector) cases. 
 $F(0,0)$ is the form factor in the limit of vanishing lattice spacing and physical masses that we wish to determine.
The factor of $(1-q_2^2/M_{\text{pole}}^2)$ allows us to adjust for the amount by which photon 2 is off-shell with a simple one-pole parameterisation which was tested in~\cite{Dudek:2006ej, Chen:2016yau}. The pole form was found to work well, with the pole mass $M_{\text{pole}}$ taking a value around the $J/{\psi}$ mass.  Here we will take $M_{\text{pole}}$ as a fit parameter, with prior 3.0(3) GeV. As discussed in Section~\ref{sec:Ctilde}, our $q_2^2$ values are close to zero here (see Table~\ref{tab:f00_metac}) and the pole factor only has a small effect, at most 2\% on our coarsest lattices in the LOCAL set-up and less than 1\% in other cases. 

Equation~\eqref{eqn:F_fit_form} takes discretisation effects into account, mainly through the first non-trivial term in the square brackets. These discretisation effects arise from the HISQ action but also through the trapezoidal integration used to determine $\tilde{C}_{\mu\nu}$ and the oscillating term contribution to that integral (Appendix~\ref{sec:osc-sum}). We allow for the size of these discretisation effects to be set by scale $\Lambda$ and include terms up $(a\Lambda^{2i_{\text{max}}})$. Since we want to fit both the LOCAL and ONE-LINK cases with the same fit form, although they differ significantly in their discretisation effects (see Fig.~\ref{fig:continuum_extrap}), we give them different $\Lambda$ parameters and choose $\Lambda^{(t)}$ using the Empirical Bayes criterion. This means varying $\Lambda^{(t)}$ in fits to the two cases (with independent fit parameters, $\kappa_{a\Lambda}^{(i,t)}$) and taking the values that maximise the Bayes factor~\cite{Lepage:2001ym}. We also allow for discretisation effects coming from the sea in the final term, including terms up to $(a\tilde{\Lambda})^4$. We take $\tilde{\Lambda} = 1 \; \mathrm{GeV}$ in both cases but use independent fit parameters, $\kappa_{\mathrm{sea},uds}^{(1/2,t)}$.

We allow for mistuning of sea and valence quark masses in Eq.~\eqref{eqn:F_fit_form} with the terms containing $\delta$. 
The mistunings of the $c$ valence and sea quark masses and the $u$, $d$ and $s$ sea quark masses are expressed as
\begin{align}
	\delta^{\mathrm{val}, c} &=\frac{a m_{c}^{\mathrm{val}}-a m_{c}^{\text {tuned }}} {am_{c}^{\mathrm{tuned}}} \label{eqn:mt_val_charm} \\
	\delta^{\mathrm{sea},c} &=\frac{a m_{c}^{\mathrm{sea}}-a m_{c}^{\text {tuned }}} {am_{c}^{\mathrm{tuned}}} \label{eqn:mt_sea_charm} \\
	\delta^{\mathrm{sea}, uds} &= \frac{2 am_{l}^{\mathrm{sea}}+am_{s}^{\mathrm{sea}}-2 am_{l}^{\mathrm{tuned}}-am_{s}^{\mathrm{tuned}}}{10 am_{s}^{\mathrm{tuned}}}
	\label{eqn:mt_sea_uds}
\end{align}
respectively. 

To tune the valence and sea $c$ quark mass, we take $am_c^{\mathrm{tuned}}$ to be
\begin{align}
	a m_{c}^{\text {tuned }}=a m_{c}^{\mathrm{val}}\left(\frac{M_{J / \psi}^{\mathrm{expt}}}{M_{J / \psi}}\right)^{1.5} \label{eqn:defn_amctuned}
\end{align}
where $M_{J / \psi}^{\mathrm{expt}}= 3.0969 \: \mathrm{GeV}$ from~\cite{Workman:2022ynf}, and lattice values for $aM_{J / \psi}$ are obtained from Table III in~\cite{Hatton:2020qhk}. The power of 1.5 is empirically chosen, based on the results from~\cite{Hatton:2020qhk} (the power is not 1, because of the binding energy inside the $J/\psi$).  

The tuned $s$ quark mass is given by 
\begin{align}
	am_{s}^{\text {tuned }}=am_{s}^{\mathrm{val}}\left(\frac{M_{\eta_{s}}^{\text {phys }}}{M_{\eta_{s}}}\right)^{2} \label{eqn:mstuned}
\end{align}
from leading-order chiral perturbation theory. 
$am_l^{\mathrm{tuned}}$ is found by dividing the value for $m_{s}^{\text {tuned }}$ in Eq.~\eqref{eqn:mstuned} by the ratio~\cite{Bazavov:2017lyh}
\begin{align}
	\frac{m_{s}^{\text {phys }}}{m_{l}^{\text {phys }}}=27.18(10).
\end{align}
For all except set 5 we use $\delta m_{uds}^{\mathrm{sea}}/m_s$ values from Table I of~\cite{Chakraborty:2014aca}.
For set 5 in Table~\ref{tab:params}, values in lattice units for $M_{\eta_{s}}$ are taken from~\cite{McLean:2019sds} and the \lq physical\rq\: value of the $\eta_s$ meson, $688.5(2.2) \: \mathrm{MeV}$, is taken from~\cite{Dowdall:2013rya}. The $\eta_s$ is an unphysical pseudoscalar $s\overline{s}$ meson whose mass can nevertheless be determined in terms of $\pi$ and $K$ masses in lattice QCD. This gives a $\delta^{\mathrm{sea}, uds}$ value of 0.0297(17) for set 5. 

Since the dependence of $F(0,0)$ on quark masses is a physical effect it should be the same (up to discretisation errors) for the LOCAL and ONE-LINK cases. We therefore take the $\kappa_{\mathrm{sea},c}$, $\kappa_{\mathrm{val},c}$ and $\kappa_{\mathrm{sea},uds}^{(0)}$ parameters to be the same in the two cases when fitting them simultaneously.

The parameters to be determined by the fit are $M_{\text{pole}}$, $\kappa_{a\Lambda}^{(i)}$ for $i=1$ to $i_{\text{max}}$, $\kappa_{\mathrm{sea}, c}$, $\kappa_{\mathrm{val}, c}$, and $\kappa_{\mathrm{sea}, u d s}^{(j)}$ for $j=0,1,2$. As discussed above, we take the parameters to be independent for our two set-ups when they correspond to discretisation effects, but otherwise take them to be the same. 
For priors, we use $F(0,0) = 0.1(1)$, $M_{\text{pole}}$ = 3.0(3) GeV, $\kappa_{a\Lambda}^{(i)} = 0(1)$ (except for $i=1$ for the ONE-LINK case where we take the prior to be 0(2) from inspection of Fig.~\ref{fig:continuum_extrap}), $\kappa_{\mathrm{val}, c} = 0(2)$ (from comparison of results from sets 3 and 3A), $\kappa_{\mathrm{sea}, uds}^{(k)} = 0(1)$ and $\kappa_{\mathrm{sea}, c} = 0.0(1)$ (since we expect the effect of $c$ in the sea to be minor). For our final fit we take the largest coefficient for discretisation effects, $i_{\mathrm{max}}=3$. Our preferred fit is a joint fit to the LOCAL and ONE-LINK data, but we obtain almost identical results from fitting simply the LOCAL results.

Figure~\ref{fig:continuum_extrap} shows our lattice results (now in physical units) for both the LOCAL and ONE-LINK set-ups as a function of squared lattice spacing. The lattice results have been adjusted to correspond to the $F(0,0)$ on-shell point, i.e. $F_{\text{latt}}(0,q_2^2)$ has been multiplied by the pole term $(1-q_2^2/M^2_{\text{pole}})$. The uncertainty on the lattice results is dominated by the correlated uncertainty in the value of the lattice spacing. Also shown is our continuum fit using Eq.~\eqref{eqn:F_fit_form} to both sets of data simultaneously. Notice that the discretisation effects are much larger in the ONE-LINK case than in the LOCAL case. Using the Empirical Bayes criterion~\cite{Lepage:2001ym} we find that the optimal $\Lambda$ is 0.10 GeV for LOCAL and 0.49 GeV for ONE-LINK. The larger discretisation effects for ONE-LINK are not surprising because the vector currents in that case are 1-link operators with tree-level discretisation errors. The LOCAL case, in contrast, uses local vector current operators that have no tree-level errors at any order in $a$. The bands plotted on Fig.~\ref{fig:continuum_extrap} correspond to the fit at tuned sea masses as a function of lattice spacing, i.e. $F(0,0)[1+\sum_i \kappa^{(i,t)}_{a\Lambda}(a\Lambda^{(t)})^{2i}]$ (see Eq.~\eqref{eqn:F_fit_form}), with $i_{\text{max}}=3$.  

The result we obtain for $F(0,0)$ in the continuum limit from the joint fit to the LOCAL and ONE-LINK results is $\fitF$ with a $\chi^2/\mathrm{dof}$ of 0.8. This confirms that the LOCAL and ONE-LINK results can readily be fit to a common continuum value, providing a test that HISQ taste effects are purely lattice artefacts. The result from fitting LOCAL alone is very similar, not surprisingly because we have the best coverage of lattice spacing and sea masses in that case and discretisation effects are smaller than for the ONE-LINK case. In Section~\ref{sec:results} we will discuss additional sources of systematic error that must be accounted for in our final result.

\begin{figure}
	\begin{center}
		\includegraphics[width=0.45\textwidth]{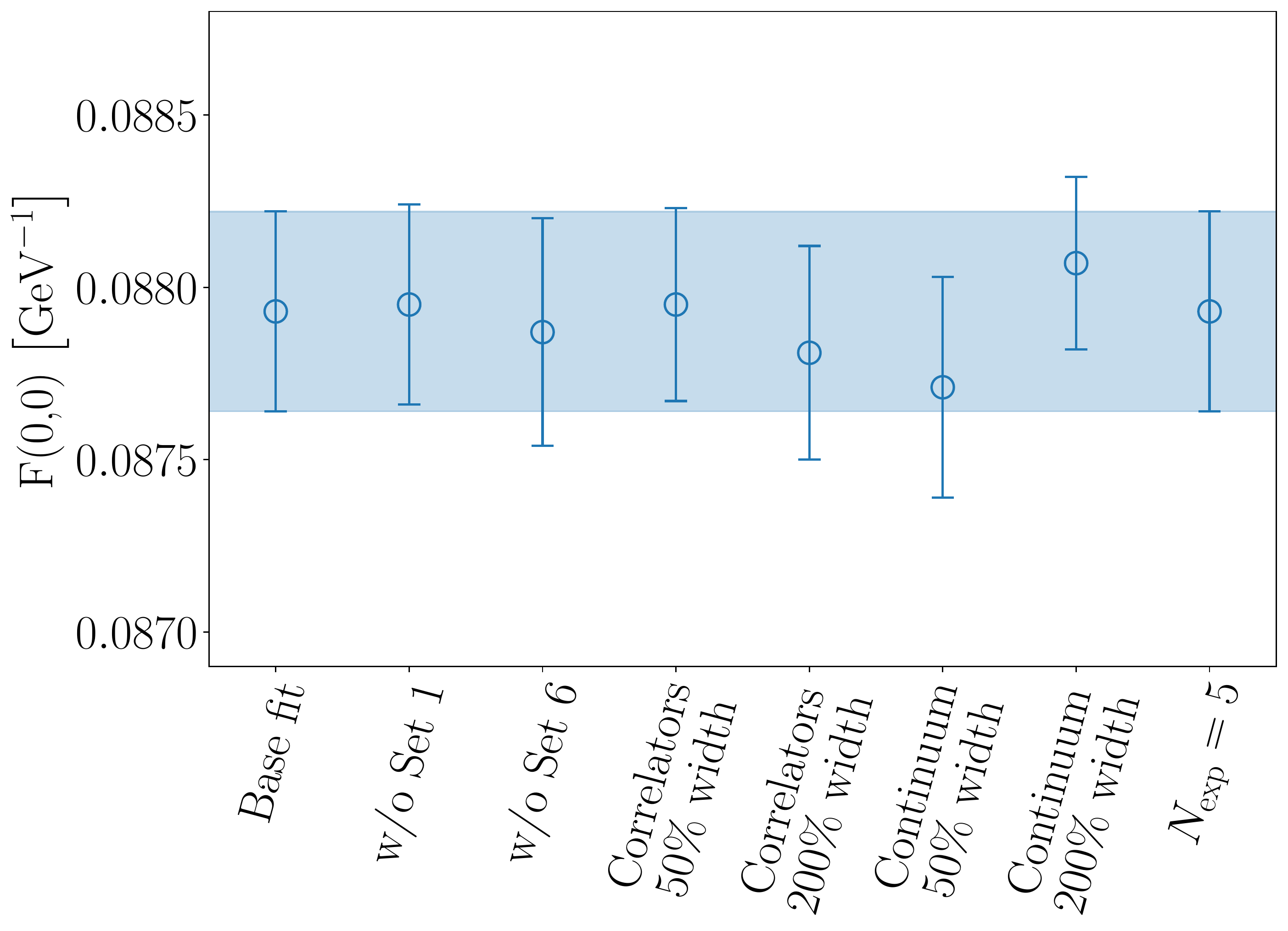}
		\caption{The value of $F(0,0)$ in the limit of vanishing lattice spacing and physical quark masses obtained from variations to our base fit. These include (from left to right) dropping the coarsest and finest datasets, changing all the prior widths in our correlator fits, changing all the prior widths in our chiral/continuum fits and adding an additional normal and oscillating exponential to our correlator fits. Note that the values for $\Lambda$ in Eq.~\eqref{eqn:F_fit_form} are fixed (see text) under these fit variations.}
		\label{fig:F00_fit_vars}
	\end{center}
\end{figure}

First we discuss the stability of our fitted value for $F(0,0)$ as we change details of our fits. This is shown in Fig.~\ref{fig:F00_fit_vars}, where we plot the value of $F(0,0)$ in the physical/continuum limit on changing some aspect of either the correlator fits or the chiral/continuum fit. The preferred (base) fit described above is given on the left. Variations include making all the priors a factor of 2 smaller or larger and dropping datasets at either end of the lattice spacing range. We see very little variation in the final answer under any of these variations, showing that our result is robust. 

\subsubsection{Additional systematic uncertainties} \label{sec:etac2gamma_uncertainties}

As discussed in Section~\ref{sec:method} our lattice QCD calculation does not include quark-line disconnected contributions or QED effects. Here we estimate the size of these and include an additional systematic uncertainty to allow for them. 

In~\cite{Hatton:2020qhk} the small difference between the mass of the $\eta_c$ obtained from a lattice QCD calculation including connected correlation functions only and the mass found in experiment was interpreted as the effect on the mass of missing quark-line disconnected correlation functions. The Wick contractions for the disconnected correlation functions include the annihilation of the $\eta_c$ into gluons and allow mixing between the $\eta_c$ and other flavour-singlet pseudoscalar states. The difference found in~\cite{Hatton:2020qhk} amounted to 0.2\% of the $\eta_c$ mass. 
A reasonable estimate of the impact of these disconnected diagrams on the $\eta_c$ wavefunction (which can be related to the decay amplitude to $\gamma\gamma$ in the nonrelativistic limit) would then also be 0.2\%, giving a 0.4\% uncertainty in the decay width. That this is reasonable is confirmed by our fit, which includes the dependence of $F(0,0)$ on the $\eta_c$ mass and returns a coefficient for $\kappa_{\text{val},c}$ close to 1. This dependence is visible in Fig.~\ref{fig:continuum_extrap} and Table~\ref{tab:f00_metac}, comparing values for tuned and mistuned $m_c$.

We can also consider the impact of another class of quark-line disconnected diagrams in which the $\eta_c$ radiates a photon before annihilating to gluons. The quark loop generated from the gluons then also annihilates to a photon. The impact of this diagram should be very small, because of suppression both by powers of $\alpha_s$ and by quark mass effects since the sum of the electric charges of the light quarks in the sea is zero. We therefore expect a contribution smaller than a relative size of $\alpha_s^2m_s^2/m_c^2\approx0.2\%$. This does not then modify our estimate of the uncertainty from missing disconnected diagrams as 0.2\% from above. 

The impact of the $c$ quark's electric charge on the decay amplitude (also missing in our calculation) can be estimated from the impact of QED on the $\eta_c$ decay constant, determined in~\cite{Hatton:2020qhk}. This effect was 0.17\%, so we allow an additional uncertainty of 0.2\% in $F(0,0)$ for this. Further QED corrections to the decay width coming from additional radiation are tiny since there are no electrically charged particles in either the initial or final states. 
By charge conjugation, the $\eta_c$ can't decay to $\gamma \gamma \gamma$, so QED corrections to the decay rate would come from $\eta_c$ to $\gamma \gamma \gamma \gamma$.
This would be suppressed by a further 2 powers of $\alpha$, and thus negligible. 

Adding these two 0.2\% systematic uncertainties in quadrature gives an additional uncertainty of 0.3\% in $F(0,0)$ and 0.6\% to the decay rate. 

\subsection{Results}
\label{sec:results}

We take our final result for the form factor for $\eta_c\to \gamma\gamma$ from the joint fit to the LOCAL and ONE-LINK cases, giving 
\begin{align}
	\finalF \label{eqn:final_F00}
\end{align}
which has a total uncertainty of $\finalError\%$.
The second uncertainty in Eq.~(\ref{eqn:final_F00}) comes from the additional systematic uncertainties that we estimate in  Section~\ref{sec:etac2gamma_uncertainties}.

\begin{table}
\centering
 \caption{Error budgets for our results for $F(0,0)$ and $\hat{V}(0)$ (discussed in Section~\ref{sec:Jpsiresults}). The top 7 entries come from our fits to Eq.~\eqref{eqn:F_fit_form} and Eq.~\eqref{eqn:Vhat_fit_form} respectively. The uncertainty labelled `$q^2$ dependence' arises from the tuning to $q^2=0$ in the $F(0,0)$ case and from capturing the $q^2$-dependence in the $\hat{V}(0)$ case. The lowest 2 entries are additional systematic uncertainties from missing quark-line disconnected correlations functions and QED effects. These are discussed in Section~\ref{sec:etac2gamma_uncertainties} for $F(0,0)$ and in Section~\ref{sec:Jpsi_uncertainties} for $\hat{V}(0)$, where the two sources of uncertainty are combined together. }
  \label{tab:errorbudget}
\begin{tabular}{lcc}
    \hline
    \hline
    & $F(0,0)$ & $\hat{V}(0)$\\
    \hline
    Statistics & 0.17 & 0.29 \\
    $w_0/a$ & 0.05  & 0.01 \\
    $w_0$ & 0.25 & 0.07 \\
    $a^2\to 0$ & 0.08 & 0.23 \\
    Valence mistuning & 0.01 & 0.03 \\
    Sea mistuning & 0.05 & 0.05 \\
    $q^2$ mistuning & 0.01 & 0.10 \\
    Missing `disconnected' correlators & 0.2 & 0.4 \\
    Missing QED & 0.2 & - \\
    \hline
    Total & 0.43 & 0.56 \\
    \hline
    \hline
  \end{tabular}
  \end{table}

The error budget for this result is given in Table~\ref{tab:errorbudget}. We see that the dominant uncertainty is that from fixing the lattice spacing using $w_0$. The next most important uncertainties come from statistics and from systematic errors from missing quark-line disconnected diagrams and QED effects. 

We use our value for $F(0,0)$ in Eq.~(\ref{eqn:final_F00}) and the formula in Eq.~(\ref{eq:decayrate}) to find the decay width $\Gamma (\eta_c \to \gamma \gamma)$.
For the $\eta_c$ mass in Eq.~(\ref{eq:decayrate}), we use the experimental value  $M_{\eta_c}^{\mathrm{exp}} = 2.9839(4) \; \mathrm{GeV}$ (the average from~\cite{Workman:2022ynf}) since this is a purely kinematic factor. We also take $1/\alpha=137.036$ from~\cite{Workman:2022ynf}, noting that the momentum scale for this decay is a relatively low one.
We obtain the decay width with an uncertainty of $\finalWidthEtacGammaGammaUncertainty$ as
\begin{align}
\label{eq:finalwidthetac}
	\finalWidthEtacGammaGamma \,.
\end{align}
Using the experimental value for the $\eta_c$ total width of 32.0(7) MeV~\cite{Workman:2022ynf}, this corresponds to a branching fraction of
\begin{align}
\label{eq:finalBretac}
	\finalBrEtacGammaGamma \,.
\end{align}
The third uncertainty here is from the experimental total width and it dominates over the lattice QCD uncertainties. 

\subsection{Discussion} 
\label{sec:etac-discussion}

\begin{figure}
	\begin{center}
		\includegraphics[width=0.45\textwidth]{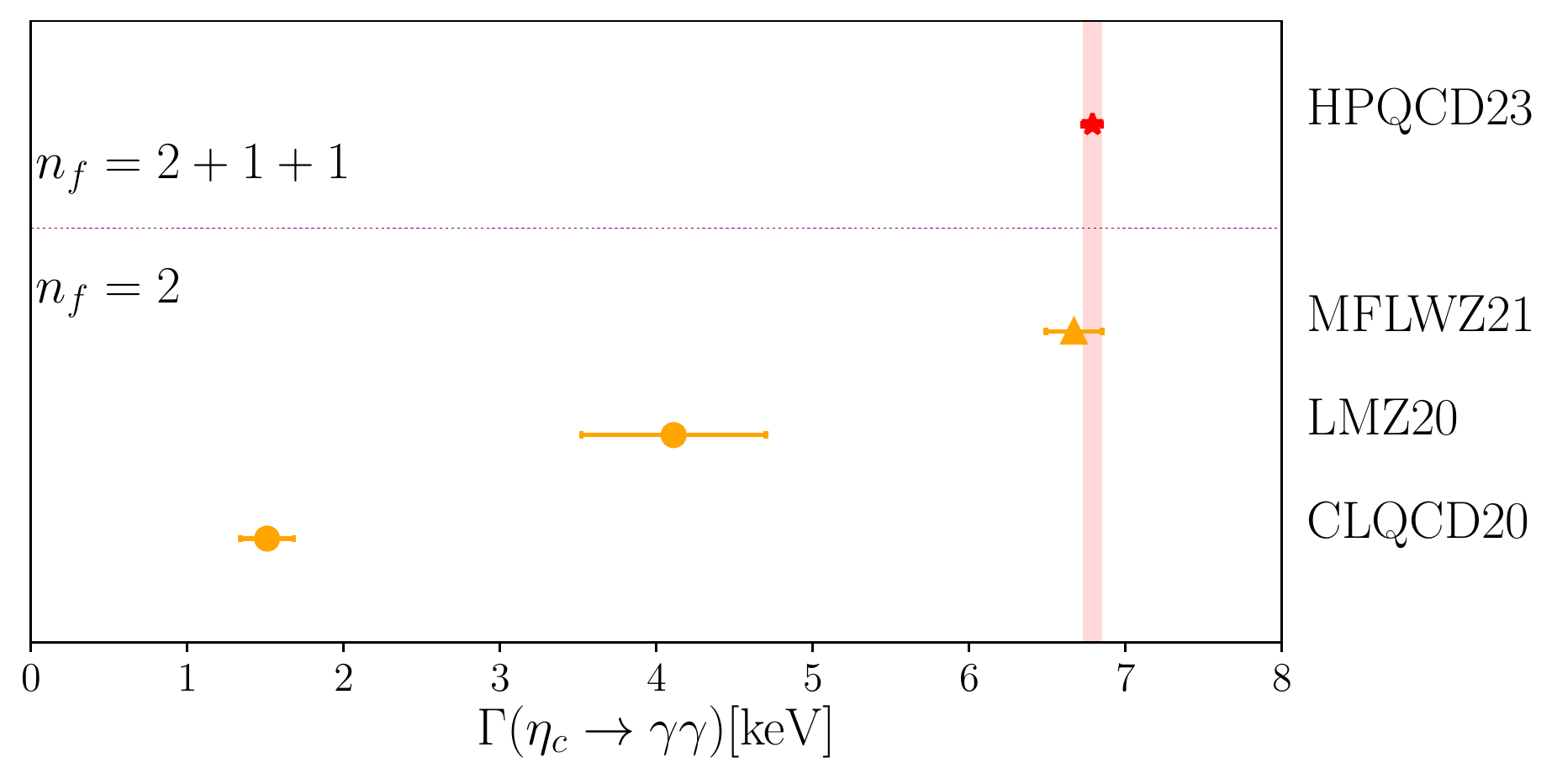}
		\caption{A comparison of full lattice QCD results for the width for $\eta_c \to \gamma \gamma $ decay. The result obtained here is denoted `HPQCD23' (red asterisk, error bar same size as symbol) and uses gluon field configurations that include $n_f=2+1+1$ sea quark flavours at four values of the lattice spacing to determine a physical result. Earlier results use $n_f=2$ gluon field configurations at two values of the lattice spacing (orange filled circles) or three values (orange filled triangle). The points denoted `MFLWZ21' from ~\cite{Meng:2021ecs} and `LMZ20' from~\cite{Liu:2020qfz} are from determinations of the rate in the continuum limit including an estimate of systematic errors, although no error for missing $s$ quarks in the sea is included. The point denoted `CLQCD20' corresponds to the value quoted at the finest lattice spacing used in~\cite{Chen:2020nar} with no extrapolation to the continuum limit. The red band carries our result down the plot for comparison.}
		\label{fig:compF}
	\end{center}
\end{figure}

\begin{figure}
		\includegraphics[width=0.45\textwidth]{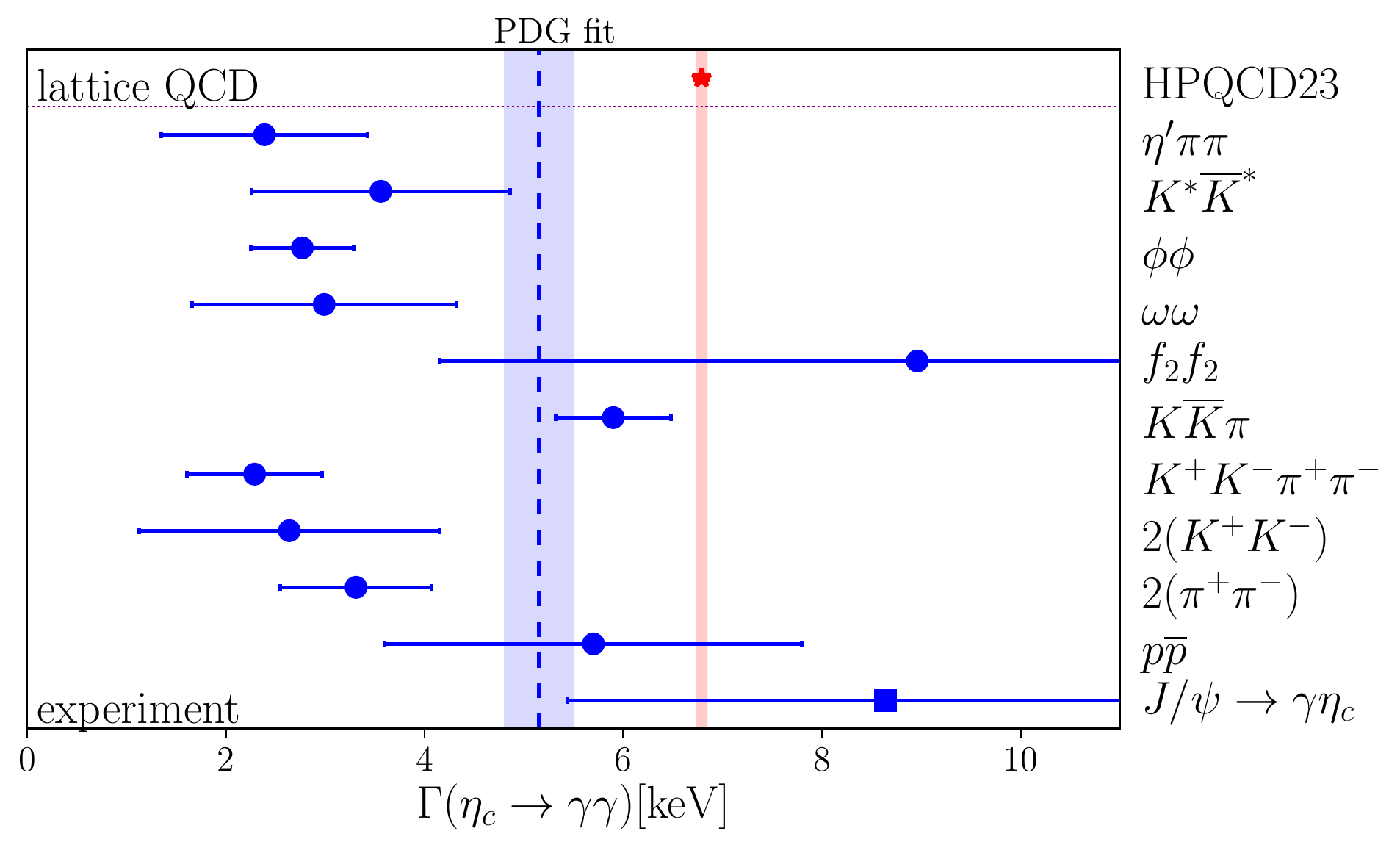}
		\caption{A comparison of our lattice QCD result (HPQCD23, red asterisk and red band) for the decay width $\Gamma(\eta_c \to \gamma \gamma)$ to experimental values taken from~\cite{Workman:2022ynf}. Blue circles denote determinations of the product $\Gamma(\eta_c \to i)\Gamma(\eta_c\to \gamma\gamma)/\Gamma_{\text{total}}(\eta_c)$, where the decay channel $i$ is listed on the right. The values plotted are derived from either single experimental results or PDG average values from several experiments for the product above, divided by the branching fraction $\mathcal{B}(i)$, again using either a single result or the PDG average. Uncertainties are combined in quadrature. The blue square denotes a determination from the product of branching fractions for $J/\psi \to \gamma \eta_c$ and $\eta_c \to \gamma\gamma$ from BESIII~\cite{BESIII:2012lxx} combined with the PDG average for the branching fraction $\mathcal{B}(J/\psi \to \gamma \eta_c)$. The blue band shows the result of the PDG fit to the experimental data shown here along with other results (e.g. for ratios of branching fractions). The fit has a $\chi^2$ of 118 for 81 degrees of freedom~\cite{Workman:2022ynf}, reflecting the inconsistencies in the experimental data seen above. 
				}
		\label{fig:compGamma-expt}
\end{figure}

Figure~\ref{fig:compF} compares our result for $\Gamma(\eta_c \to \gamma \gamma)$ from Eq.~\eqref{eq:finalwidthetac} with earlier lattice QCD results obtained on gluon field configurations that include sea quarks. It is more natural to compare results for $F(0,0)$ from lattice calculations but some of the earlier results do not include this information. The values plotted come from~\cite{Meng:2021ecs} (MFLWZ21),~\cite{Liu:2020qfz} (LMZ20) and~\cite{Chen:2020nar} (CLQCD20) that all work with twisted mass quarks on $n_f=2$ gluon field configurations (i.e. including $u$ and $d$ quarks in the sea but no $s$ quarks) at two or three values of the lattice spacing. We do not include the earlier value from CLQCD in~\cite{Chen:2016yau} that is superseded by~\cite{Chen:2020nar}. CLQCD20 does not include a continuum extrapolation; we take the value as that quoted for their finest lattice. The uncertainty quoted does not include systematic errors. We plot the continuum results for the $\eta_c\to\gamma\gamma$ decay rate quoted by MFLWZ21 and LMZ20, with systematic uncertainties combined in quadrature. Because the sea content is unphysical for the $n_f=2$ case, those results will not necessarily agree with ours; no uncertainty is included in CLQCD20, LMZ20 or MFLW21 for missing $s$ quarks in the sea. Our result is larger than the two earlier values with a tension exceeding $3\sigma$, but agrees well with that from MFLWZ21.  We note that the three $n_f=2$ results do not agree well with each other, however. Our value is obtained on $n_f=2+1+1$ gluon field configurations (i.e. with a realistic sea quark content) at four values of the lattice spacing and includes systematic errors for missing quark-line disconnected diagrams and QED. Our total uncertainty is smaller than that of the earlier values.  Further lattice calculations with a full sea quark complement and uncertainties comparable to ours are needed. 

Figure~\ref{fig:compGamma-expt} compares our result for $\Gamma(\eta_c\to \gamma\gamma)$ to results from experiment. There is a lot of experimental information that has a bearing on $\Gamma(\eta_c \to \gamma\gamma)$, generally obtained as products of branching fractions depending on the method of $\eta_c$ production and the decay mode observed. The PDG~\cite{Workman:2022ynf} combines this information to yield a fit value for $\Gamma(\eta_c\to\gamma\gamma)$ of 5.15(35) keV. This is shown by the blue band in Fig.~\ref{fig:compGamma-expt}. It is lower than our lattice QCD result by \diffpdgsigma$\sigma$, where we have combined lattice and PDG fit uncertainties in quadrature (the PDG fit uncertainty dominates) to obtain $\sigma$. This denotes very significant tension between the Standard Model and experiment, which could be taken as an indication of new physics. The PDG fit returns a large $\chi^2$ of 118 for 81 degrees of freedom~\cite{Workman:2022ynf}, however, and this calls into question the reliability of both their central fit value and its uncertainty. We note that an alternative global fit of $\eta_c$ data which does not include any pre-1995 results, has a better $\chi^2$ and gives a larger partial width for $\eta_c\to \gamma\gamma$ decay of ${5.43} {{+0.41}\atop{-0.38}}$ keV~\cite{Wang:2021dxw}. This shows a reduced, but still sizeable, tension of 3.3$\sigma$ with our lattice QCD result.

Individual experimental results have much larger uncertainties and a large spread of central values. Figure~\ref{fig:compGamma-expt} shows values derived from results listed in~\cite{Workman:2022ynf}. The blue circles use values quoted in the section headed `$\Gamma(i)\Gamma(\gamma\gamma)/\Gamma(\text{total})$', where the $\eta_c$ is produced via two-photon fusion in $e^+e^-$ collisions and detected through its decay to channel $i$ listed on the right of Fig.~\ref{fig:compGamma-expt}. These values are determined without reference to the PDG fit. They use either the PDG average value (where one is given) or the single experimental value (if there is no average) for the product above. We then divide by the branching fraction for channel $i$ again using either the PDG average or the single experimental value quoted, if there is no average. Uncertainties are combined in quadrature. We see a large spread of experimental values in Fig.~\ref{fig:compGamma-expt}, several of which are in significant ($4\sigma$) tension with the PDG fit result. This is not surprising given the $\chi^2$ value for the fit. Some of the low values for $\Gamma(\eta_c\to \gamma\gamma)$ seen are in disagreement with our result; the values from the $\phi\phi$ and $K^+K^-\pi^+\pi^-$ channels differ by an amount exceeding $6\sigma$. On the other hand, the experimental result using the $K\overline{K}\pi$ channel is in good agreement with our value, within $2\sigma$. The $K\overline{K}\pi$ channel has been studied by several experiments because it has a relatively large branching fraction. The value plotted comes from an average of 8 different experimental results for the product of rates (the average is dominated by results from CLEO~\cite{CLEO:2003gwz}  and BaBar~\cite{BaBar:2010siw}) and 10 for the branching fraction (where the average is dominated by results from BESIII~\cite{BESIII:2019eyx}). This gives a final result for $\Gamma(\eta_c \to \gamma\gamma)$ of 5.90(58) keV. The 10\% uncertainty is the smallest relative uncertainty for any channel. 

We conclude that the experimental picture of $\eta_c$ decay is not yet a very coherent one. Further experimental results, with small uncertainties, will be needed to resolve the issue of whether or not there is tension between experiment and lattice QCD/the Standard Model for $\eta_c \to \gamma\gamma$ decay. 

The filled blue square in Fig.~\ref{fig:compGamma-expt} comes from a determination of the product of branching fractions for $J/\psi \to \gamma\eta_c$ and $\eta_c \to \gamma\gamma$ by BESIII~\cite{BESIII:2012lxx} combined with the PDG average for the branching fraction $\mathcal{B}(J/\psi \to \gamma \eta_c)$~\cite{Workman:2022ynf}. The value agrees with our result, but has a large uncertainty. The PDG gives an average value for the branching ratio of $\eta_c \to \gamma\gamma$ by combining this result with an earlier one for the same product of branching fractions from CLEO~\cite{CLEO:2008qfy}\footnote{Note that the CLEO result is incorrectly quoted in~\cite{Workman:2022ynf}}. This gives an average branching fraction of $2.2 {+0.9 \atop -0.6} \times 10^{-4}$\cite{PDGpriv}. This also agrees with our value within its large uncertainties. We will discuss these results further in Sec.~\ref{sec:conclusions}. 

\begin{figure}
		\includegraphics[width=0.45\textwidth]{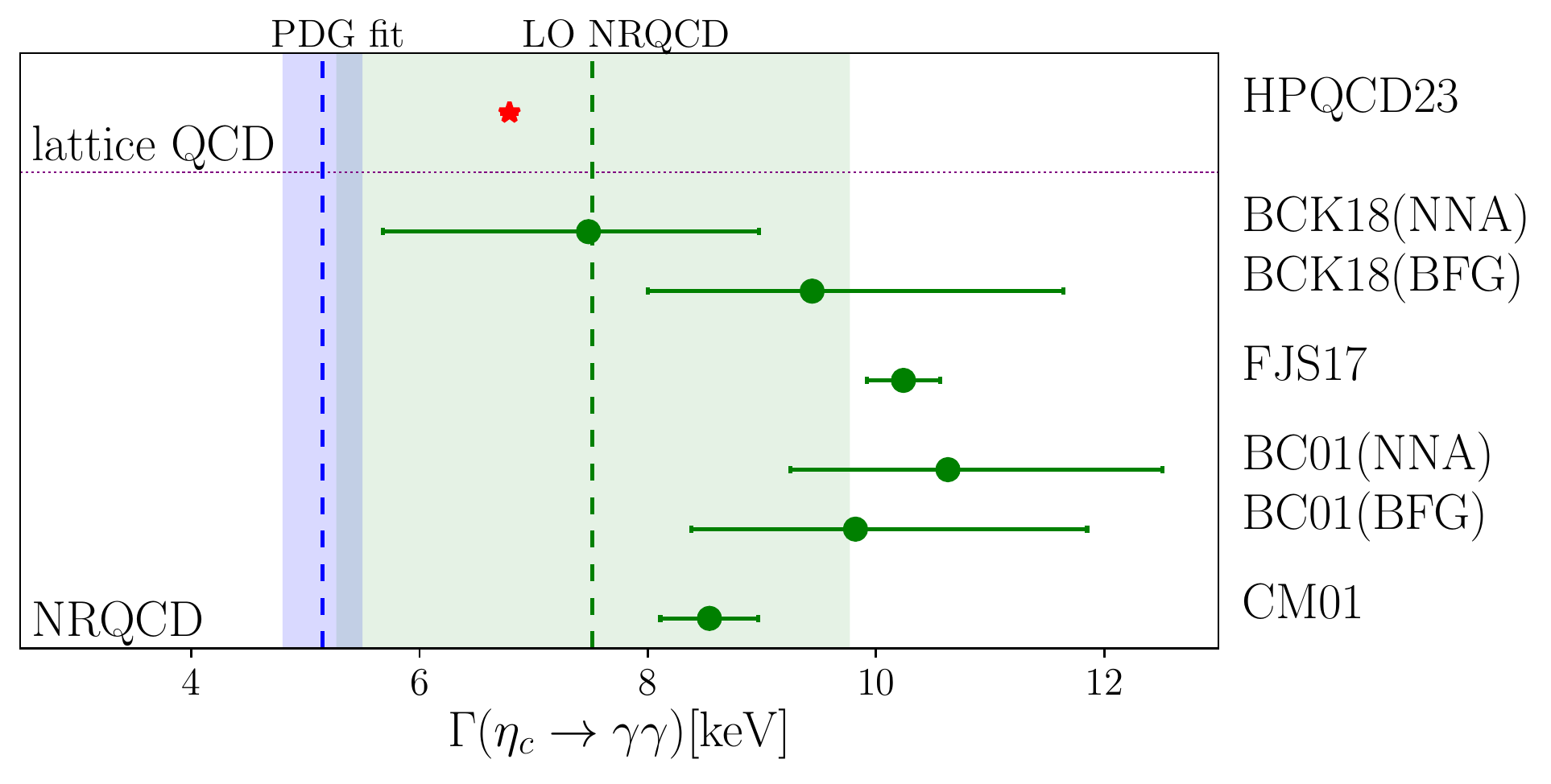}
		\caption{A comparison of our lattice QCD result (HPQCD23, red asterisk) for the decay width $\Gamma(\eta_c \to \gamma \gamma)$ to values obtained from theory calculations using NRQCD. The green dashed line is the leading order NRQCD result from Eq.~\eqref{eq:LONRQCD}, i.e. $4 \Gamma(J/\psi \to e^+e^-)/3$ where we have taken the $J/\psi$ leptonic width from lattice QCD~\cite{Hatton:2020qhk} (which agrees well with experiment). The uncertainty in the LO NRQCD value from higher order corrections is $\pm$ 30\% and denoted by the green band. The green points give results from higher-order calculations. CM01 is from~\cite{Czarnecki:2001zc}, determining QCD and velocity-expansion corrections to the ratio of $\Gamma(J/\psi\to e^+e^-)/\Gamma(\eta_c \to \gamma\gamma)$. FJS17~\cite{Feng:2017hlu} calculates NNLO QCD corrections to the $\eta_c\to \gamma\gamma$ branching fraction through $v^2$ order in the NRQCD velocity expansion. BC01~\cite{Bodwin:2001pt} and BCK18~\cite{Brambilla:2018tyu} calculate the inverse branching fraction resumming QCD corrections in the large $n_f$ limit. They give results from two resummation methods, na\"{i}ve non-Abelianisation (NNA) and background-field gauge (BFG). We use the PDG average~\cite{Workman:2022ynf} for the $\eta_c$ total width to convert their branching fractions into a width for $\eta_c \to \gamma \gamma$. The blue band repeats the PDG fit result shown in Fig.~\ref{fig:compGamma-expt}.  }
		\label{fig:compGamma-nrqcd}
\end{figure}

We can also explore what our results imply about the nonrelativistic nature of the $c$ quarks inside the $\eta_c$. As discussed in Section~\ref{sec:intro} there is a very simple relationship between $\Gamma(\eta_c \to\gamma \gamma)$ and $\Gamma(J/\psi \to e^+e^-)$ in LO NRQCD (Eq.~\eqref{eq:LONRQCD}). In Fig.~\ref{fig:compGamma-nrqcd} we compare this LO result, shown as a green dashed line, to our lattice QCD value (red asterisk). The central value for the LO result (7.5 keV) uses $\Gamma(J/\psi \to e^+ e^-) = 5.637(49)\,\text{keV}$ from lattice QCD+QED~\cite{Hatton:2020qhk} (this lattice QCD+QED result agrees well with the experimental average of 5.53(10) keV~\cite{Workman:2022ynf} but has a smaller uncertainty). The LO NRQCD central value is then 10\% above our lattice QCD result. Fig.~\ref{fig:compGamma-nrqcd} shows a $\pm$30\% error band in green around the LO NRQCD central value to allow for sub-leading corrections. The lattice QCD result, incorporating the full relativistic dynamics of the $c$ quarks, falls well within this 30\% band showing that the LO nonrelativistic approximation works well here.

The green points with error bars show results from two different calculations in continuum NRQCD, going beyond LO. In CM01~\cite{Czarnecki:2001zc} higher-order QCD and relativistic corrections to Eq.~\eqref{eq:LONRQCD} were added. The authors find substantial $\mathcal{O}(30\%)$ corrections from the two sources but also see a large amount of cancellation between them. They conclude with an estimate of a 10\% upward shift of $\Gamma(\eta_c\to \gamma\gamma)$ compared to the LO expression, with a 5\% uncertainty. They warn, however, that missing higher-order corrections might be substantial. The comparison with our result in Fig.~\ref{fig:compGamma-nrqcd} shows this to be the case because their 10\% shift has taken them in the wrong direction from the LO result and their 5\% uncertainty is insufficient to cover the gap. Refs~\cite{Penin:2004ay,Kiyo:2010jm} extend the analysis of this approach but do not quote final values, concluding that the theoretical uncertainties are very large for the charmonium case. 

FJS17~\cite{Feng:2017hlu} calculates NNLO QCD corrections to the $\eta_c\to \gamma\gamma$ branching fraction through $v^2$ order in the NRQCD velocity expansion and concludes that NRQCD factorisation does not work well for this case, given the large value that they obtain for the branching fraction. BC01~\cite{Bodwin:2001pt} and BCK18~\cite{Brambilla:2018tyu} resum QCD corrections to the inverse branching fraction in the large $n_f$ limit each using two different approaches, and allowing uncertainties for missing color-octet contributions. We use the PDG average~\cite{Workman:2022ynf} for the $\eta_c$ total width to convert their branching fractions into a width for $\eta_c \to \gamma \gamma$. Given the large uncertainties they have, all of their results agree and both are in reasonable agreement with our value. It is clear, however, that accurate results for $\Gamma(\eta_c \to \gamma\gamma)$ are only currently available using lattice QCD. 

\begin{figure}
		\includegraphics[width=0.45\textwidth]{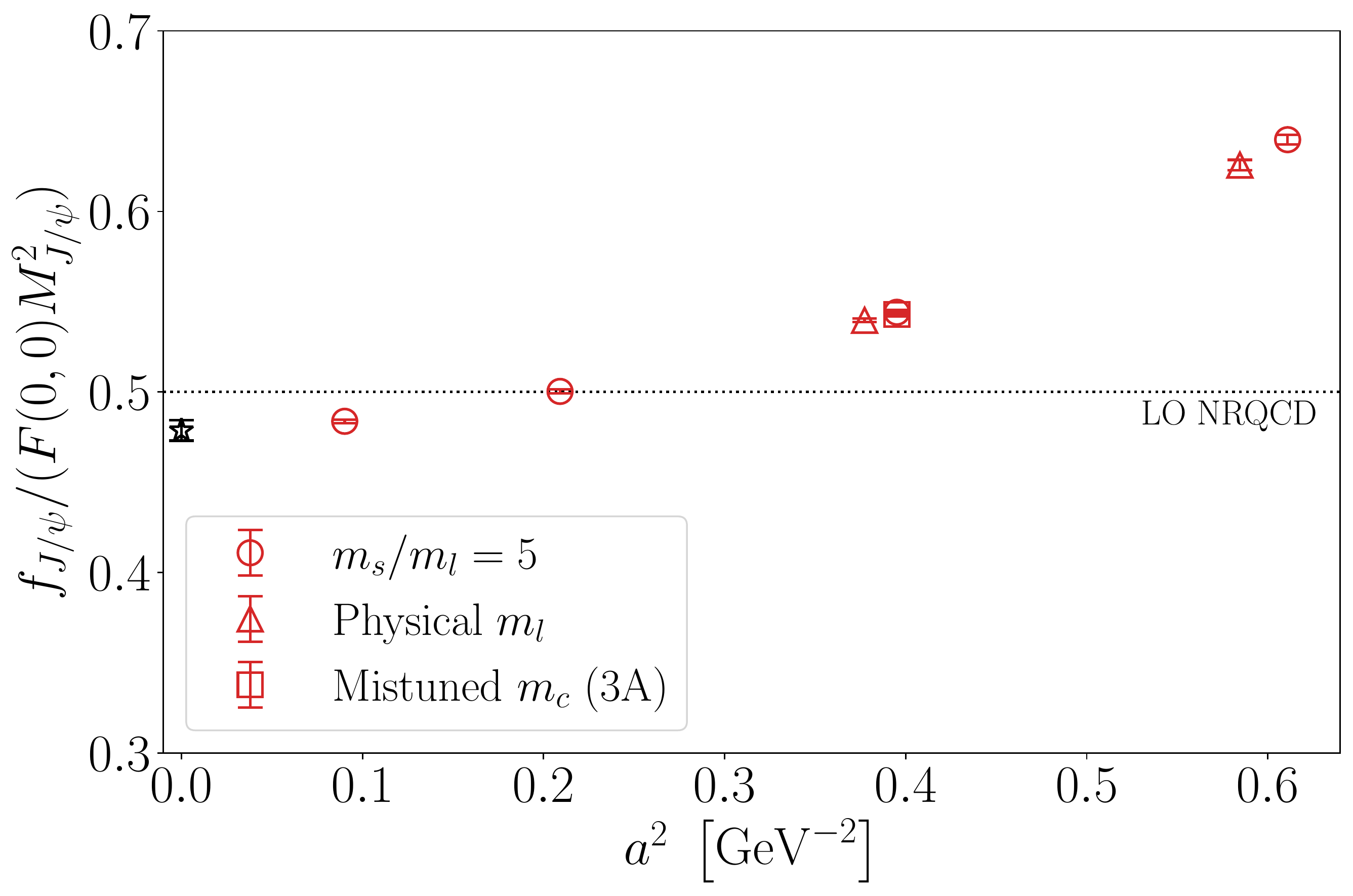}
		\caption{The ratio of $J/\psi$ decay constant to $F(0,0)$ multiplied by the square of the $J/\psi$ mass (red points) as a function of squared lattice spacing using our $F(0,0)$ values for the LOCAL set-up (with a local vector current) and results from~\cite{Hatton:2020qhk} for $f_{J/\psi}$ and $M_{J/\psi}$. The values for $F$ have been corrected for the slight off-shellness of one photon (see text). The blue band gives our continuum fit (see text) and the black dotted line gives the LO NRQCD result of 0.5 (Eq.~\eqref{eq:frat2}). }
		\label{fig:fFM2}
\end{figure}

We can extend our comparison to the LO NRQCD expectation by converting Eq.~\eqref{eq:LONRQCD} into a relationship between the hadronic parameters $F(0,0)$ and the $J/\psi$ decay constant, $f_{J/\psi}$. Using 
\begin{equation}
\label{eq:psi-leptwidth}
\Gamma(J/\psi \to e^+e^-) = \frac{4\pi}{3}\alpha^2Q_c^2\frac{f^2_{J/\psi}}{M_{J/\psi}}
\end{equation}
along with Eq.~\eqref{eq:decayrate} and Eq.~\eqref{eq:LONRQCD} we see that the LO NRQCD result implies
\begin{equation}
\label{eq:frat}
 \frac{f_{J/\psi}}{F(0,0)M_{\eta_c}\sqrt{M_{\eta_c}M_{J/\psi}}}= \frac{1}{2} (1+\mathcal{O}(\alpha_s) +\mathcal{O}(v^2/c^2))\, ,
\end{equation}
independent of $Q_c$.
Since there is no distinction between $M_{\eta_c}$ and $M_{J/\psi}$ at this order in NRQCD, we can rewrite this as
\begin{equation}
\label{eq:frat2}
R_{fF} \equiv \frac{f_{J/\psi}}{F(0,0)M^2_{J/\psi}}= \frac{1}{2}(1+\mathcal{O}(\alpha_s) +\mathcal{O}(v^2/c^2)) \, .
\end{equation}
$R_{fF}$ is now a simple combination of hadronic parameters that we can calculate directly in lattice QCD. 

Figure~\ref{fig:fFM2} shows our results for $R_{fF}$ as a function of lattice spacing using $F(0,0)$ values for the LOCAL set-up (which corresponds to our largest set of results in Fig.~\ref{fig:continuum_extrap}). We take values of $M_{J/\psi}$ and $f_{J/\psi}$ from~\cite{Hatton:2020qhk}, not including the QED effects that are calculated there. We also determine values for the mistuned $m_c$ case (set 3A) from those results. 

We obtain a value for the ratio in the continuum limit at physical quark masses by performing a fit of the same form as that in Eq.~\eqref{eqn:F_fit_form} to our lattice results for $f_{J/\psi}/[(1-q_2^2/M_{\text{pole}}^2)F(0,q_2^2)M_{J/\psi}^2 ]$.  The values show strong lattice-spacing dependence coming from the decay constant results. This is because the annihilation of a $J/\psi$ meson is a very short-distance process (shorter distance than $\eta_c \to \gamma\gamma$ where the energy is shared between two photons). Using the Empirical Bayes criterion we find a value for $\Lambda$ for this fit of 0.86 GeV, larger than that seen in either of the earlier fits shown in Fig.~\ref{fig:continuum_extrap}. Because of the larger discretisation effects we take $i_{\text{max}}=5$ (see Eq.~\eqref{eqn:F_fit_form}) for this fit. The result in the continuum limit does not change between $i_{\text{max}}=4$ or 5. Notice also that the results vary little between the tuned and mistuned $c$ quark mass in contrast to what was seen in Fig.~\ref{fig:continuum_extrap}. From NRQCD we expect the result for the ratio in Eq.~\eqref{eq:frat2} to depend on the (heavy) quark mass only through sub-leading terms in the velocity expansion. 

The result for the ratio of Eq.~\eqref{eq:frat2} that we obtain in the continuum limit and at physical quark masses is 
\begin{equation}
\label{eq:fratresult}
\frac{f_{J/\psi}}{F(0,0)M^2_{J/\psi}}= \fitrat \, .
\end{equation}
Here we have added a second uncertainty of 0.3\% to allow for additional systematic errors in $F(0,0)$ as discussed in Section~\ref{sec:etac2gamma_uncertainties}, giving a total uncertainty of 1.2\%. The $\chi^2/\mathrm{dof}$ for the fit is 0.27. 
The result of Eq.~\eqref{eq:fratresult} for $R_{fF}$ is (only) 4.5(1.2)\% below the LO NRQCD value of 0.5 given in Eq.~\eqref{eq:frat2}. 

This test of LO NRQCD is slightly different, although equally valid, to that given in Fig.~\ref{fig:compGamma-nrqcd}. This is why they both result in a LO NRQCD value that is above the lattice QCD result, even though the quantities tested are inversely related to each other. The differences arise from small effects that are ignored in LO NRQCD. These include the differences in the scale of $\alpha$ used in $\Gamma(J/\psi \to e^+ e^-)$\footnote{The lattice QCD+QED result~\cite{Hatton:2020qhk} for $\Gamma(J/\psi \to e^+e^-)$ uses $\alpha=1/134.02$, taking the scale of $\alpha$ to be $M_{J/\psi}$, appropriate to $J/\psi$ annihilation. See~\cite{Hatton:2020qhk} for a discussion of how the scale of $\alpha$ affects the agreement with experiment.} and $\Gamma(\eta_c \to \gamma\gamma)$ for the comparison in Fig.~\ref{fig:compGamma-nrqcd} and the simplification of the combination of meson masses in going from Eq.~\eqref{eq:frat} to Eq.~\eqref{eq:frat2}. Once both of these effects are taken into account, the two comparisons in Figs.~\ref{fig:compGamma-nrqcd} and~\ref{fig:fFM2} are consistent. Looking at either figure, we conclude that LO NRQCD is better than might have been expected as an approximation here. This only becomes clear with accurate lattice QCD results.

\section{Calculating $\Gamma(J/\psi \to \gamma \eta_c)$} \label{sec:JpsiGammaEtac}

\subsection{Method}
\label{sec:methodjpsi}

The $J/\psi \to\gamma \eta_c$ decay is an M1 electromagnetic meson-to-meson transition. The rate is straightforward to determine in lattice QCD. Using 3-point and 2-point correlation functions, to be discussed below, we can calculate the matrix element of the $c$ electromagnetic current, $j_c^{\mu} = \bar{c} \gamma^{\mu} c$, between the initial $J/\psi$ and final $\eta_c$ states. Note that, in keeping with Section~\ref{sec:etac2gamma}, we do {\it{not}} include a factor of the electric charge in the current. 
The matrix element can be parametrised by the form factor $\hat{V}(q^2)$ as
\begin{align}
	\left\langle\eta_{c}\left(p^{\prime}\right)\left| j_c^{\mu} \right| J / \psi(p)\right\rangle=\frac{ 2\hat{V}\left(q^{2}\right)}{M_{J / \psi}+M_{\eta_{c}}} \varepsilon^{\mu \alpha \beta \sigma} p_{\alpha}^{\prime} p_{\beta} \epsilon^{J / \psi}_{\sigma} \label{eqn:defn_matElt_Jpsi}
\end{align}
where $\epsilon^{J / \psi}_{\sigma}$ is the polarisation of the $J/\psi$ meson and $q = p - p'$ is the 4-momentum transfer.

The decay width is then given by~\cite{Dudek:2006ej}
\begin{align}
	\Gamma\left(J / \psi \rightarrow  \gamma\eta_{c}\right)= \alpha Q_c^2 \frac{16}{3} \frac{|\boldsymbol{k}|^{3} }{\left(M_{\eta_{c}}+M_{\psi}\right)^{2}} |\hat{V}(0)|^{2} \label{eqn:defn_V}
\end{align}
where $\hat{V}$ is evaluated at $q^2=0$ for an on-shell photon. 
$|\boldsymbol{k}|$ takes the value
\begin{align}
	\left|\boldsymbol{k}\right|=\frac{\left(M_{\eta_{c}}+M_{J / \psi}\right)\left(M_{J / \psi}-M_{\eta_{c}}\right)}{2 M_{J / \psi}} \label{eqn:mag3mom_onShell}
\end{align}
corresponding to the spatial momentum of the $\eta_c$ (or photon) in the $J/\psi$ rest frame at $q^2=0$. 

It is convenient for us to use $\hat{V} (q^2)$ rather than the more conventional $V(q^2)\equiv 2\hat{V}(q^2)$ because in our lattice QCD calculation we will compute only 
one of the two diagrams that contribute identically. A photon can be emitted by either the $c$ or $\overline{c}$ constituent quarks of the initial $J/\psi$ meson and we will evaluate one of these two cases.

Whilst the decay width to a real photon is a function of $\hat{V}(q^2=0)$, we can also map out the $q^2$ dependence of $\hat{V}$ up to $q^2_{\mathrm{max}} = (M_{J/\psi} - M_{\eta_c})^2$.  This is needed to compute the width of the Dalitz decay $\Gamma (J/\psi \to \eta_c e^+ e^-)$.
We do this by defining the ratio 
\begin{align}
\label{eq:Rdef}
	R_{e e\gamma} = \frac{\mathcal{B}\left(J/\psi \to \eta_c e^+ e^-\right)}{\mathcal{B}\left(J/\psi \rightarrow \eta_c \gamma \right)}.
\end{align}
The derivative of $R_{e e\gamma}$ with respect to $q^2$, the squared 4-momentum of the virtual photon in $J/\psi \to \eta_c e^+ e^-$, can be written in terms of the form factor $\hat{V}(q^2)$ as~\cite{Landsberg:1985gaz}
\begin{align}
	\frac{d R_{e e\gamma}}{d q^{2}} &=\frac{\alpha}{3 \pi q^{2}}\left|\frac{\hat{V}(q^2)}{\hat{V}(0)}\right|^{2}\left(1-\frac{4 m_{e}^{2}}{q^{2}}\right)^{\frac{1}{2}}\left(1+\frac{2 m_{e}^{2}}{q^{2}}\right) \nonumber \\
	& \times\left(\left(1+\frac{q^{2}}{M_{J/\psi}^2 - M_{\eta_c}^2}\right)^2-\frac{4 M_{J/\psi}^{2} q^{2}}{\left(M_{J/\psi}^2 - M_{\eta_c}^2\right)^2}\right)^{\frac{3}{2}} .\label{eqn:dReedqSq}
\end{align}
By multiplying $R_{ee\gamma}$ by $\Gamma (J/\psi \to \gamma\eta_c)$ (determined through Eq.~(\ref{eqn:defn_V})), we can then also find the decay width $\Gamma(J/\psi \to \eta_c e^+ e^-)$.

\subsection{Lattice Calculation}

We use the same gluon field ensembles and quark mass parameters for this calculation as for $\eta_c \to \gamma\gamma$. These are given in Table~\ref{tab:params}. 
\begin{figure}
	\begin{center}
		\includegraphics[width=0.45\textwidth]{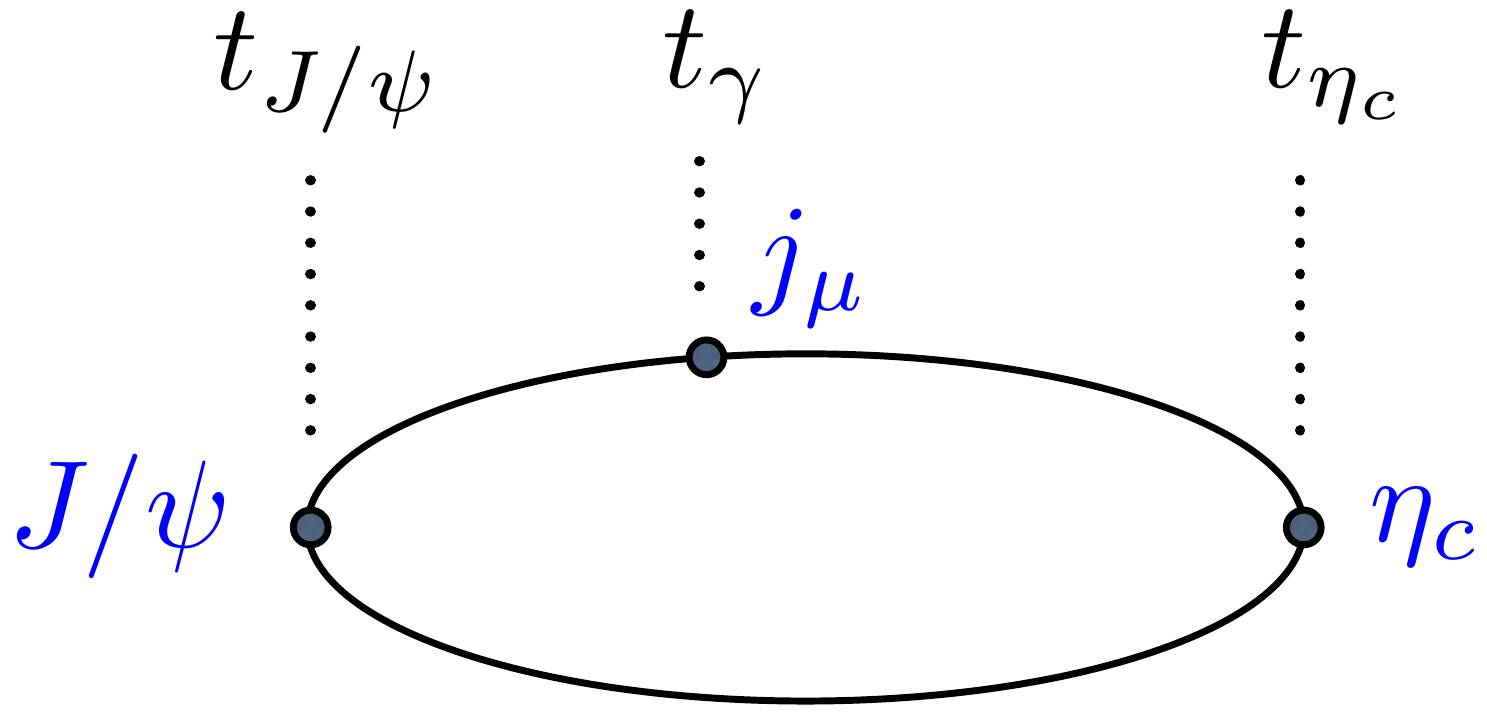}
		\caption{Schematic diagram for the connected 3-point correlation function for the $J/\psi$ transition to $\eta_c$ via an electromagnetic current. The lines between the operators represent quark propagators at the mass of the charm quark.}
		\label{fig:3pt_Jpsi}
	\end{center}
\end{figure}

\subsubsection{Correlation functions} \label{sec:lm_Jpsi}

The matrix element given in Eq.~(\ref{eqn:defn_matElt_Jpsi}) can be obtained by computing an appropriate 3-point correlation function on the lattice.
This correlation function, summarised diagrammatically in Fig.~\ref{fig:3pt_Jpsi}, is similar in appearance to the the correlator given in Fig.~\ref{fig:3pt_etacgg} used in the calculation for $\eta_c \to \gamma \gamma$; the difference is that one of the vector operators now couples to a $c \overline{c}$ vector meson.
We consider only the connected contribution to the decay $J/\psi \to \eta_c \gamma^{(*)} $ which comes from an insertion of a $\overline{c} \gamma^{\mu} c$ current onto one of the two quark lines between the initial and final states.
There are two such diagrams which contribute equally;
 we compute only one of them.
We expect quark-line disconnected contributions to be very small here because of the large mass of the $c$ quark~\cite{Dudek:2006ej}.
We will discuss the uncertainty associated with neglecting these diagrams in Section~\ref{sec:Jpsi_uncertainties}.

Our 3-point functions have three operator insertions, two vector and one pseudoscalar, just as in the 3-point function in Section~\ref{sec:etac2gamma_corrfs}.
We use different spin-taste combinations here, however.
We take the $\gamma_5 \otimes \gamma_5$ interpolator for the $\eta_c$ meson, $\gamma_x \gamma_t \otimes \gamma_5 \gamma_z$ to interpolate the $J/\psi$ meson, and the local $\gamma_z \otimes \gamma_z$ operator for the electromagnetic vector current insertion.\footnote{ See~\cite{Donald:2012ga} for a comparison of different taste configurations for this calculation.}
As discussed in Section~\ref{sec:ZV}, this allows us to use the very precise values from~\cite{Hatton:2019gha} for the multiplicative renormalisation $Z_V$ for the local HISQ-HISQ vector current.

Spatial momentum is applied to the charm propagator between the $\eta_c$ operator and the vector current insertion to give momentum to the $\eta_c$ meson while the $J/\psi$ meson is always at rest on the lattice.
As well as the three-momentum necessary to achieve $q^2 = 0$, the case of an on-shell photon, we calculate correlation functions for various other three-momenta of the $\eta_c$ meson so that we obtain results for a range of $q^2$ values around $q^2=0$. In addition we calculate 2-point correlation functions corresponding to the operator used for the $J/\psi$ (at rest) and for the $\eta_c$ (at the spatial momentum values used). 

As for $\eta_c\to \gamma\gamma$ (see Sec.~\ref{sec:etac2gamma_corrfs}), we insert spatial momentum by using a twist angle. The twist values used for each set are tabulated in Table~\ref{tab:twists_Jpsi}.
The twists are chosen to broadly cover the full range $0\leq q^2 \leq q^2_{\mathrm{max}} = (M_{J/\psi} - M_{\eta_c})^2$. Given that we have chosen $x$ and $z$ polarisations for the $J/\psi$ and the vector current respectively, then Eq.~\eqref{eqn:defn_matElt_Jpsi} requires the spatial momentum to have a component in the $y$ direction. In fact we take the momentum to be parallel to the $y$-axis. 
\begin{table}
	\centering
	\caption{Parameters for the correlation functions for the decay $J/\psi \to \gamma \eta_c$. The second column gives the set of twists $\theta$ used for each lattice implementing twisted boundary conditions in the $(0 \: 1 \: 0)$ direction. The corresponding 3-momentum component is given by $q^y = \theta \pi / aN_x$ (see Eq.~\eqref{eq:q1theta}). Given the $x$ and $z$ polarisations chosen for the $J/\psi$ and vector currents respectively this gives a non-zero matrix element in Eq.~\eqref{eqn:defn_matElt_Jpsi}. The twists are chosen to cover the kinematic range from zero recoil to $q^2 = 0$. The third column gives the different time separation between source and sink ($T=t_{J/\psi}-t_{\eta_c}$, see Fig.~\ref{fig:3pt_Jpsi}) of the 3-point correlation function. }
	\begin{tabular}{c c c} 
		\hline\hline
		Set & $\theta$ & $T$ \\ [0.1ex] 
		\hline
        1 & $0.1750, 0.3031, 0.3914, 0.4631$ & $15,16,17$ \\
		2 & $0.6406, 0.9060, 1.1097$ & $13,14,15,16$ \\
		3 & $0.2111,0.2986,0.3657,0.4223,$ & $16,17,18,19,20,21$  \\
		& $0.4721,0.5172,0.5586,0.5972$ & \\
		4 & $0.7719, 1.0918, 1.3373$ & $17,18,19,20$ \\
		5 & $0.2423, 0.3427, 0.4197,$ & $21, 24, 27, 30$ \\
		& $0.4846, 0.5418, 0.5936$ \\
		6 & $0.3774, 0.5338, 0.6538$ & $33,36,39,42$ \\
		\hline\hline
	\end{tabular}
	\label{tab:twists_Jpsi}
\end{table}

We calculate 3-point correlation functions for several different values of the source-sink separation, $T = t_{J/\psi}- t_{\eta_c}$ (see Fig.~\ref{fig:3pt_Jpsi}). This improves  the determination of the ground-state to ground-state matrix elements that we are interested in by giving a better handle on the excited states present in the correlation function. The values of $T$ that we use are also listed in Table~\ref{tab:twists_Jpsi}. 

As in Section~\ref{sec:etac2gamma_corrfs}, we fit the 2-point correlation functions to the form in Eq.~(\ref{corrfitform_2pt}).
The 3-point functions (for all $T$ and momenta) are simultaneously fit to the standard form
\begin{align} \label{corrfitform_3pt}
   & C_{\text{3pt}} (t=t_{\gamma}-t_{J/\psi},T=t_{J/\psi}-t_{\eta_c}) = \\
   & \sum_{i,j}^{N_{\mathrm{n}}, N_{\mathrm{n}}}  a^{\gamma_x \gamma_t \otimes \gamma_5 \gamma_z}_{i} e^{-E_{i}t}  V_{ij} a^{\gamma_5 \otimes \gamma_5}_{j} e^{-E_{j} (T-t)} \nonumber 
\end{align}
with the addition of terms that oscillate in time and involve opposite parity states~\cite{Donald:2012ga} that are not shown here. 
The energies of the states with overlap onto $\gamma_x \gamma_t \otimes \gamma_5 \gamma_z$ (the $J/\psi$ and its excitations) are indexed by $i$, and the energies of the states with overlap onto $\gamma_5 \otimes \gamma_5$ (the $\eta_c$ and its excitations) are indexed by $j$. The energies and the amplitudes $a_{i/j}$ are also parameters for the 2-point correlator fits and this enables us to extract the parameters $V_{ij}$. 

\begin{table}[!t]
\caption{Results for the masses of the $J/\psi$ and $\eta_c$ mesons from our combined two- and three-point correlator fits aimed at determining the form factor for $J/\psi \to \gamma \eta_c$ decay. The $\eta_c$ meson here has spin-taste $\gamma_5\otimes\gamma_5$ and so has a different mass to the values given in Table~\ref{tab:f00_metac}. }
 \begin{tabular}{ccc}
  \hline
  \hline
  Set & $aM^{\text{latt}}_{J/\psi}$ & $aM^{\text{latt}}_{\eta_c}$ \\
  \hline
   1  & $2.43418(38)$  & $2.331959(86)$    \\
   2  & $2.38671(15)$  & $2.287729(36)$    \\
   3  & $1.94525(21)$  & $1.876352(68)$    \\
   3A & $1.92625(21)$  & $1.857007(69)$    \\
   4  & $1.90225(7)$   & $1.833910(22)$    \\
   5  & $1.41616(13)$  & $1.366980(46)$    \\
   6  & $0.92973(12)$  & $0.896714(42)$    \\
  \hline
\end{tabular}  \label{tab:Jpsimasses}
\end{table}

\begin{table*}[!t]
\caption{Our lattice results for $\hat{V}_{\text{latt}}(q^2)$ on each ensemble from Table~\ref{tab:params} and at each value of the momentum used. The corresponding values of $q^2$ are also listed, in lattice units. $q^2$ is calculated from $(M^{\text{latt}}_{J/\psi}-E^{\text{latt}}_{\eta_c})^2-{\mathbf q}^2$ where the ground-state $J/\psi$ mass and $\eta_c$ energy, $M^{\text{latt}}_{J/\psi}$ and $E^{\text{latt}}_{\eta_c}$, are obtained from the correlator fits and ${\mathbf q}^2$ is determined from the twist angles imposed (see Table~\ref{tab:twists_Jpsi}). $M^{\text{latt}}_{J/\psi}$ and $M^{\text{latt}}_{\eta_c}$ values are given in Table~\ref{tab:Jpsimasses}.}
 \begin{tabular}{c|c|cccccc}
    \hline
    \hline
     \multirow{2}{*}{Set 1}& $a^2q^2$ & $0.009220(74)$ & $0.006762(74)$ & $0.004301(74)$ & $0.001843(73)$ \\
    & $\hat{V}_{\text{latt}}(q^2)$ & $1.834(38)$ & $1.834(23)$ & $1.833(18)$ & $1.831(15)$  \\
     \hline
     \multirow{2}{*}{Set 2}& $a^2q^2$ & $0.005680(28)$ & $0.001564(28)$ & $-0.002551(27)$ & - \\
    & $\hat{V}_{\text{latt}}(q^2)$ & $1.858(12)$ & $1.8520(91)$ & $1.8478(76)$ & -  \\
    \hline
    \multirow{4}{*}{Set 3}& $a^2q^2$ & $0.003956(27)$ & $0.003165(26)$ & $0.002374(26)$ & $0.001583(26)$ \\
    & $\hat{V}_{\text{latt}}(q^2)$ & $1.887(53)$ & $1.881(38)$ & $1.878(31)$ & $1.875(27)$  \\
    \cline{2-6}
    & $a^2q^2$ & $0.000793(26)$ & $0.000001(26)$ & $-0.000789(26)$& $-0.001580(26)$  \\
    & $\hat{V}_{\text{latt}}(q^2)$ & $1.873(25)$ & $1.872(23)$ & $1.870(21)$ & $1.869(20)$ \\
    \hline
    \multirow{4}{*}{Set 3A}& $a^2q^2$ & $0.004002(27)$ & $0.003211(27)$ & $0.002419(27)$ & $0.001628(27)$ \\
    & $\hat{V}_{\text{latt}}(q^2)$ & $1.888(54)$ & $1.880(38)$ & $1.877(32)$ & $1.874(28)$  \\
    \cline{2-6}
    & $a^2q^2$ & $0.000837(27)$ & $0.000045(27)$ & $-0.000745(27)$& $-0.001536(26)$  \\
    & $\hat{V}_{\text{latt}}(q^2)$ & $1.872(25)$ & $1.870(23)$ & $1.869(21)$ & $1.867(20)$ \\
    \hline
     \multirow{2}{*}{Set 4}& $a^2q^2$ & $0.0020259(86)$ & $-0.0006190(85)$ & $-0.0032636(84)$ & - \\
    & $\hat{V}_{\text{latt}}(q^2)$ & $1.8558(47)$ & $1.8530(34)$ & $1.8501(29)$ & -  \\
     \hline
    \multirow{4}{*}{Set 5}& $a^2q^2$ & $0.001833(12)$ & $0.001247(12)$ & $0.000661(12)$ & $0.000075(12)$ \\
    & $\hat{V}_{\text{latt}}(q^2)$ & $1.854(14)$ & $1.856(10)$ & $1.8565(83)$ & $1.8562(73)$  \\
    \cline{2-6}
    & $a^2q^2$ & $-0.000510(12)$ & $-0.001097(12)$ & -&-  \\
    & $\hat{V}_{\text{latt}}(q^2)$ & $1.8555(66)$ & $1.8547(62)$ & -&- \\
    \hline
    \multirow{2}{*}{Set 6}& $a^2q^2$ & $0004578(76)$ & $-0.0001746(75)$ & $-0.0008068(74)$ & - \\
    & $\hat{V}_{\text{latt}}(q^2)$ & $1.873(13)$ & $1.8723(93)$ & $1.8704(79)$ & -  \\
    \hline
    \hline
  \end{tabular}  \label{tab:Jpsiresults}
\end{table*}

Our fits account for four normal vector and pseudoscalar charmonium states in Eq.~\eqref{corrfitform_3pt} (i.e. $N_n=4$) and four additional oscillating states for each. As for the fit to $C_{\eta_c}$ described in Section~\ref{sec:etacggfits}, we drop 2-point correlator data below $t_{\mathrm{min}}/a=N_t/8$ when fitting. We also drop 3-point correlator data below a $t_{\mathrm{min}}/a$ (and $(T-t)_{\mathrm{min}}/a$) value of between 2 and 4, increasing as the lattice spacing falls. 

The ground-state to ground-state matrix element that characterises the $J/\psi \to \eta_c$ transition is related to the parameter $V_{00}$ of Eq.~\eqref{corrfitform_3pt} via
\begin{align}
	\frac{\left\langle\eta_{c}|j_c^{\mu}| J / \psi\right\rangle}{2} =  Z_V \sqrt{2 M^{\text{latt}}_{J / \psi}} \sqrt{2 E^{\text{latt}}_{\eta_{c}}} V_{00}, \label{eqn:mat_elt_Vnn}
\end{align}
where $M^{\text{latt}}_{J / \psi} \equiv E_{{i=0}} $ and $ E^{\text{latt}}_{\eta_c} \equiv E_{{j=0}} $ and the $\sqrt{}$ factors are the relativistic normalisations of the states $\ket{J/\psi}$ and $\ket{\eta_c}$ respectively. $Z_V$ is the multiplicative renormalisation factor for the local $\gamma_z \otimes \gamma_z$ vector current. The factor of $1/2$ present on the left-hand side above accounts for the second Wick contraction (that contributes equally) where the vector current insertion is placed onto the other quark line connecting the initial and final states.

$\hat{V}(q^2)$ is then obtained from $V_{00}$ using 
\begin{equation}
\label{eq:VhatV00}
\hat{V}(q^2) =  \frac{M^{\text{latt}}_{J/\psi}+M^{\text{latt}}_{\eta_c}}{M^{\text{latt}}_{J/\psi}q^y}Z_V \sqrt{2 M^{\text{latt}}_{J / \psi}} \sqrt{2 E^{\text{latt}}_{\eta_{c}}} V_{00}
\end{equation}
where $q^y$ is the component of the $\eta_c$ momentum in the $y$-direction, orthogonal to both the vector current and $J/\psi$ polarisations (see Table~\ref{tab:twists_Jpsi} for the values used). 

Table~\ref{tab:Jpsimasses} lists the values of the $J/\psi$ and $\eta_c$ masses in lattice units given by our correlator fits on each ensemble.  These are the values used to determine $\hat{V}$ in Eq.~\eqref{eq:VhatV00}.
In Fig.~\ref{fig:etac_taste_split} we compare the mass of the $\gamma_x \gamma_t \otimes \gamma_5\gamma_z$  $J/\psi$ meson that we use for the study of $J/\psi  \to \gamma \eta_c$ (from Table~\ref{tab:Jpsimasses}) with the $J/\psi$ mass from the $\gamma_z \otimes \gamma_z$ interpolator from~\cite{Hatton:2020qhk}. The mass splitting between these two different tastes of $J/\psi$ is small (less than 20 MeV even on the coarsest lattices) and disappears in the continuum limit, with the mass of the $\gamma_x \gamma_t \otimes \gamma_5\gamma_z$ $J/\psi$ being larger. 
Fig.~\ref{fig:etac_taste_split}, also compares the masses of the different tastes of $\eta_c$ used for our calculations. The $J/\psi\to\gamma\eta_c$ study uses the $\gamma_5\otimes \gamma_5$ (Goldstone) $\eta_c$ and the $\eta_c \to \gamma\gamma$ analysis uses two different tastes, both of which are heavier. Here the taste-splittings are 2--3 times larger than for the vector meson but again disappear rapidly as $(am_c)^2$ falls towards to the continuum limit. 

Table~\ref{tab:Jpsiresults} then gives our results for $\hat{V}(q^2)$ on each ensemble and for each value of the spatial momentum inserted. 
In the next section we describe how we fit these results to obtain the function $\hat{V}(q^2)$ in the $a\to 0$ continuum limit at physical quark masses.

\subsubsection{Taking the physical-continuum limit} \label{sec:Jpsi_cont_limit}

We fit our lattice results for $\hat{V}$, which we denote $\hat{V}_{\mathrm{latt}}$ (see Table~\ref{tab:Jpsiresults}), to the following function:
\begin{align}
 \label{eqn:Vhat_fit_form}
	&\hat{V}_{\mathrm{latt}} (q^2) = \sum_{k=0}^{2} A^{(k)}\Big( \frac{q^2}{(M^{\text{latt}}_{J/\psi})^2} \Big)^{k} \times  \\
	& \hspace{0mm} \Bigg[1+\sum_{i=1}^{i_{\mathrm{max}}} \kappa_{a\Lambda}^{(i,k)}\left(a \Lambda\right)^{2 i} + \kappa_{\mathrm{val}, c}^{(k)}  \delta^{\mathrm{val}, c} +\kappa_{\mathrm{sea}, c}^{(k)} \delta^{\mathrm{sea}, c}  \nonumber \\
	& \hspace{3mm} + \kappa_{\mathrm{sea}, u d s}^{(0,k)} \delta^{\mathrm{sea}, u d s}\left\{1+\kappa_{\mathrm{sea}, u d s}^{(1,k)}(\tilde{\Lambda} a)^{2}+\kappa_{\mathrm{sea}, u d s}^{(2,k)}(\tilde{\Lambda} a)^{4}\right\}\Bigg]\, . \nonumber
\end{align}
This takes the same form as that for $F(0,0)$ in Eq.~\eqref{eqn:F_fit_form}, except that we must allow here for dependence on $q^2$. 
We do this through a truncated Taylor series in $q^2/M^2_{J/\psi}$, $M_{J/\psi}$ being the appropriate mass for a form factor induced by a $c\overline{c}$ vector current. The coefficients $A^{(k)}$ for each term allow the fit to adjust the mass away from $M_{J/\psi}$ if required. 
The $q^2$ range is very small here (relative to $M^2_{J/\psi}$), so we expect the form factor to be very flat as a function of $q^2$ and do not need many terms in the polynomial. We include terms up to $q^4$. The $J/\psi$ mass used in the $q^2/M^2_{J/\psi}$ term is that obtained from the fit (see Table~\ref{tab:Jpsimasses}).
The fit form allows for independent discretisation effects and quark mass mistuning terms for each power of $q^2/M^2_{J/\psi}$ (denoted by $k$). 
The form of these terms is the same as in the fit function for $F(0,0)$ in Eq.~(\ref{eqn:F_fit_form}) and
definitions for $\delta$ can be found in Eqs.~(\ref{eqn:mt_val_charm}),~(\ref{eqn:mt_sea_charm}) and~(\ref{eqn:mt_sea_uds}).

We take priors of $A^{(k)} = 2(1)$. The choice is informed by leading-order NRQCD where the rate for the radiative transition depends on the wavefunction overlap between $J/\psi$ and $\eta_c$ modulated by a Bessel function. For small momentum transfer and ignoring relativistic corrections that generate spin-dependent differences in the wavefunction, this wavefunction overlap is 1. Then we expect $\hat{V}(0) \approx 2$ from comparing Eq. (5) of~\cite{Eichten:2007qx} to Eq.~\eqref{eqn:defn_V} and taking $m_c \approx M_{\eta_c}/2 \approx M_{J/\psi}/2$. Our lattice QCD calculation includes relativistic effects fully so it will give a much more accurate result for $\hat{V}(0)$ than this leading-order nonrelativistic argument. How different the result is, we will see below. The priors for $A^{(k>0)}$ encompass $q^2$-dependence following a pole form $(1-q^2/M^2_{J/\psi})^{-1}$. 

As with the fit of $F(0,0)$ in Section~\ref{sec:fit_form_F}, we test the size of discretisation effects using the Empirical Bayes approach. We do not expect large discretisation effects here based on the nonrelativistic arguments above which tell us that discretisation effects can only enter through the small momentum transfer between $J/\psi$ and $\eta_c$ and through small spin-dependent effects on the wavefunction overlap. We find that the Bayes factor is maximised by $\Lambda = 0.12\,\mathrm{GeV}$. This is close to the largest $|\bf{q}|$ value in the kinematic range of the decay and so seems a reasonable value to set the scale for discretisation effects.  

Following Section~\ref{sec:fit_form_F} we take the priors $\kappa_{a\Lambda}^{(i,k)} = 0(1)$, $\kappa_{\mathrm{val}, c}^{(k)} = 0(1)$ (little difference is seen between our results on sets 3 and 3A), $\kappa_{\mathrm{sea}, c}^{(k)} = 0.0(1)$ and $\kappa_{\mathrm{sea}, uds}^{(j,k)} = 0(1)$ for $j=0, 1, 2$. Our preferred fit takes the number of discretisation terms included, $i_{\mathrm{max}}=3$. 

In the limit of vanishing lattice spacing and physical quark masses, the form factor is then given by
\begin{align}
\label{eq:Vhatcont}
\hat{V} (q^2) = \sum_{k=0}^{2} A^{(k)}\Big( \frac{q^2}{M_{J/\psi}^2} \Big)^{k}.
\end{align}
The value of $M_{J/\psi}$ used here is that from experiment~\cite{Workman:2022ynf}. Note that our $c$ quark mass is tuned so that our $J/\psi$ masses match that value (see Eq.~\eqref{eqn:defn_amctuned}). 
In particular the form factor at $q^2 = 0$ needed to determine $\Gamma (J/\psi \to \gamma \eta_c)$ via Eq.~(\ref{eqn:defn_V}), $\hat{V} (0)$, is simply given by the parameter $A^{(0)}$.

\begin{figure}
	\begin{center}
		\includegraphics[width=0.45\textwidth]{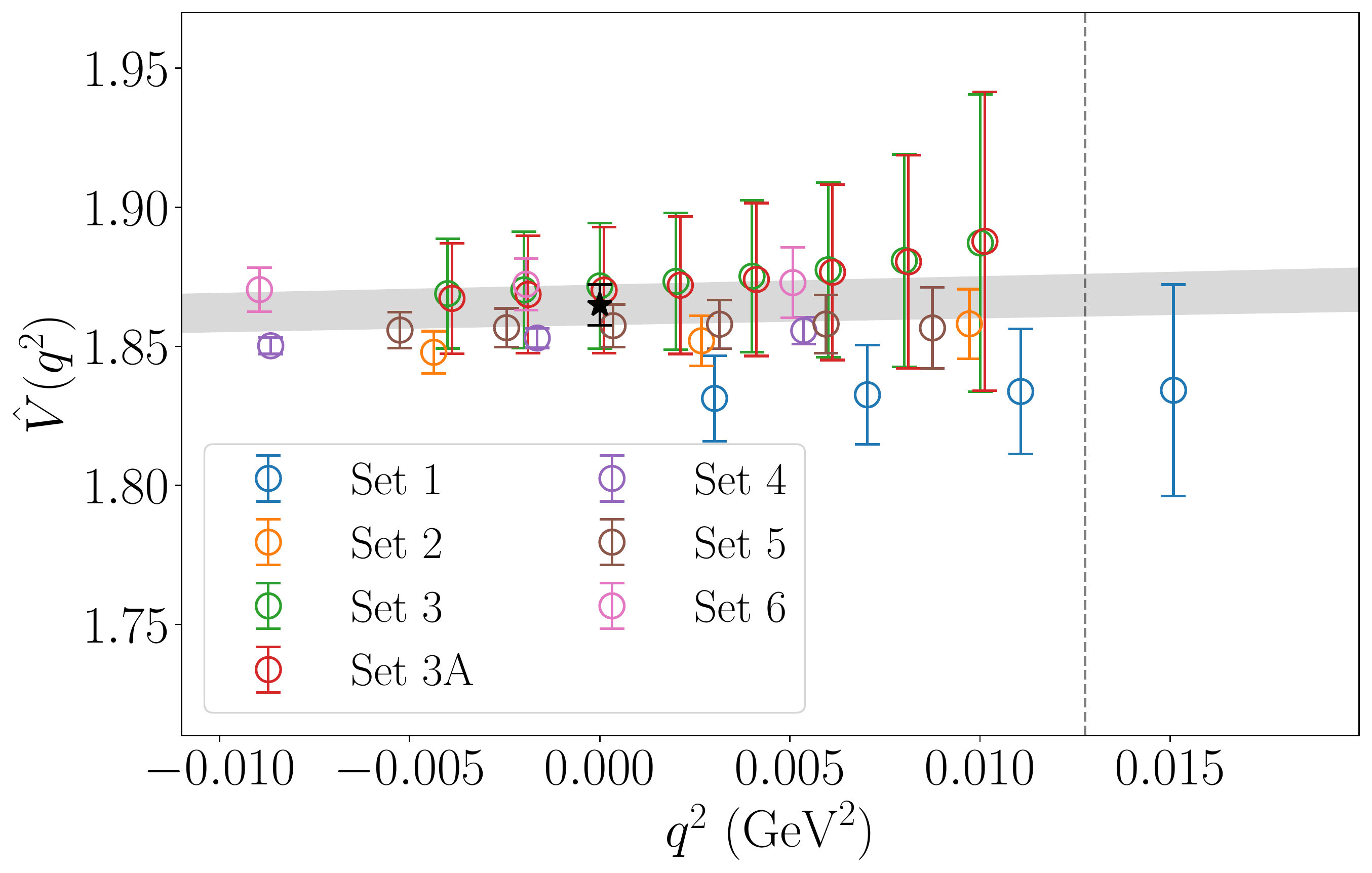}
				\includegraphics[width=0.45\textwidth]{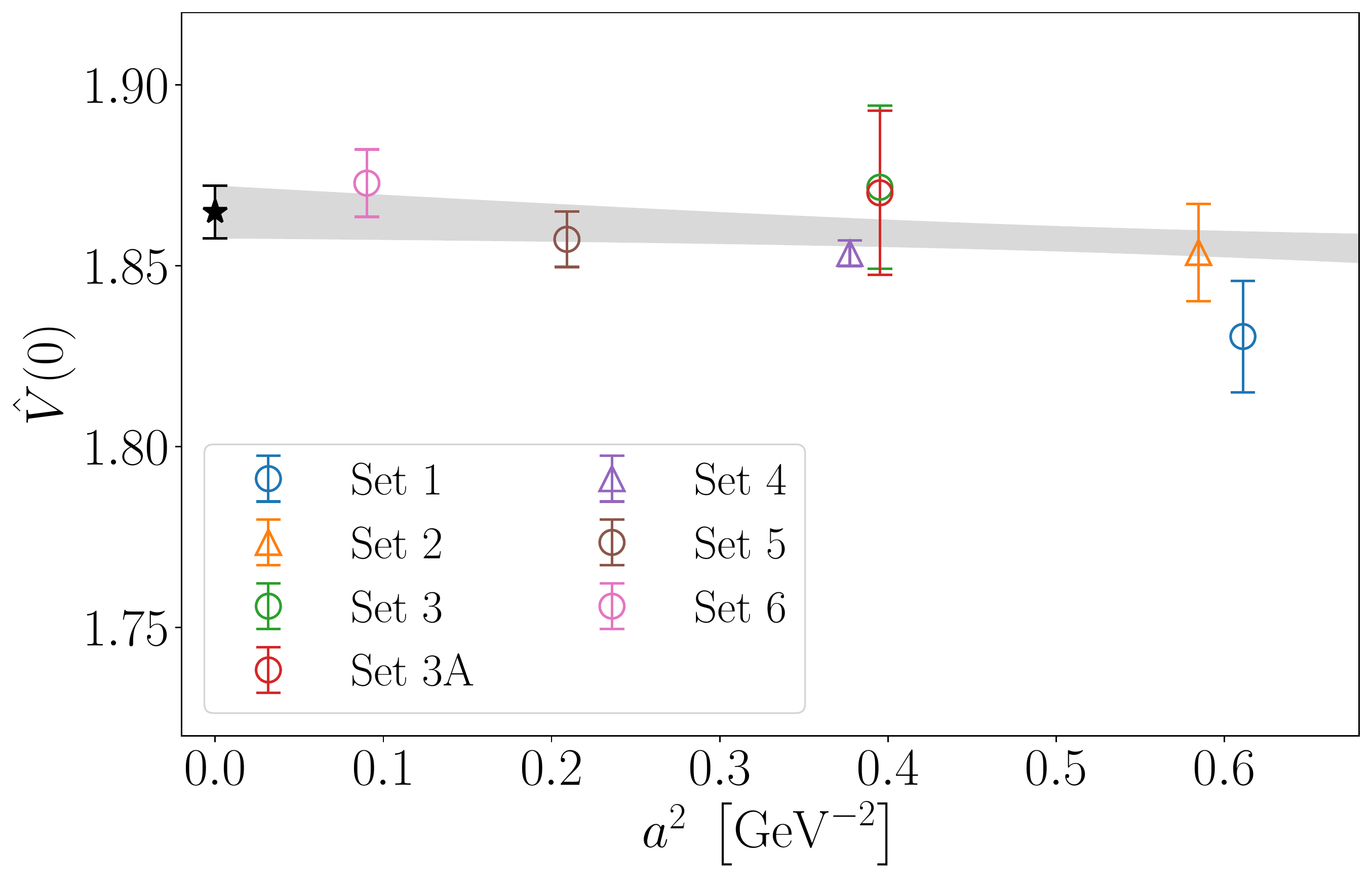}
		\caption{Upper plot: Lattice results for $\hat{V}(q^2)$ plotted against $q^2$. The different colours denote the different ensembles used (see Table~\ref{tab:params}). The grey band gives the result, with $\pm 1\sigma$ error bars, from the fit to Eq.~\eqref{eqn:Vhat_fit_form} evaluated in the continuum limit at physical quark masses. The dashed line corresponds to the maximum physical $q^2$ value, $(M_{J/\psi}-M_{\eta_c})^2$. Lower plot: Lattice results on each ensemble interpolated to $q^2=0$ using the continuum fit result.  The grey band gives the result from the fit to Eq.~\eqref{eqn:Vhat_fit_form}  as a function of lattice spacing at physical quark masses.}
		\label{fig:Vhat_data}
	\end{center}
\end{figure}

In Fig.~\ref{fig:Vhat_data}, we plot our lattice results for $\hat{V}_{\mathrm{latt}} (q^2)$ from Table~\ref{tab:Jpsiresults} against $q^2$ for each ensemble (upper plot). The results show little dependence on $q^2$, as expected. Note that the statistical uncertainties on the points are smaller for values at low $q^2$ in comparison to those at zero recoil.  The statistical noise of the $\eta_c$ correlators increases with spatial momentum, but this is offset by the increased value of $V_{00}$ so that its relative uncertainty falls. We also show the fit band in grey that corresponds to the fit of Eq.~\eqref{eqn:Vhat_fit_form} evaluated in the continuum limit with physical quark masses (see Eq.~\eqref{eq:Vhatcont}). 

The lower plot of Fig.~\ref{fig:Vhat_data} shows the lattice results interpolated to $q^2=0$ on each ensemble and plotted against the square of the lattice spacing. The interpolation in $q^2$ is done using the continuum result for the $q^2$ dependence of $\hat{V}$. The figure shows, again as expected, very little dependence on the lattice spacing. 

The value that we obtain from our fit in the continuum limit and at physical quark masses is 
\begin{align}
\label{eqn:fit_Vhat}
\fitVhat,
\end{align}
with an uncertainty of $\fitVhatUncertainty$. The fit has a $\chi^2/\mathrm{dof}$ value of 0.19. Our result is clearly distinguishable from the leading-order NRQCD expectation of 2, indeed it differs by 7.2(4)\%. Before discussing additional systematic uncertainties that need to be included, we first discuss the stability of our result for $\hat{V}(0)$ under changes to the parameters of our fits. 

\begin{figure}
	\begin{center}
		\includegraphics[width=0.45\textwidth]{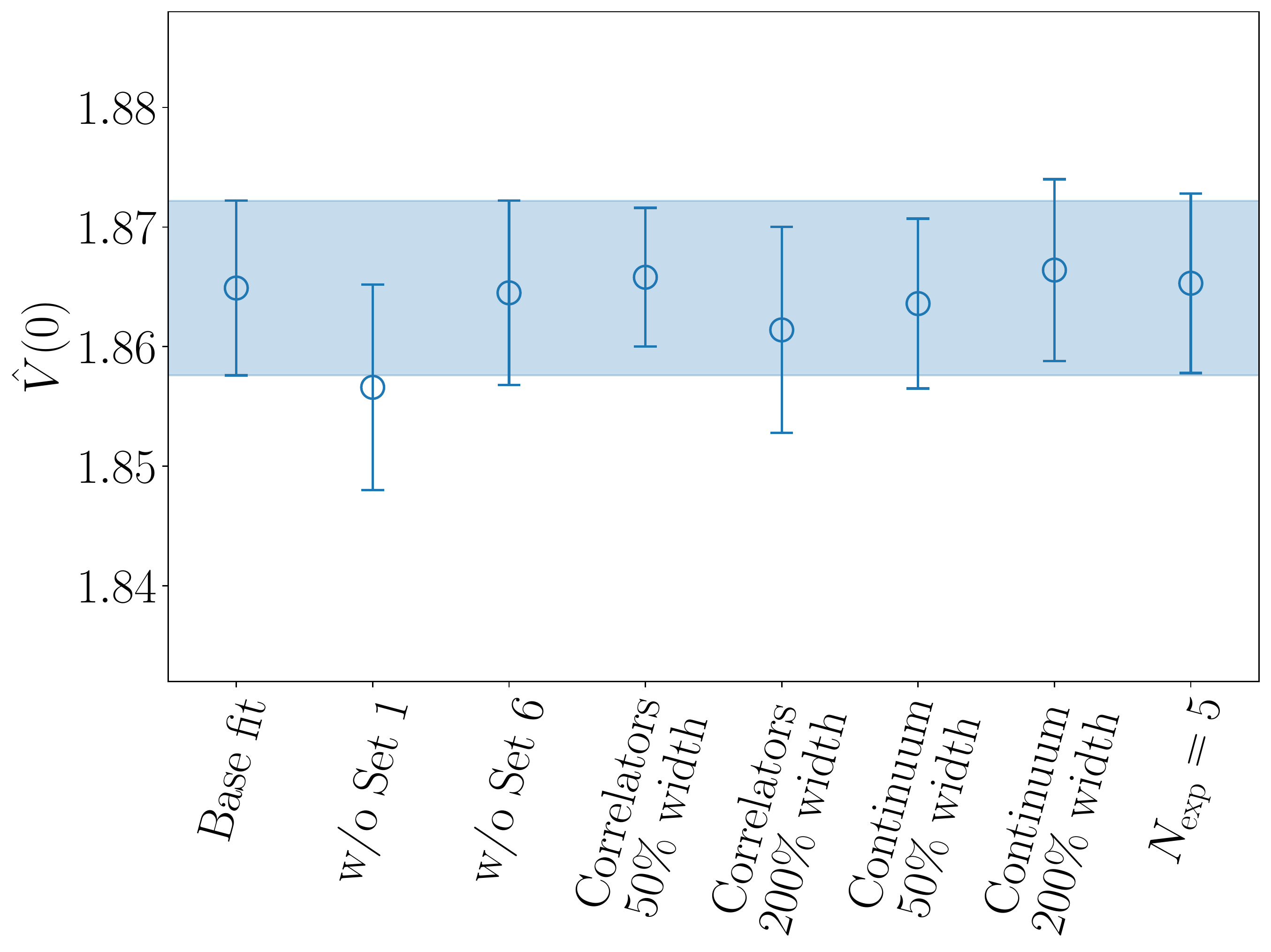}
		\caption{The value of $\hat{V}(0)$ in the limit of vanishing lattice spacing and physical quark masses obtained from variations to our base fit. These include (from left to right) dropping the coarsest and finest datasets, changing all the prior widths in our correlator fits, changing all the prior widths in our chiral/continuum fits and adding an additional normal and oscillating exponential to our correlator fits. Note that the values for $\Lambda$ in Eq.~\eqref{eqn:Vhat_fit_form} are fixed (see text) under these fit variations.}
		\label{fig:Vhat_stability}
	\end{center}
\end{figure}

We plot the impact on the value of $\hat{V}(0)$ of changes to either our correlator and continuum/chiral fits in Fig.~\ref{fig:Vhat_stability}. Our preferred (base) fit described above is given on the left. Variations include dropping datasets and changing the priors by factors of 0.5 or 2.0. We see very little variation in the final answer under any of these variations, confirming that our result is robust. 

\subsubsection{Additional systematic uncertainties} \label{sec:Jpsi_uncertainties}

In Section~\ref{sec:etac2gamma_uncertainties} we estimated the additional systematic uncertainty on $F(0,0)$ for $\eta_c \to \gamma \gamma$ from missing quark-line disconnected correlation functions and from missing QED effects. Here we do the same for $\hat{V}(q^2)$. 

As discussed in Section~\ref{sec:etac2gamma_uncertainties}, the missing disconnected correlation functions mean that there is a 7.3 MeV mismatch between the $\eta_c$ mass determined on the lattice in the continuum limit (tuning the $c$ quark mass so that the $J/\psi$ mass is correct) and that determined in experiment~\cite{Hatton:2020qhk}.
In Section~\ref{sec:Jpsi_cont_limit} we described how leading-order NRQCD gives a result of 2 for $\hat{V}(0)$ because it ignores the spin-dependent differences between the $J/\psi$ and $\eta_c$ wavefunctions. Our results, using a fully relativistic approach, show that $\hat{V}(0)$ differs from 2 by 7.2(4)\% . However, the missing disconnected correlation functions mean that we are missing a small part of the effects that generate a difference between the $J/\psi$ and $\eta_c$ (and result in their `wavefunction-overlap' differing from 1). The 7.3 MeV shift is 6\% of the mass difference between the $J/\psi$ and $\eta_c$ mesons (the hyperfine splitting). We might therefore expect that the missing disconnected correlation functions could generate a shift in $\hat{V}$ of 6\% of the 7.2\% difference from 2, i.e. a 0.4\% shift of $\hat{V}$. 

An additional effect to be considered is the identification of the value at $q^2=0$. Because the lattice $\eta_c$ mass does not exactly match that in experiment, the $q^2=0$ point will correspond to a slightly (6\%) incorrect value for the $\eta_c$ spatial momentum, $|\bf{q}|$. The $q^2$ dependence of $\hat{V}$ is so small, however, that shifting the $q^2$ value at which we determine $\hat{V}$ has negligible effect. Note also that our results on sets 3 and 3A for $\hat{V}$ (see Table~\ref{tab:Jpsiresults}) show almost no difference when we change the $\eta_c$ mass by 1\% (see Table~\ref{tab:Jpsimasses}). A 1\% shift in $\eta_c$ mass is much larger than the 0.2\% shift coming from missing quark-line disconnected correlation functions. 

We conclude that a 0.4\% systematic uncertainty in $\hat{V}$ from missing quark-line disconnected correlation functions is reasonable. 
This systematic uncertainty includes the effect of the fact that the $c$ quarks in charmonium carry an electric charge. The hyperfine splitting was calculated in~\cite{Hatton:2020qhk} in lattice QCD+QED and the 7.3 MeV shift between the lattice and experimental hyperfine splittings quoted above includes this QED effect. The impact of final-state QED interactions should be negligible for $\Gamma(J/\psi \to \gamma \eta_c)$ because there is no electric charge in the final state. We allow an additional uncertainty of $\mathcal{O}(\alpha/\pi) = 0.2\%$ for higher-order QED corrections when calculating $\Gamma(J/\psi \to \gamma \eta_c)$. 

The systematic uncertainty from missing quark-line disconnected correlation functions discussed above will be largely independent of $q^2$, since $\hat{V}(q^2)$ is such a flat function over the kinematic range of the decay. This means that it will cancel almost entirely in the ratio $R_{ee\gamma}$ of Eq.~\eqref{eq:Rdef}. 
When we come to consider  $\Gamma(J/\psi \to \eta_c e^+ e^-)$, however, we must allow a systematic uncertainty for final-state QED interactions because of the charged particles produced.  We will take an $\mathcal{O}(\alpha) \approx 1\%$ systematic uncertainty for this.

\subsection{Results} \label{sec:Jpsiresults}

Combining our fit result of Eq.~\eqref{eqn:fit_Vhat} with the additional systematic error discussed in Section~\ref{sec:Jpsi_uncertainties} we obtain a final result of 
\begin{align}
\label{eq:final_Vhat}
\finalVhat.
\end{align}
The total uncertainty here is $\finalVhatUncertainty$. The error budget is given in Table~\ref{tab:errorbudget} and can be compared to that previously discussed for $F(0,0)$. The main sources of uncertainty are somewhat different reflecting the fact that this is a dimensionless quantity, less sensitive to $w_0$, but with larger statistical errors from fitting three-point correlation functions and larger uncertainties from $a^2$ and $q^2$ dependence. The relative total uncertainty is similar in the two cases.

We can use this value in Eq.~\eqref{eqn:defn_V} to determine the decay width. For the kinematic factors of masses on the right-hand side of the equation we use experimental average values from~\cite{Workman:2022ynf}. These give $|{\bf{k}}|=110.9(4)\,\mathrm{MeV}$. Taking $\alpha=1/137.036$~\cite{Workman:2022ynf}, appropriate to the low moment-transfer here, we obtain
\begin{align}
\label{eqn:result_dw_JpsiGammaEtac}
	\finalWidthJpsiToGammaEtac 
\end{align}
The third error here is from the experimental hyperfine splitting that appears in $|{\bf k}|$. Since $|{\bf k}|$ is raised to the third power in $\Gamma$ this gives the largest single uncertainty in the final answer. The fourth uncertainty is from additional QED effects in the rate, as discussed in Section~\ref{sec:Jpsi_uncertainties}. Our total uncertainty on $\Gamma$ is \finalGammaUncertainty , adding the four contributions in quadrature. Taking the total $J/\psi$ width as 92.6(1.7) keV from~\cite{Workman:2022ynf} gives a branching fraction of 
\begin{align}
\label{eqn:result_br_JpsiGammaEtac}
	\finalBrJpsiToGammaEtac 
\end{align}
We have combined the two lattice uncertainties (from the fit and from the additional systematics) into the first uncertainty here. The second comes from the experimental hyperfine splitting, the additional QED uncertainty and from the $J/\psi$ total width, combined in quadrature. It is dominated by the error in the $J/\psi$ total width. 

\begin{figure}
	\begin{center}
		\includegraphics[width=0.45\textwidth]{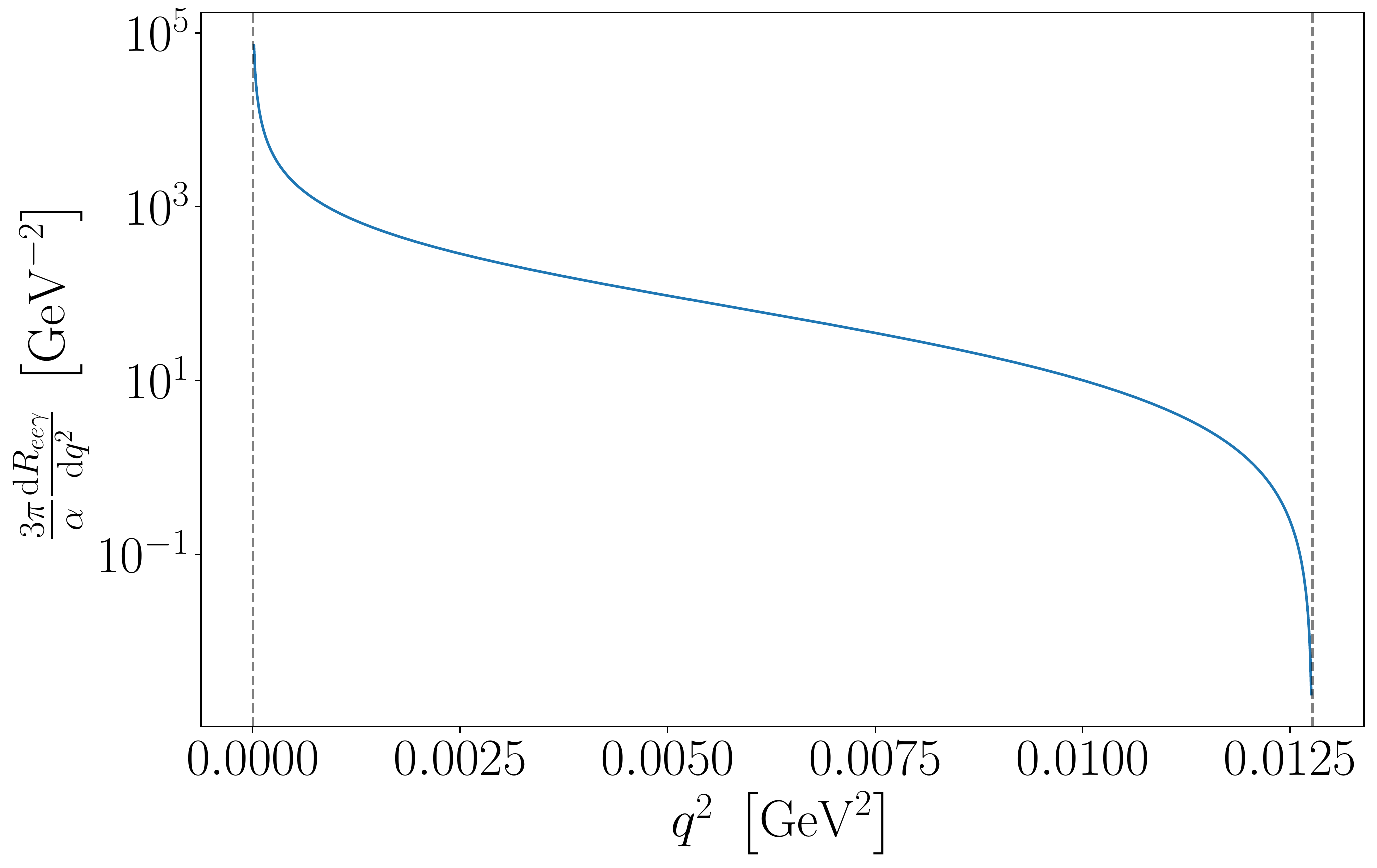}
		\caption{$3\pi/\alpha \times d R_{ee\gamma}/dq^2$ from Eq.~(\ref{eqn:dReedqSq}) plotted against $q^2$. Note the log scale on the $y$-axis. The vertical dashed lines mark the kinematic end-points of the integral at $4m_e^2$ and $(M_{J/\psi}-M_{\eta_c})^2$. The integrand for the case where $\hat{V}(q^2)$ is set equal to $\hat{V}(0)$, i.e. the form factor is taken to be completely flat in $q^2$, is visually indistinguishable from this.}
		\label{fig:Ree}
	\end{center}
\end{figure}

Turning now to $J/\psi \to \eta_c e^+e^-$, we plot the expression in Eq.~(\ref{eqn:dReedqSq}) for $d R_{ee\gamma} / d q^2$ as a function of $q^2$ in Fig.~\ref{fig:Ree}, taking out the normalising factor of $3\pi/\alpha$.
The integrand is very singular at low $q^2$ values and requires some care to integrate. It is cut off at the lower kinematic end-point, $4m^2_e$. We find the integrated value to be (using $\alpha$ = 1/137.036)
\begin{align}
 \label{eqn:Ree}
\finalRee.
\end{align}
From Fig.~\ref{fig:Vhat_data}, we know that the form factor $\hat{V}$ is very flat with respect to $q^2$ and we find that the integrand is virtually indistinguishable if we replace the factor of $\hat{V}(q^2)/\hat{V}(0)$ with 1.0. 
If we integrate Eq.~(\ref{eqn:Ree}) setting $\hat{V}$ to a constant we obtain a value of \ReenoVhat which differs from that above in Eq.~\eqref{eqn:Ree} by \finalDiffReeFromVhatUnity.  

Combining our results for $\Gamma(J/\psi \rightarrow \gamma \eta_c)$ and $R_{ee}$ in Eqs.~(\ref{eqn:result_dw_JpsiGammaEtac}) and~(\ref{eqn:Ree}) respectively, we find the decay width
\begin{align}
	\finalDecayWidthJpsiEtacee. \label{eqn:finalDecayWidthJpsiEtacee}
\end{align}
The first uncertainty comes from the combined `fit+syst' lattice uncertainty on $\Gamma(J/\psi \to \gamma \eta_c)$ from Eq.~\eqref{eqn:result_dw_JpsiGammaEtac} and the second uncertainty from the experimental contribution to that. The third uncertainty is the additional systematic error from QED effects such as final-state interactions discussed in Section~\ref{sec:Jpsi_uncertainties}. Combining the value from Eq.~\eqref{eqn:finalDecayWidthJpsiEtacee} with the $J/\psi$ total width gives a branching fraction
\begin{align}
	\finalDecayBrJpsiEtacee. \label{eqn:finalDecayBrJpsiEtacee}
\end{align}
The third, experimental, uncertainty is dominated by that from the $J/\psi$ total width. 

\subsection{Discussion} \label{sec:Jpsidiscussion}

\begin{figure}
	\begin{center}
		\includegraphics[width=0.45\textwidth]{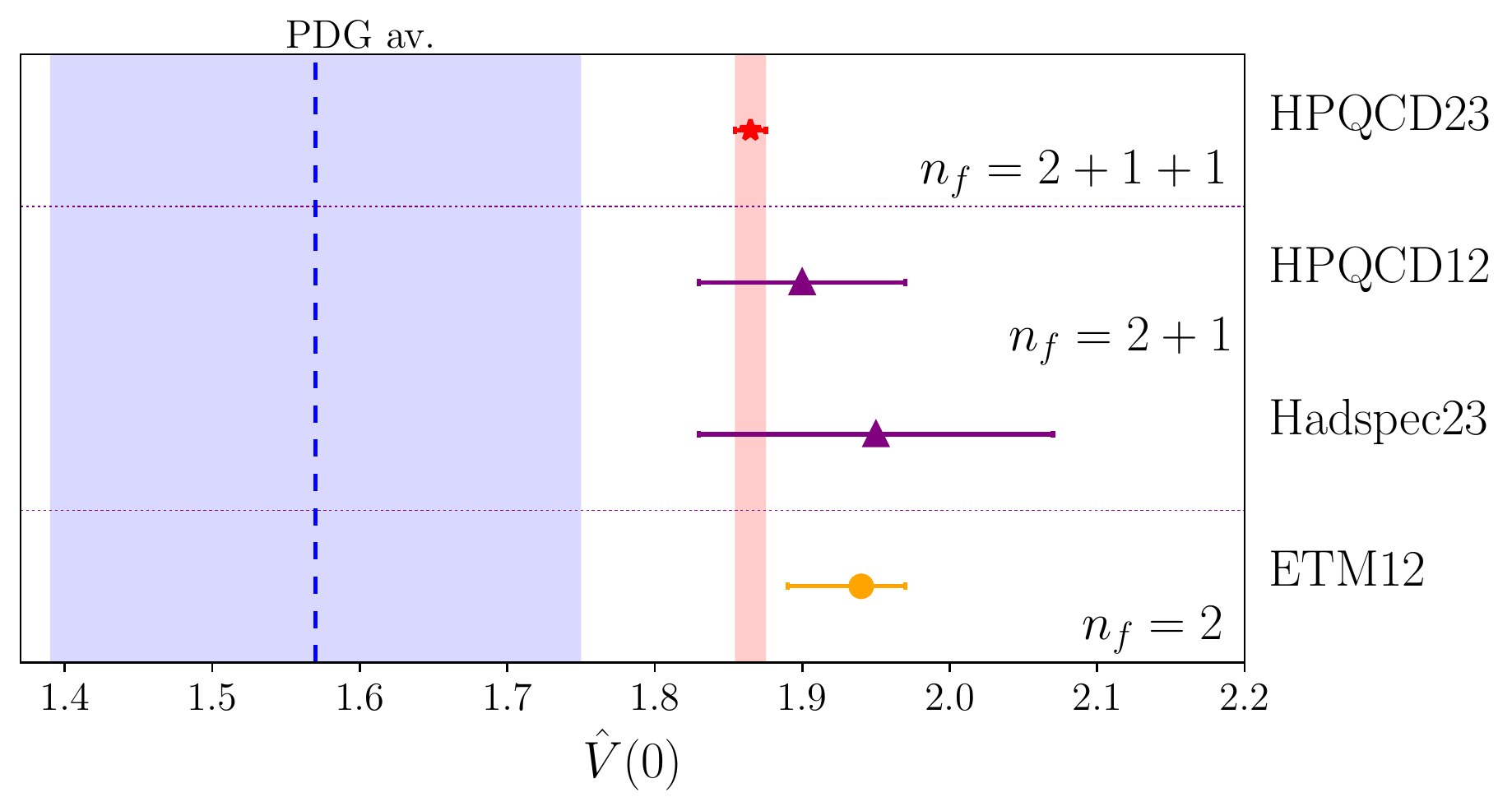}
		\caption{A comparison of values of $\hat{V}(0)$ from lattice QCD. The result from the work here is labelled `HPQCD23' (red asterisk) and includes $u$, $d$, $s$ and $c$ quarks in the sea with results at multiple lattice spacing values.  The results with $u$, $d$ and $s$ quarks are denoted with purple filled triangles. `HPQCD12'  used HISQ $c$ quarks on gluon field configurations including sea asqtad staggered quarks and two values of the lattice spacing~\cite{Donald:2012ga}. `Hadspec23' used clover quarks on anisotropic lattices at one value of the lattice spacing but include an estimate of systematic errors in their quoted uncertainty.  `ETM12' (filled orange circle) used the twisted mass formalism with gluon field configurations including $u$ and $d$ sea quarks only and four values of the lattice spacing~\cite{Becirevic:2012dc}. The blue band shows the value for $\hat{V}(0)$ inferred from the average experimental branching fraction~\cite{Workman:2022ynf}. The red band carries our result down the plot for comparison.}
		\label{fig:compare_Vhat}
	\end{center}
\end{figure}
\begin{figure}
	\begin{center}
		\includegraphics[width=0.45\textwidth]{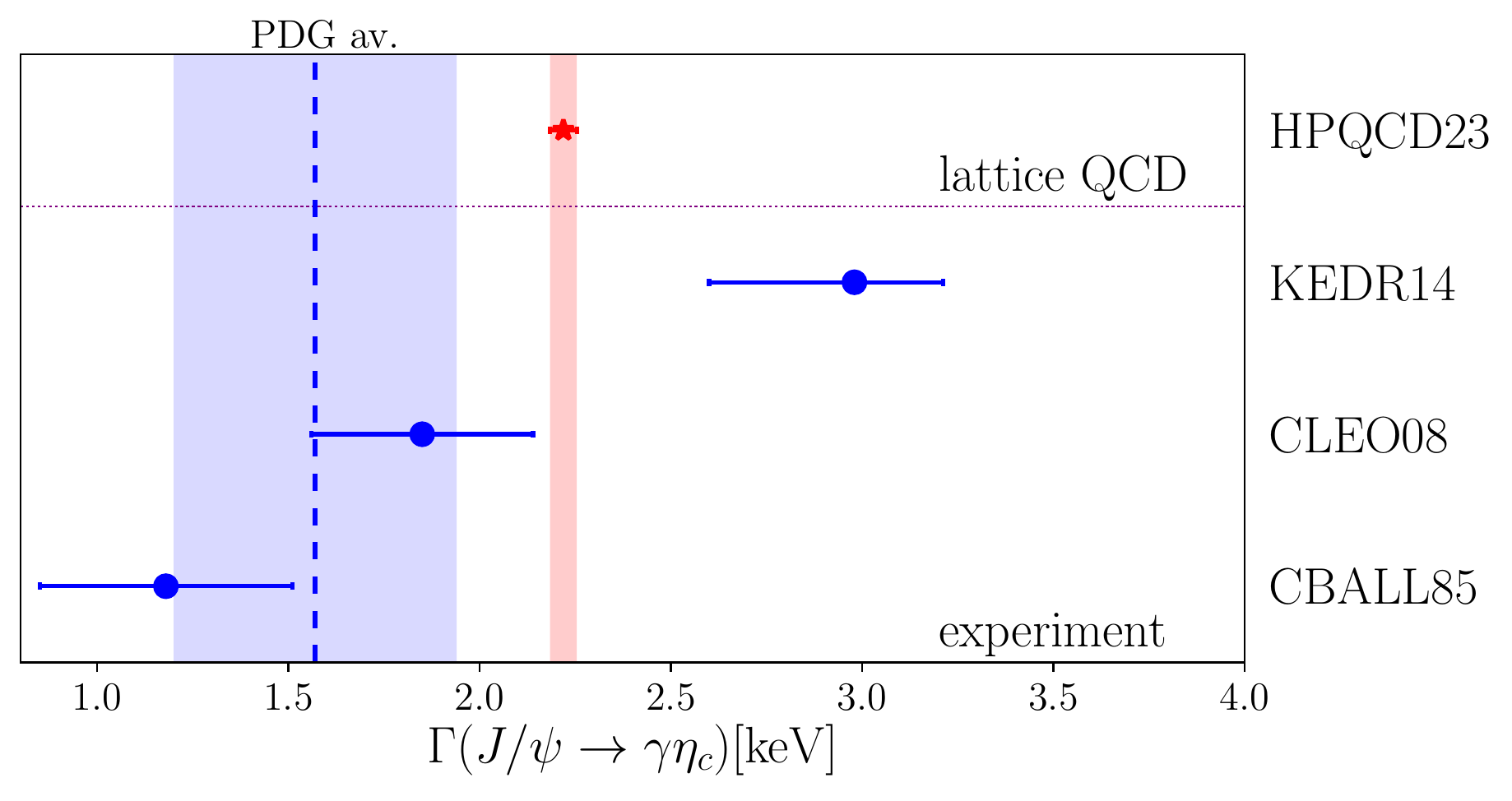}
		\caption{A comparison of our result for $\Gamma(J/\psi \to \gamma \eta_c)$ from lattice QCD to values from experiment. The result from the work here is labelled `HPQCD23' (red asterisk). The filled blue circles are results from individual experiments: `CBALL85' is from the Crystal Ball~\cite{Gaiser:1985ix}, `CLEO08' is from CLEO~\cite{CLEO:2008pln} and `KEDR14' is from KEDR~\cite{Anashin:2014wva} (plotting the quantity denoted $\Gamma^0_{\gamma\eta_c}$). The blue band shows the average of the CLEO and Crystal Ball results~\cite{Workman:2022ynf}. The red band carries our result down the plot for comparison.}
		\label{fig:compare_GammaJpsi}
	\end{center}
\end{figure}

Fig.~\ref{fig:compare_Vhat} compares our final result from Eq.~\eqref{eq:final_Vhat} for the form factor $\hat{V}(0)$ needed to determine the rate for $J/\psi \to \gamma \eta_c$ decay to earlier lattice QCD calculations including different numbers of flavours of sea quarks. Our result is a big improvement in accuracy over the earlier calculations, as well as having the most realistic sea quark content with $u$, $d$, $s$ and $c$ quarks in the sea. We also show, as a shaded blue band, the value of $\hat{V}(0)$ of 1.57(18) inferred from the average branching fraction for $J/\psi \to \gamma \eta_c$ of 1.7(4)\% and $J/\psi$ width of 92.6(1.7) keV~\cite{Workman:2022ynf} using Eq.~\eqref{eqn:defn_V}. We see good agreement between the lattice results but they are all higher than the value for $\hat{V}(0)$ inferred from the experimental rate. The experimental average branching fraction has a large uncertainty, inflated by a scale factor of 1.5 because of poor agreement between experiments. This means that the tension with our lattice QCD result is $1.6\sigma$ where $\sigma$ comes from the experimental value. 

We show more detail of the experimental picture in Fig.~\ref{fig:compare_GammaJpsi}. There we plot our result for $\Gamma(J/\psi \to \gamma \eta_c)$ (from Eq.~\eqref{eqn:result_dw_JpsiGammaEtac}) along with the three most recent experimental results, from KEDR~\cite{Anashin:2014wva}, CLEO~\cite{CLEO:2008pln} and Crystal Ball~\cite{Gaiser:1985ix}. Uncertainties quoted on the experimental values are combined in quadrature. We also include the PDG average value~\cite{Workman:2022ynf} which is an average of the branching fractions from CLEO and Crystal Ball (that we multiply by the average total $J/\psi$ width). We see that the Crystal Ball result is the lowest, over $3\sigma$ below our value from lattice QCD. The KEDR result is much higher, $2\sigma$ above our value. The KEDR analysis includes some model dependence and they quote their result as $\Gamma^0_{\gamma\eta_c}$, which is the value we plot in Fig.~\ref{fig:compare_GammaJpsi}. The CLEO result is in between and in good agreement (within $1.5\sigma$) with our result. 

\begin{figure}
	\begin{center}
		\includegraphics[width=0.45\textwidth]{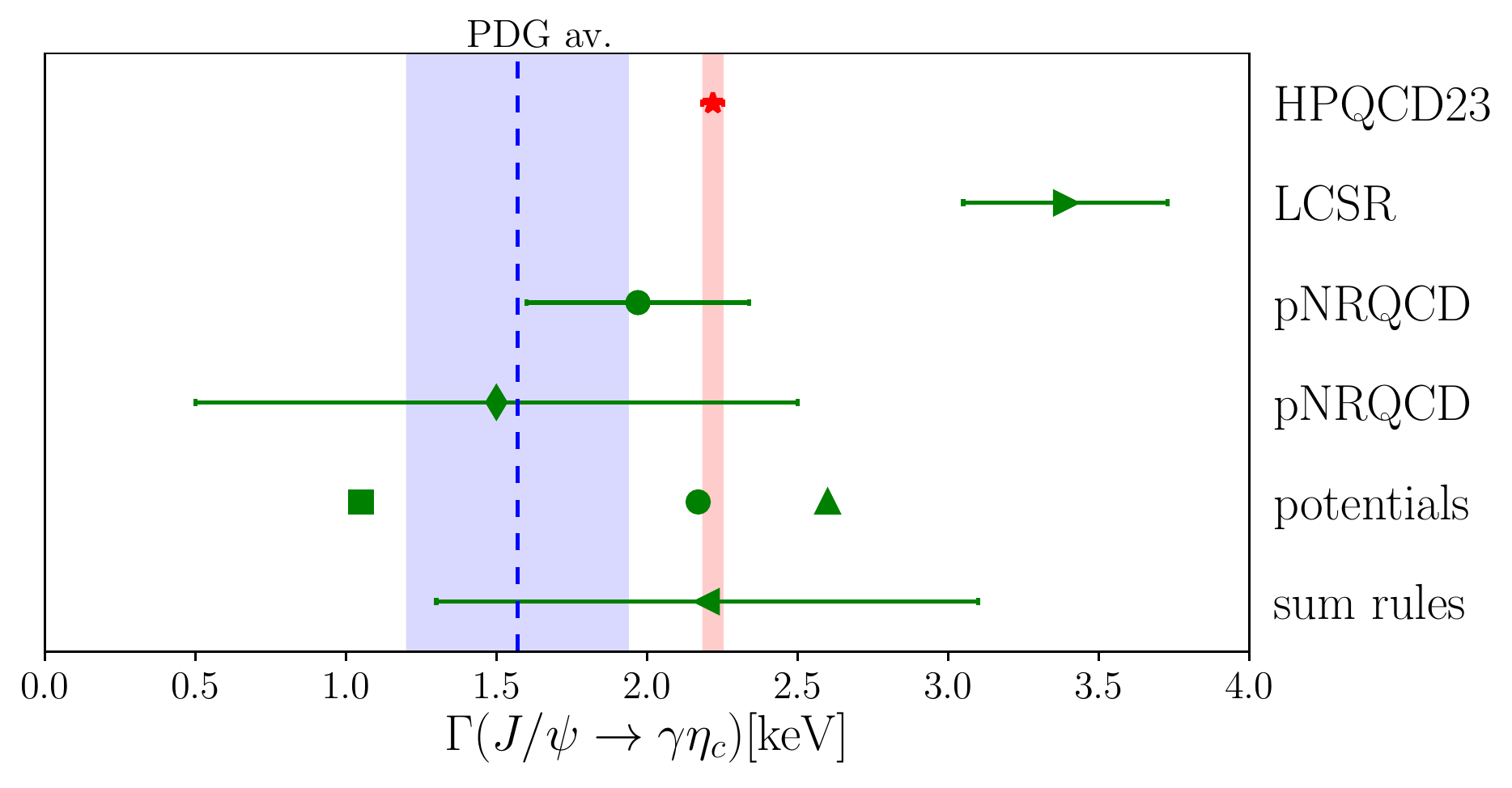}
		\caption{A comparison of our result for $\Gamma(J/\psi \to \gamma \eta_c)$ from lattice QCD to a selection of values from other theoretical approaches. The result from the work here is labelled `HPQCD23' (red asterisk). The left-pointing green triangle labelled sum rules is an early result with this method from~\cite{khodjamirian:1983gd}.  Three values are given in the row labelled 'potentials' to encompass the range of results. The rightmost point (green triangle) is from~\cite{Eichten:2007qx}, the middle point (green circle) is the starting point for a pNRQCD analysis from~\cite{pineda:2013lta}. The numbers plotted for these two cases have been updated to use the current value for $|{\bf k}|$ (see text). The leftmost point (green square) is from a relativistic quark model~\cite{Ebert:2002pp}. Two pNRQCD analyses are shown; the lower value (green diamond) is from~\cite{Brambilla:2005zw} using the weak-coupling limit and the upper one (green circle) from~\cite{pineda:2013lta} including corrections through $\alpha_s^2$ and $v^2$ (this value has also been corrected to use the current value of $|{\bf k}|$). The right-pointing green triangle labelled `LCSR' is a recent result using light-cone sum rules from~\cite{Guo:2019xqa}.The blue band shows the experimental average of the CLEO and Crystal Ball results~\cite{Workman:2022ynf}. The red band carries our result down the plot for comparison.}
		\label{fig:compare_GammaJpsi_theory}
	\end{center}
\end{figure}

Figure~\ref{fig:compare_GammaJpsi_theory} shows a comparison between our lattice result for $\Gamma(J/\psi \to \gamma \eta_c)$ and a selection (not intended to be exhaustive) of earlier theoretical results using different techniques. It is clear that lattice QCD is able to provide a much more accurate result for this decay width than previous approaches. A comparison is nevertheless useful to allow an assessment of these other approaches for use in cases that are not as amenable to lattice QCD calculations. Indeed here the experimental picture is not at all clear and lattice QCD results can substitute for an accurate experimental value in this assessment, as for $\Gamma(\eta_c \to \gamma \gamma)$ in Section~\ref{sec:etac-discussion}. 

We see in Figure~\ref{fig:compare_GammaJpsi_theory} that theory results for $\Gamma(J/\psi \to \gamma \eta_c)$ have covered a wide range (compared to the lattice QCD results of Figure~\ref{fig:compare_Vhat}). A traditional approach has been to use a nonrelativistic potential, but often these results are quoted with no error bars. In Fig.~\ref{fig:compare_GammaJpsi_theory} we give two results from nonrelativistic potentials. One, from~\cite{Eichten:2007qx}, takes the wavefunction-overlap between $J/\psi$ and $\eta_c$ to be 1 and the other, from~\cite{pineda:2013lta}, is the leading-order result for a pNRQCD analysis. To make a fair comparison with our value, we have corrected both of these results to use the value for $|{\bf k}|$ (see Eq.~\eqref{eqn:mag3mom_onShell}) obtained from current average experimental masses~\cite{Workman:2022ynf} rather than the values they used from earlier PDG reports. Since the experimental average hyperfine splitting has changed by several percent over time and $|{\bf k}|$ appears cubed in $\Gamma$, this has some impact. These two nonrelativistic potential model results differ by 20\%, bracketing our value, and this reflects reasonably the range of results (compare, for example, more recent values in~\cite{Deng:2016stx}). The pNRQCD approach systematically adds correction terms to this~\cite{Brambilla:2005zw,pineda:2013lta}. We can see how this works most clearly in the result of~\cite{pineda:2013lta} since the size of corrections through $\mathcal{O}(\alpha_s^2)$ and $\mathcal{O}(v^2)$ are tabulated. The $v^2$ corrections are large but of opposite sign to those at $\mathcal{O}(\alpha_s)$ and $\mathcal{O}(\alpha_s^2)$. Disappointingly we see, by comparing the two green circles in Fig.~\ref{fig:compare_GammaJpsi_theory}, that the net effect of these corrections is to move the result further from our value rather than towards it. Higher-order corrections are still likely to be sizeable and the good news is that the uncertainty estimates attached to the pNRQCD results means that they are in good agreement with our value. In contrast a recent result from light-cone sum rules~\cite{Guo:2019xqa} has significant tension, over 3$\sigma$, with our value. The result from a relativistic quark model~\cite{Ebert:2002pp} also looks in disagreement. 

We now turn to a test of the relationship between $\Gamma(J/\psi \to \gamma \eta_c)$, $\Gamma(\eta_c \to \gamma\gamma)$ and $\Gamma(J/\psi \to e^+e^-)$ suggested by Shifman many years ago and given in Eq.~\eqref{eq:shifman}. 
The accurate results that we now have from lattice QCD for these decay widths enables us to see how well this approximate relationship works. It is easiest to do this by converting the expression of Eq.~\eqref{eq:shifman} into a connection between the hadronic parameters, $\hat{V}(0)$, $F(0,0)$ and $f_{J/\psi}$. Eq~\eqref{eq:shifman} becomes 
\begin{equation}
\hat{V}(0) = \frac{F(0,0)}{2f_{J/\psi}} M_{J/\psi}(M_{J/\psi}+M_{\eta_c})(1 + \mathcal{O}(\alpha_s))\, .
\end{equation}
In terms of the ratio $R_{fF}$ defined in Eq.~\eqref{eq:frat2} this reduces to 
\begin{equation}
\label{eq:shifman3}
\hat{V}(0)R_{fF} = \frac{M_{J/\psi}+M_{\eta_c}}{2M_{J/\psi}}(1 + \mathcal{O}(\alpha_s))\, .
\end{equation}

Our results for $\hat{V}(0)$ from Eq.~\eqref{eq:final_Vhat} and $R_{fF}$ from Eq.~\eqref{eq:fratresult} yield
\begin{equation}
\label{eq:shifman4}
\hat{V}(0)R_{fF} = \finalshif \, .
\end{equation}
The expectation from the ratio of masses on the right-hand side of Eq.~\eqref{eq:shifman3} gives 0.982 using experimental averages from~\cite{Workman:2022ynf}. This differs from our result for $\hat{V}(0)R_{fF}$ by 10\%, which is well within the leeway provided on the 0.982 by possible $\mathcal{O}(\alpha_s)$ corrections. In the nonrelativistic limit, the mass ratio would simply be 1.0, and 0.982 is a slight improvement on this, in terms of being closer to the full lattice QCD value that we obtain in Eq.~\eqref{eq:shifman4}. Going further in the nonrelativistic direction, Eq.~\eqref{eq:shifman3} would reduce to the expectation that $\hat{V}(0)=2$ (when $R_{fF}$ is assumed to take the value 1/2), which in fact works just as well in comparison to our results (since 2.0 differs from our value for $\hat{V}(0)$ by 7\%). 

We conclude from the comparison of our results to other theory values that nonrelativistic approaches to $J/\psi \to \gamma \eta_c$ work at the 10--20\% level at leading-order, but it is hard to improve on this by adding corrections. 

Finally, we note that our result for $R_{ee\gamma}$ in Eq.~\eqref{eqn:Ree} agrees with that given in~\cite{Gu:2019qwo} using a simple $\psi^{\prime}$ pole model for $\hat{V}(q^2)$.  This is not surprising since, as discussed in Section~\ref{sec:Jpsiresults}, the result is insensitive to details of $\hat{V}(q^2)$ over the short $q^2$ range of the decay. 

\section{Conclusions} \label{sec:conclusions}

We give improved lattice QCD results for the hadronic matrix elements needed to determine $\Gamma(\eta_c \to \gamma \gamma)$, $\Gamma(J/\psi \to \gamma \eta_c)$ and $\Gamma(J/\psi \to \eta_c e^+e^-)$, These results include $u$, $d$, $s$ and $c$ quarks in the sea for the first time. The HISQ action gives us small discretisation errors and we use a wide range of lattice spacing values and sea $u/d$ quark mass values for good control of the physical-continuum limit. Our lattice QCD results now have smaller uncertainties than the corresponding experimental values. This means that new experimental results with improved uncertainties could have considerable impact as stringent tests of QCD. 

Our results for $\eta_c \to \gamma\gamma$ transform the theoretical picture for this decay. Our determination of the hadronic form factor $F(0,0)$, defined in Eq.~\eqref{eq:Fdef}, is (repeating Eq.~\eqref{eqn:final_F00})
\begin{align}
	\finalF \label{eqn2:final_F00} \, .
\end{align}
This gives a decay width of 
\begin{align}
\label{eq2:finalwidthetac}
	\finalWidthEtacGammaGamma \, ,
\end{align}
repeating Eq.~\eqref{eq:finalwidthetac}. The first uncertainty comes from the lattice calculation and the second is from remaining systematic errors (see Section~\ref{sec:etac2gamma_uncertainties}) from missing quark-line disconnected diagrams and QED effects.

As discussed in Section~\ref{sec:etac-discussion}, our result has \diffpdgsigma$\sigma$ tension with the PDG fit result of 5.15(35) keV~\cite{Workman:2022ynf}. The PDG fit has a poor $\chi^2$ and the difficulty of determining a reliable fit value and uncertainty from the wide range of indirect experimental results that exist is clear in Fig.~\ref{fig:compGamma-expt}. We believe that this fit needs to be revisited. Instead our result for the decay width agrees within 2$\sigma$ with the value for the width of 5.90(58) keV obtained from the PDG average~\cite{Workman:2022ynf} of a more restricted set of experimental results, those for $\eta_c$ production via 2-photon fusion using the $\eta_c$ decay mode to $K\overline{K}\pi$. This channel also has the advantage of having the smallest relative uncertainty.

Our results, along with earlier 0.4\%-accurate HPQCD calculations of the $J/\psi$ decay constant~\cite{Hatton:2020qhk}, allow us to test how well earlier expectations using NRQCD work when confronted with the results from a fully-relativistic QCD calculation. Leading-order NRQCD gives a simple relationship between $F(0,0)$ for $\eta_c\to\gamma\gamma$ decay and the $J/\psi$ decay constant given in Eq.~\eqref{eq:frat}. We determine the ratio (repeating Eq.~\eqref{eq:fratresult})
\begin{equation}
\label{eq2:fratresult}
R_{fF} \equiv \frac{f_{J/\psi}}{F(0,0)M^2_{J/\psi}}= \fitrat \, .
\end{equation}
The leading-order NRQCD expectation of 0.5 is not far from this number, suggesting that there is significant cancellation of the NRQCD higher-order corrections, expected to be individually of order 30\%~\cite{Czarnecki:2001zc,Penin:2004ay,Kiyo:2010jm}. The comparison of NRQCD calculations to our result is shown in Fig.~\ref{fig:compGamma-nrqcd}.

The relative success of the leading-order NRQCD expectation for the ratio above suggests that it may also be used to predict $\eta_b \to \gamma\gamma$. We therefore expect that 
\begin{equation}
\label{eq:LONRQCDb}
\frac{\Gamma (\Upsilon \rightarrow e^+e^-)}{\Gamma (\eta_b \rightarrow \gamma\gamma )} = \frac{1}{3Q_b^2}(1+\mathcal{O}(\alpha_s) +\mathcal{O}(v^2/c^2)) \approx 3.
\end{equation}
Using HPQCD's lattice QCD results for $\Gamma (\Upsilon \rightarrow e^+e^-)$ of 1.292(37)(3) keV~\cite{Hatton:2021dvg} we obtain the prediction 
\begin{equation}
\label{eq:etab-prediction}
\Gamma (\eta_b \rightarrow \gamma\gamma ) = 431(12)(86) \mathrm{eV} \, .
\end{equation}
The second uncertainty allows for 20\% variation from missing higher-order radiative and relativistic corrections. For the $b$ case we expect the $\mathcal{O}(\alpha_s\approx 0.2)$ corrections to be larger than the relativistic corrections ($\mathcal{O}(v^2 \approx 0.1)$) and the cancellation of these may then not work as well as in the $c$ case. Indeed the authors of CM01~\cite{Czarnecki:2001zc} are quoted in~\cite{ALEPH:2002xdz} as determining a width $\Gamma (\eta_b \rightarrow \gamma\gamma ) = 570(50)\,$eV including corrections through $\alpha_s^2$ in continuum NRQCD to Eq.~\eqref{eq:LONRQCDb}. A more recent analysis~\cite{Kiyo:2010jm} gives a similar central value but larger uncertainty, with a decay width of 540(150)\,eV. This would represent a larger correction to the leading-order NRQCD result than seen in the $c$ case but going in the same (positive) direction. We saw from Fig.~\ref{fig:compGamma-nrqcd}, however, that the full lattice QCD result for 
$\Gamma (\eta_c \rightarrow \gamma\gamma )$ is below the LO NRQCD result rather than above.

Ultimately the SM value  for $\Gamma (\eta_b \rightarrow \gamma\gamma)$ will be resolved by an accurate calculation in lattice QCD. Now that we have demonstrated the accuracy possible for the $\eta_c$ calculation using HISQ quarks we envisage increasing the mass up towards the $b$ and applying the techniques used in~\cite{Hatton:2021dvg} to do this. A prediction ahead of possible experimental results from Belle II would be timely.

\begin{figure}
	\begin{center}
		\includegraphics[width=0.45\textwidth]{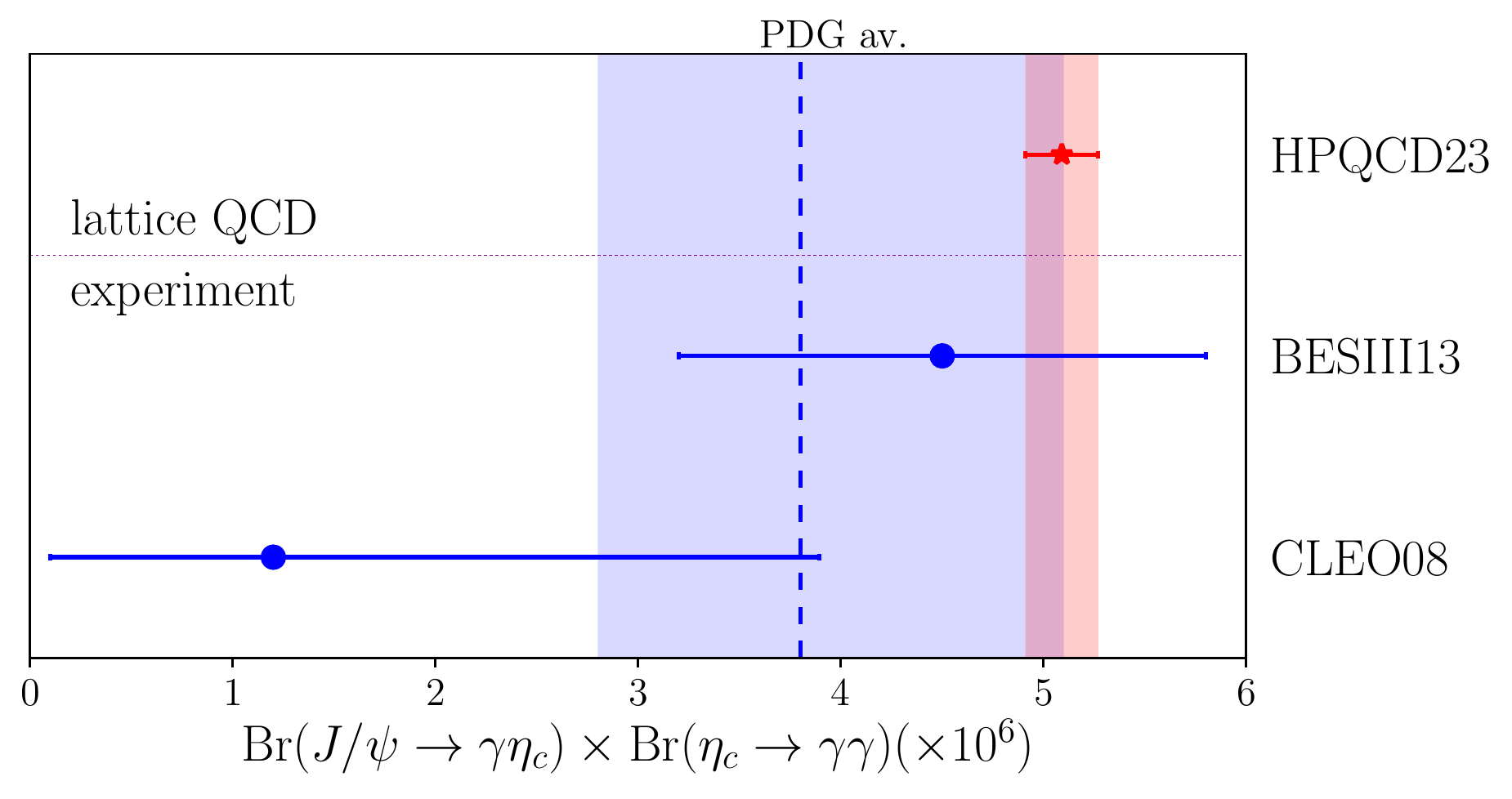}
		\caption{A comparison of our result for the product of branching fractions $\text{Br}(J/\psi \to \gamma \eta_c)\times\text{Br}(\eta_c\to \gamma\gamma)$ from lattice QCD to values from experiment. The result from the work here is labelled `HPQCD23' (red asterisk). The filled blue circles are from CLEO~\cite{CLEO:2008qfy} (labelled `CLEO08') and BESIII~\cite{BESIII:2012lxx} (labelled `BESIII13'). The blue band shows the average of the CLEO and BESIII results~\cite{Workman:2022ynf}. The red band carries our result down the plot for comparison.}
		\label{fig:compare_BrJpsi3g}
	\end{center}
\end{figure}

Our results for the form factor for the M1 radiative transition $J/\psi \to \gamma \eta_c$ also represent a step up in accuracy over previous lattice QCD results. We find a value for the form factor at $q^2=0$, repeating Eq.~\eqref{eq:final_Vhat},
\begin{align}
\label{eq:final_Vhat2}
\finalVhat.
\end{align}
This gives a decay width, repeating Eq.~\eqref{eqn:result_dw_JpsiGammaEtac}
\begin{align}
\label{eqn:result_dw_JpsiGammaEtac2}
	\finalWidthJpsiToGammaEtac 
\end{align}
The first uncertainty here comes from the lattice fit, the second from additional systematic errors from missing quark-line disconnected correlation functions, the third uncertainty is from the experimental average hyperfine splitting that enters the kinematic factors converting the squared form factor into a decay width (see Eq.~\eqref{eqn:defn_V}) and the fourth from additional QED effects in the rate. For the related Dalitz decay, $J/\psi \to \eta_c e^+e^-$ we predict 
\begin{align}
	\finalDecayWidthJpsiEtacee. \label{eqn:finalDecayWidthJpsiEtacee2}
\end{align}repeating Eq.~\eqref{eqn:finalDecayWidthJpsiEtacee}.

Our result for $\Gamma(J/\psi \to \gamma \eta_c)$ joins earlier lattice QCD values in being higher than the experimental average value~\cite{Workman:2022ynf} obtained from averaging results from Crystal Ball~\cite{Gaiser:1985ix} and CLEO~\cite{CLEO:2008pln}. The small uncertainty of our result makes the tension more compelling than before. Our result is in 3$\sigma$ tension with that from Crystal Ball but agrees within 1.5$\sigma$ with that from CLEO. See Fig.~\ref{fig:compare_GammaJpsi} for the comparison. 

We can use our results to calibrate other theoretical approaches (see Fig.~\ref{fig:compare_GammaJpsi_theory}) and also to test the suggested~\cite{Shifman:1979nx} simple relationship between $\hat{V}(0)$, $F(0,0)$ and $f_{J/\psi}$, by determining (repeating Eq.~\eqref{eq:shifman4}) 
\begin{equation}
\label{eq:shifman4-2}
\hat{V}(0)R_{fF} = \finalshif \, .
\end{equation}
This is to be compared with 0.982 from Eq.~\eqref{eq:shifman3}, or 1.0 in the nonrelativistic limit. Once again we find that the simple expectations work fairly well, at the $\sim 10\%$ level, but this does not of course provide the kind of accuracy available now from lattice QCD, as we have shown here. 

Finally, for an alternative perspective on the experimental situation, we multiply together the branching fractions we have determined for $J/\psi \to \gamma \eta_c$ and $\eta_c \to \gamma\gamma$ (in Eqs.~\eqref{eqn:result_br_JpsiGammaEtac} and~\eqref{eq:finalBretac}) and compare to direct experimental determination of this product by CLEO~\cite{CLEO:2008qfy} and BESIII~\cite{BESIII:2012lxx}. This comparison is shown in Fig.~\ref{fig:compare_BrJpsi3g}. We see good agreement between the lattice QCD result and the experimental values, particularly that from BESIII. The experimental determinations have large uncertainties at present but improvements here to provide more stringent tests against lattice QCD would be very useful, as we have stressed throughout this paper.

\section{Acknowledgements} \label{sec:ack}

We thank the MILC collaboration for making publicly available their gauge configurations and their code MILC-7.7.11 \cite{MILCgithub}. We are grateful to Gordon Donald and Dan Hatton for many useful discussions. We thank Jim Simone for providing the $u_0$ values listed in Table~\ref{tab:ZV} and  the Particle Data Group for providing a corrected average for the $\eta_c \to \gamma \gamma$ branching fraction. 
This work was performed using the Cambridge Service for Data Driven Discovery (CSD3), part of which is operated by the University of Cambridge Research Computing Service on behalf of the Science and Technology Facilities Council (STFC) DiRAC HPC Facility. The DiRAC component of CSD3 was funded by BEIS capital funding via STFC capital grants ST/P002307/1 and ST/R002452/1 and STFC operations grant ST/R00689X/1. DiRAC is part of the National e-Infrastructure. 
We are grateful to the CSD3 support staff for assistance.
Funding for this work came from STFC grant ST/T000945/1.

\appendix

\section{Trapezoidal integration and oscillating contributions} \label{sec:osc-sum}

Here we discuss the accuracy of representing the integral over $t_{\gamma_1}$ in the continuum with a sum over lattice times in Eq.~\eqref{eq:Ctilde} and the impact on this of the presence of oscillating terms arising from the use of a staggered quark formalism. 

We want to approximate the integral
\begin{equation}
\label{eq:fullint}
\int_{-\infty}^{\infty} f(t) dt
\end{equation}
with a sum
\begin{equation}
\label{eq:staggsum}
a\sum_{n=-\infty}^{n=+\infty} \left( f(na) + (-1)^n g(na) \right) 
\end{equation}
that includes a term oscillating in time through the factor $(-1)^n$. $f$ and $g$ are smooth functions, continuous at $t=0$, that vanish in a well-behaved way as $t \to \infty$. 

For the sum over $f$ Eq.~\eqref{eq:staggsum} is using the standard trapezoidal rule, which has $a^2$ errors. 
For the sum over $g$ we combine three adjacent terms, reducing the sum to even values of $n$ only,
\begin{eqnarray}
\label{eq:sum-on-n-g}
a\sum_{n=-\infty, n\,{\text{even}}}^{n=+\infty, n\,{\text{even}}} \frac{ g(na) - 2g(na+a) + g(na+2a)}{2} && \\
&&\hspace{-15em}= a\sum_{n=-\infty, n\,{\text{even}}}^{n=+\infty, n\,{\text{even}}}  \left[\frac{a^2}{2}{g^{\prime\prime}(na+a)} + \ldots \right] \nonumber
\end{eqnarray}
Splitting the sum into two pieces, for positive and negative $n$, we have for positive $n$
\begin{eqnarray}
\label{eq:intgpp}
a\sum_{n=0,2,4}^{n=+\infty}  \left[\frac{a^2}{2}{g^{\prime\prime}(na+a)} + \ldots \right] &=& \frac{a^2}{4}\int dt \, g^{\prime\prime} (t) \nonumber \\
&=& -\frac{a^2}{4}g^{\prime}(0^+)
\end{eqnarray}
if $g$ and $g^{\prime}$ vanish at $t\to \infty$. Negative $n$ gives a similar result, with opposite sign, so that we have a total for the sum over $g$ in Eq.~\eqref{eq:staggsum} of 
\begin{equation}
\label{eq:sum-on-g}
\frac{a^2}{4} \left[ g^{\prime}(0^-) - g^{\prime}(0^+) \right] \, .
\end{equation}
The result is a discretisation effect, proportional to the discontinuity in the derivative of $g$ at $t=0$ but vanishing as $a \to 0$ as $a^2$. 

We conclude that summing over lattice time slices to obtain $\tilde{C}_{\mu\nu}$ (Eq.~\eqref{eq:Ctilde}) as an approximation to the time integral that sets the photon on-shell introduces discretisation errors proportional to $a^2$ at leading order. These come both from the trapezoidal integration implied by the sum and from the oscillating terms in the correlation function that arise from the use of staggered quarks. Such discretisation errors are taken into account by our fit to the results for the form factor as a function of lattice spacing and removed in our result for $F(0,0)$ in the continuum limit. 

A toy model illustrates this further. We take $f$ and $g$ to be single exponentials, $f=\exp(-M_\mathrm{n} |t|)$ and $g=\exp(-M_\mathrm{o} |t|)$. For $f$ we have
\begin{align}
\label{eq:testint}
	\int_{-\infty}^{+\infty} dt \: e^{-M_{\mathrm{n}}|t|} &\approx a \sum_{j=-\infty}^{+\infty}  e^{-aM_{\mathrm{n}}j} \nonumber \\
	& = a \Big(-1 +  2\sum_{j =0}^{\infty} e^{-aM_\mathrm{n}  j}\Big) 
\end{align}
on the lattice. The left-hand side of Eq.~\eqref{eq:testint} evaluates to $2/M_{\mathrm{n}}$.
The sum on the right-hand side is a geometric series, so we have 
\begin{align}
	\mathrm{RHS} &= a \Big(-1 +  \frac{2}{1 - e^{-aM_\mathrm{n} }}  \Big)  \\
	&= \frac{2}{M_{\mathrm{n}}} \Big( 1 + \frac{(aM_{\mathrm{n}})^2}{12} + \mathcal{O}((aM_{\mathrm{n}})^4) \Big) \, .\nonumber
\end{align}
As expected the integral obtained in this manner is accurate up to $(aM_{\mathrm{n}})^2$ errors. 

We repeat this exercise for the oscillating exponential to obtain
\begin{align}
	a\Big(-1 +  2\sum_{j =0}^{\infty} (-1)^{j} e^{-aM_\mathrm{o}  j} \Big) &= a\Big(-1 +  \frac{2}{1 + e^{-aM_\mathrm{o} }} \Big) \\
	&\hspace{-5em}=\frac{1}{M_{\mathrm{o}}} \Big( \frac{(aM_{\mathrm{o}})^2}{2} + \mathcal{O}((aM_{\mathrm{o}})^4) \Big). \nonumber
\end{align}
As expected the impact of the oscillatory contributions vanish in the continuum limit as $a^2$ and the result matches that from Eq.~\eqref{eq:sum-on-g}. 

We reach the same conclusions about discretisation error from the toy model as from the more general $f$ and $g$ functions of Eq.~\eqref{eq:staggsum}. 

\section{Fitting a subset of $\eta_c \to \gamma \gamma$ data} \label{sec:T-test}
 \begin{figure}
	\begin{center}
		\includegraphics[width=0.45\textwidth]{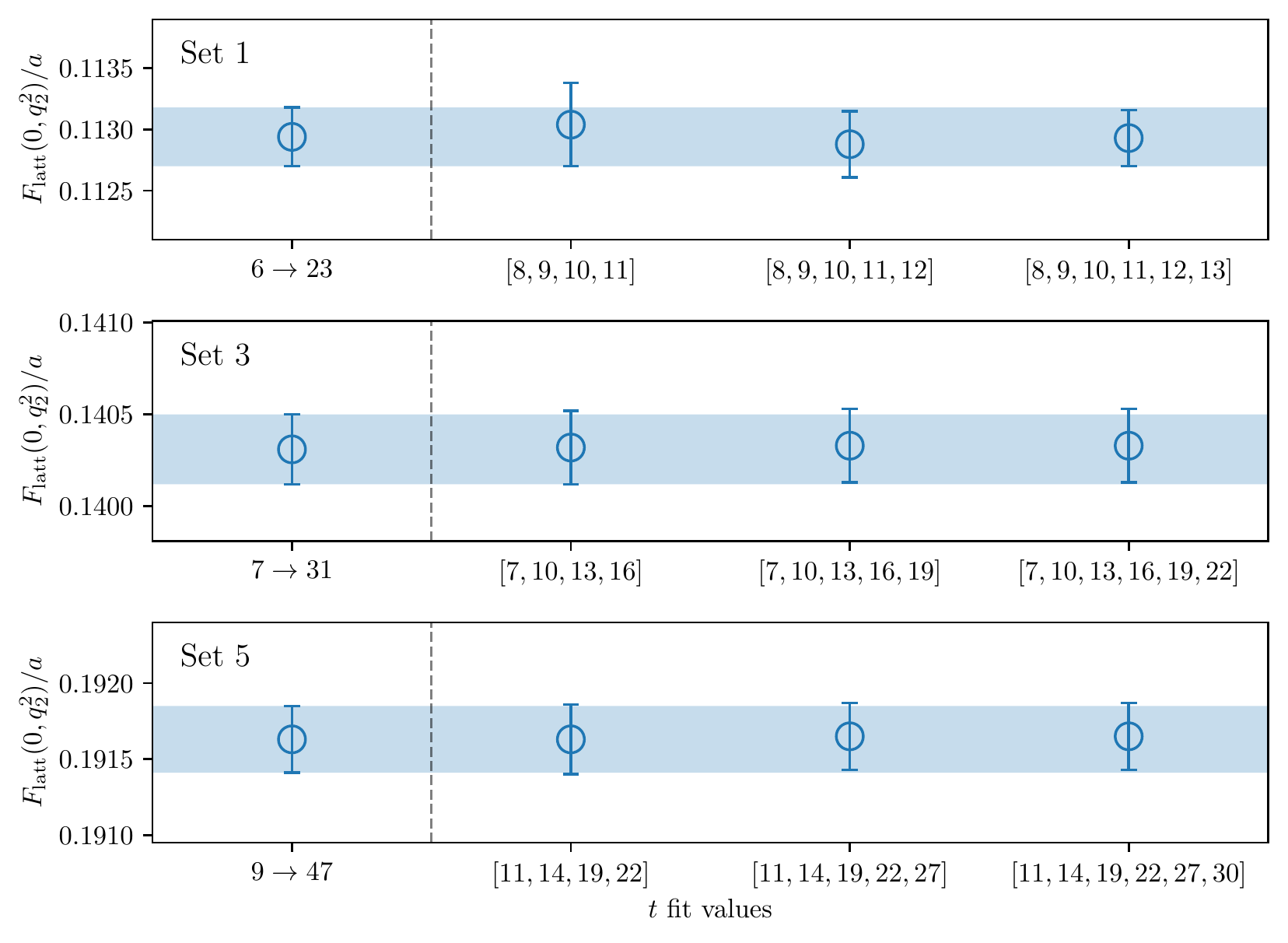}
		\caption{Fitted results for $F_{\text{latt}}(0,q_2^2)$, comparing values obtained from our full fit with those that use a subset of time separations, $t$, between source and sink in $\tilde{C}_{\mu\nu}(t)$. Results are given for the LOCAL set-up on set 1 (very-coarse, $a \approx$ 0.15 fm), set 3 (coarse, $a \approx$ 0.12 fm, fit simultaneously with set 3A) and 5 (fine, $a \approx$ 0.09 fm); all values are in lattice units. The values on the left give results from the full fit to all $t$ values except those discarded at early times (below $t_{\text{min}}$, see text in section~\ref{sec:etacggfits}). The values on the right give fit results for a selected 4, 5 or 6 $t$ values. We see that uncertainties close to that of the full fit are possible even with 4 $t$ values. Even and odd $t$ values are needed and the spacing between $t$ values and range covered is adjusted as a function of lattice spacing.  }
		\label{fig:Ttest}
	\end{center}
\end{figure}

As remarked in Section~\ref{sec:etacggfits}, it is possible to obtain accurate results for the $\eta_c\to \gamma\gamma$ form factor with a lot less numerical work than we have expended here. We showed, in Fig.~\ref{fig:vary-width}, that our results converge very rapidly to their final value as a function of the $t_{\text{width}}$ region over with the 3-pt function is integrated to obtain $\tilde{C}_{\mu\nu}$. Here we provide another test that significantly reduces the amount of computation needed. 

Figure~\ref{fig:Ttest} shows the results obtained if we restrict the fit of the two-point function $\tilde{C}_{\mu\nu}(t)$ to a set of specific $t\equiv t_{\eta_c}-t_{\gamma_2}$ values rather than fitting the full $t$ range. The left-hand point shows the full fit that we use here and the right-hand points show the results from selecting 4, 5 or 6 specific $t$ values. We need a mix of even and odd $t$ values for an optimal fit because of the oscillating terms from opposite parity states (see Eq.~\eqref{corrfitform_2pt}). We must also adjust the separation in lattice units between $t$ values as the lattice spacing changes. On the very-coarse lattices a separation of 1 is appropriate, whereas for coarse and fine lattices a separation of 3 or 5 gives a better range of $t$ to reproduce the full fit. We conclude from Fig.~\ref{fig:Ttest} that we could obtain similar uncertainties to our full fit by calculating correlation functions at only 4 or 5 time separations $t_{\eta_c}-t_{\gamma_2}$ if these were well-chosen. This is possible because lattice data is correlated as a function of $t$. Usually lattice two-point functions are calculated at all $t$ values because there is no significant time saving in making a $t$ selection. Here, because we actually calculate a three-point function, working with selected $t$ values reduces the computational cost considerably and we will make use of this in future. Here, however, we use the results from our full fit. 

\bibliography{bib}{}
\bibliographystyle{apsrev4-1}

\end{document}